\documentclass[twocolumn]{aastex631}
\usepackage{natbib}
\usepackage{url}
\usepackage{color}
\usepackage{mathtools}
\usepackage{amsmath}
\usepackage{tablefootnote}
\definecolor{red}{rgb}{1,0,0}
\definecolor{orange}{rgb}{1,0.5,0}
\definecolor{green}{rgb}{0,127,0}
\definecolor{grey}{rgb}{0.6627,0.6627,0.6627}
\definecolor{skyblue}{rgb}{0.53,0.808,0.98}

\newcommand{\lya}{Ly$\alpha$}

\newcommand{\ha}{H$\alpha$}
\newcommand{\hb}{H$\beta$}

\newcommand{\oii}{[\ion{O}{2}]}

\newcommand{\oiii}{[\ion{O}{3}]}

\newcommand{\se}{\texttt{Source Extractor}}
\newcommand{\galfit}{\texttt{\textsc{Galfit}}}
\newcommand{\eazy}{\texttt{EAZY}}
\newcommand{\vband}{$V_{606}$}
\newcommand{\iaband}{$I_{600}$}
\newcommand{\ibband}{$I_{814}$}
\newcommand{\yaband}{$Y_{098}$}
\newcommand{\ybband}{$Y_{105}$}
\newcommand{\jbband}{$J_{110}$}
\newcommand{\jband}{$J_{125}$}
\newcommand{\jhband}{$JH_{140}$}
\newcommand{\hband}{$H_{160}$}

\newcommand{\hst}{\textit{HST}}
\newcommand{\jwst}{\textit{JWST}}
\newcommand{\spitzer}{\textit{Spitzer}}
\newcommand{\borg}{BoRG[$z$8]}
\newcommand{\hippies}{HIPPIES GO 12286}
\newcommand{\rhalf}{$r_{1/2}$}

\newcommand{\sersic}{S\'{e}rsic}
\newcommand{\zsamp}{$z\sim9-10$}

\defcitealias{bernard2016}{B16}

\newcommand\myvdots{\vbox{\baselineskip=1pt \lineskiplimit=0pt \kern10pt \hbox{.}\hbox{.}\hbox{.}}} 

\shorttitle{Bright $z\sim9$ Galaxies}
\shortauthors{Bagley et al.}

\begin{document}

\title{Bright $z\sim9$ Galaxies in Parallel: The Bright End of the 
Rest-UV Luminosity Function from \hst\ Parallel Programs}

\correspondingauthor{Micaela B. Bagley}
\email{mbagley@utexas.edu}

\author[0000-0002-9921-9218]{Micaela B. Bagley}
\affil{Department of Astronomy, The University of Texas at Austin, 
Austin, TX, USA}

\author[0000-0001-8519-1130]{Steven L. Finkelstein}
\affil{Department of Astronomy, The University of Texas at Austin, 
Austin, TX, USA}

\author[0000-0003-2349-9310]{Sof\'ia Rojas-Ruiz}\altaffiliation{Fellow of the International Max Planck Research School for Astronomy and Cosmic Physics at the University of Heidelberg (IMPRS--HD)}
\affiliation{Max-Planck-Institut f\"{u}r Astronomie, K\"{o}nigstuhl 17, D-69117, Heidelberg, Germany}

\author{James Diekmann}
\affil{Department of Astronomy, The University of Texas at Austin, 
Austin, TX, USA}

\author[0000-0003-0792-5877]{Keely D. Finkelstein}
\affil{Department of Astronomy, The University of Texas at Austin, 
Austin, TX, USA}

\author[0000-0002-8442-3128]{Mimi Song}
\affil{University of Massachusetts, Amherst, MA, USA}

\author[0000-0001-7503-8482]{Casey Papovich}
\affil{George P. and Cynthia Woods Mitchell Institute for Fundamental 
Physics and Astronomy, Department of Physics and Astronomy, Texas A\&M 
University, College Station, TX, USA}

\author{Rachel S. Somerville}
\affil{Center for Computational Astrophysics, Flatiron Institute, NY, USA}

\author[0000-0003-0556-2929]{Ivano Baronchelli}
\affil{INAF- Istituto di Radioastronomia - Italian ALMA Regional Centre, via Gobetti 101, 40129 Bologna, Italy}

\author[0000-0002-7928-416X]{Y.Sophia Dai}
\affiliation{Chinese Academy of Sciences South America Center for Astronomy (CASSACA), National Astronomical Observatories of China (NAOC), 20A Datun Road, Beijing, 100012, China}

%%%%%%%%%%%%%%%%%%%%%%%%%%%%%%%%%%%%%%%%
\begin{abstract}
The abundance of bright galaxies at $z>8$ can provide key constraints on 
models of galaxy formation and evolution, as the predicted abundance varies 
greatly when different physical prescriptions for gas cooling and star 
formation are implemented. 
We present the results of a search for bright \zsamp\ galaxies selected 
from pure-parallel \textit{Hubble Space Telescope} imaging programs. We include 
132 fields observed as part of the Brightest of Reionizing Galaxies survey,
the Hubble Infrared Pure Parallel Imaging Extragalactic Survey, and the 
WFC3 Infrared Spectroscopic Parallel survey. These observations cover 
a total of 620 arcmin$^2$, about 70\% of which is also covered with 
\textit{Spitzer Space Telescope} infrared imaging.
We identify thirteen candidate galaxies in the range $8.3 < z < 11$
with $24.5 < m_{H} < 26.5$ ($-22.9 < M_{UV} < -21.2$). This sample 
capitalizes on the uncorrelated nature of pure parallel observations to 
overcome cosmic variance and leverages a full multi-wavelength selection 
process to minimize contamination without sacrificing completeness. 
We perform detailed completeness and contamination analyses, and present 
measurements of the bright end of the UV luminosity function using a 
pseudo-binning technique.
We find a number density consistent with results from \citet{finkelstein2022a}
and other searches in \hst\ parallel fields. These bright candidates 
likely reside in overdensities, potentially representing some 
of the earliest sites of cosmic reionization. These new candidates are 
excellent targets for follow-up with \jwst, and four 
of them will be observed with the NIRSpec prism in Cycle 1. 
\end{abstract}

%\keywords{High-redshift galaxies (734) --- Early universe (435) ---
%Galaxy evolution (594) --- Lyman-break galaxies (979) --- 
%Luminosity function (942)}

%%%%%%%%%%%%%%%%%%%%%%%%%%%%%%%%%%%%%%%%%%%%%%%%%%%%%%%%%%%%%%%%%%%%%%%
\section{Introduction} \label{sec:intro}

The study of galaxies in the very early universe, and particularly of
UV-bright galaxies, provides key input to models of galaxy formation.
The number density of UV-bright galaxies is affected by factors including 
feedback processes \citep[e.g., ][]{somerville2008,bower2012},
dust attenuation \citep[e.g., ][]{finkelstein2012b,vogelsberger2020a},
the build-up of dark matter halos, and star formation efficiency.
Constraining the abundance of UV-luminous galaxies thus directly constrains 
the fundamental physics of star formation in the early Universe. By probing
ever-larger volumes, recent ground and space-based surveys are providing a 
progressively more complete analysis of the abundance of bright galaxies.
These studies are showing that the characteristic luminosity ($L^*$) of the 
rest-frame UV luminosity function only shallowly evolves (if at all) in the 
range $z=4-8$
\citep{bowler2014,bowler2015,bowler2020,bouwens2015a,finkelstein2015a}.
As the UV luminosity function probes recent star formation, these results
imply that star-formation rates evolve more slowly than the halo mass 
function, which would indicate that galaxies are efficient 
at converting gas into stars even out to these redshifts
\citep[e.g., ][though see \citealt{stefanon2021b}]{behroozi2015,finkelstein2015b,yung2019a,yung2019b}. 
If this observational trend continues to higher redshifts, it will provide a 
significant challenge to current galaxy formation models.

Considerable effort has been devoted to modeling galaxy evolution at early 
cosmic times ($z\gtrsim6$), via methods such as hydrodynamical 
simulations \citep[e.g.,][]{gnedin2016,wilkins2017}, (semi-)analytic models 
\citep[e.g.,][]{somerville2015b,yung2019a,yung2019b}, and empirical models
\citep[e.g.,][]{mason2015b,behroozi2020}. Yet predictions for the density of 
galaxies at $z>9$ can differ dramatically. As the predicted number of 
UV-bright galaxies is sensitive to the assumed relationship between 
molecular-gas density and the star-formation-rate density in cosmological 
simulations, robust observations in this epoch will provide powerful 
constraints on these physical processes. Analyses at $z>8$ continue to push 
the limits of what is possible with the current data. The 
majority of candidates at these redshifts 
\citep[e.g.,][]{coe2013,oesch2013,oesch2014,bouwens2015a,bouwens2016a,bouwens2019a,finkelstein2016} were selected from a small set of 
well-studied fields such as the Cosmic Assembly Near-infrared Deep 
Extragalactic Legacy Survey \citep[CANDELS;][]{grogin2011,koekemoer2011}, 
the Hubble Ultra Deep Field \citep[HUDF;][]{beckwith2006}, and the 
Hubble Frontier Fields \citep{lotz2017}.
Results measured from a few independent pointings are susceptible to 
cosmic variance, which can dominate the uncertainties at the bright end of 
the UV luminosity function.
Thus, the density of UV-bright galaxies is presently ill-constrained 
at $z> 8$, despite its importance for understanding high-redshift star 
formation and galaxy evolution.

Pure parallel programs -- where random field pointings result in completely 
independent, uncorrelated observations -- provide a key opportunity to address 
these outstanding questions. Often shallower than targeted surveys, pure 
parallel programs are well-suited to identifying bright high-redshift 
galaxies: a population of rare sources that may not be fully sampled by 
surveys covering the same area in contiguous fields. Parallel observing can 
therefore provide better constraints on the bright end of the UV luminosity 
function and is complementary to deeper surveys that probe the fainter 
population.

Pure parallel programs such as the Brightest of Reionizing Galaxies 
(BoRG, PI: Trenti; General Observer [GO] programs 11700, 12572) and the Hubble Infrared Pure Parallel 
Imaging Extragalactic Survey (HIPPIES, PI: Yan; GO 11702, 12286) were indeed 
designed for just this purpose. Using WFC3 imaging in at least four filters to 
probe the Lyman break at $z\sim8$, these surveys drastically increased the 
known sample of bright galaxies and provided crucial constraints on the 
luminosity function \citep{trenti2011,trenti2012,bradley2012,schmidt2014}. 
The independent pointings significantly reduce the effect of cosmic variance 
from $\gtrsim$20\% to $<$1\% \citep{trenti2008,bradley2012}, providing 
comparable constraining power as a similar-depth survey with twice the area 
but in a single pointing \citep{bradley2012}. Studies using the BoRG[z910] 
fields, which included \jhband-band imaging, find a relatively high number 
density of bright $z\sim9-10$ galaxies
\citep{calvi2016,morishita2018,rojas-ruiz2020}, potentially indicating more 
efficient star formation at these early times 
\citep[similar results have been seen in the CANDELS fields;][]{finkelstein2022a}.
Yet significant scatter also remains, demonstrating the systematic 
uncertainty resulting from different sample 
selection techniques (i.e., strict color cuts versus photometric redshift 
selection). While \jwst\ will quickly shed light on this tension, it is not 
optimized to conduct wide-field surveys for new bright, high-redshift galaxies.
The power of \jwst\ will lie in targeted follow-up for redshift confirmation 
and source characterization.

To that end, we have performed a new independent selection for $z = 9-10$ 
galaxies in a subset of available \hst\ pure parallel fields, improving on 
previous efforts by 
including all available \hst\ and \textit{Spitzer}/IRAC photometry in our 
photometric-redshift selection. We also included in our search 45 fields ($\sim$160 arcmin$^2$) from the WFC3 Infrared Spectroscopic Parallel survey (WISP, PI: Malkan; GO 12283, 12902, 13352, 13517, 14178), the first time these fields have been searched for $z\sim9$ candidates.
While many have performed searches for $z>9$ galaxies in the BoRG/HIPPIES 
fields \citep{bernard2016,calvi2016,morishita2018,bridge2019,rojas-ruiz2020,morishita2020a,morishita2021a,roberts-borsani2021b}, the results at the bright 
end of the UV luminosity function remain uncertain due in part to varied 
selection techniques probing close to the image detection limits. In fact, 
these teams find different samples of galaxy candidates \emph{in the same 
fields}, highlighting how sensitive \hst\ results at $z\gtrsim9$ are to 
photometric measurements, noise characterizations, and sample selections. 
In our search, we employ a full multi-wavelength selection (independent of 
a simple \jband$-$\hband\ color cut) including \spitzer/IRAC photometry where 
available, with detailed noise calculations and source vetting to remove 
contaminants -- resulting in a larger, more complete sample. This sample 
includes some of the brightest candidates at these redshifts, potentially 
representing some of the most massive galaxies to form $\lesssim$500 Myr 
after the Big Bang.

This paper is organized as follows. 
In Section~\ref{sec:obs}, we present the pure parallel imaging datasets and
describe our photometric measurements. 
We describe our sample selection criteria in 
Section~\ref{sec:sample}, including a detailed discussion of how we vetted 
each candidate, performing a thorough check for cases of 
persistence (Section~\ref{sec:persistence}) and a careful visual inspection 
(Section~\ref{sec:inspection}). We incorporate available 
\spitzer/IRAC imaging of each candidate in Section~\ref{sec:irac}, and check 
for stellar contamination in Section~\ref{sec:spex}. 
In Section~\ref{sec:results}, we present our 13 high-redshift candidate 
galaxies and compare them to previous studies in the same \hst\ fields.
We address the possibility of low-redshift contamination in 
Section~\ref{sec:contam}, and use our sample to infer the density of galaxies 
at the bright end of the UV luminosity function in Section~\ref{sec:lf}.
Finally, we discuss implications for our sample of bright, high-redshift 
galaxies in Section~\ref{sec:discussion} and summarize  in 
Section~\ref{sec:summary}.
Throughout this paper we assume a $\Lambda$CDM cosmology with 
$\Omega_M = 0.3$, $\Omega_{\Lambda}=0.7$, and $H_0 = 70$~km~s$^{-1}$~Mpc$^{-1}$
and express all magnitudes in the AB system \citep{oke1983} 
unless otherwise noted. 

%%%%%%%%%%%%%%%%%%%%%%%%%%%%%%%%%%%%%%%%%%%%%%%%%%%%%%%%%%%%%%%%%%%%%%%
\section{Observations} \label{sec:obs}
We include observations from multiple programs, all obtained 
as part of parallel imaging programs with 
WFC3\footnote{\url{www.stsci.edu/hst/wfc3}} \citep{kimble2008} while another 
\hst\ instrument was in use. 
Typically these parallel observations are taken while either the Cosmic 
Origins Spectrograph \citep[COS;][]{froning2009} or the Space Telescope 
Imager and Spectrograph \citep[STIS;][]{kimble1998} are engaged in long 
integrations of a primary target. 
Without control of the telescope pointing, pure parallel 
programs (unaffiliated with the primary programs) can accrue observations 
of independent and uncorrelated fields during these long primary integrations. 
In this paper, we combine data from a subset of the Brightest of Reionizing 
Galaxies
survey \citep[BoRG;][]{trenti2011}, the Hubble Infrared Pure Parallel Imaging 
Extragalactic Survey \citep[HIPPIES;][]{yan2011}, the WFC3 Infrared 
Spectroscopic Parallel survey \citep[WISP;][]{atek2010}, and additional 
coordinated parallels from the COS Guaranteed Time Observer (GTO) program.

The programs include imaging with both the UVIS and IR cameras on WFC3, 
as well as some optical imaging with the Advanced Camera for Surveys
\citep[ACS\footnote{\url{www.stsci.edu/hst/acs}};][]{clampin2000} for HIPPIES 
observations from Cycle 18. 
All fields are observed with the F160W filter, which we use for detection,
and three additional filters covering wavelengths from $\sim$$0.6-1.4\micron$. 
True $z\gtrsim9$ galaxies will be detected in F160W and at most 
one additional filter (either F125W or F110W depending on the survey in 
question), and the optical imaging is crucial in detecting 
the Lyman-$\alpha$ (\lya) break at 1216\AA.
By $z$$\sim$10 F160W is the only
filter available in these programs that is redward of the Lyman break.
The WFC3/IR camera has a field of view ($123\arcsec \times 134\arcsec$) that
is smaller than both of the optical cameras (WFC3/UVIS: 
$162\arcsec \times 162\arcsec$, and ACS/WFC: $202\arcsec \times 202\arcsec$) 
and therefore determines the total area covered in each field.
The native pixel scales of the cameras are $0\farcs13$/pixel (IR),
$0\farcs 04$/pixel (UVIS), and $0\farcs05$/pixel (WFC). 
The filter coverage and depth achieved in each field depends on the survey 
strategies of each program and the length of each parallel opportunity. 
The fields we assemble here therefore have inconsistent depths and filter 
coverages, which we handle with careful field-specific simulations to 
assess the completeness and effective volume probed in each field.

\begin{deluxetable}{c|ccc}
\centering
\tablecaption{Dataset Summary \label{tab:surveys}}
\tablehead{
\colhead{Dataset} & \colhead{$N_{\mathrm{fields}}$} 
& \colhead{$A_{\mathrm{total}}$}
& \colhead{Pixel scale} \\
\colhead{} & \colhead{} & \colhead{(arcmin$^2$)} & \colhead{(arcsec/pixel)}
}
\startdata
\borg\ & 64 & 308.8 & 0.08 \\
HIPPIES (GO 12286) & 23 & 108.3 & 0.10 \\
WISP & 45 & 203.9 & 0.13
\enddata
\end{deluxetable}

\begin{figure}
\epsscale{1.2}
\plotone{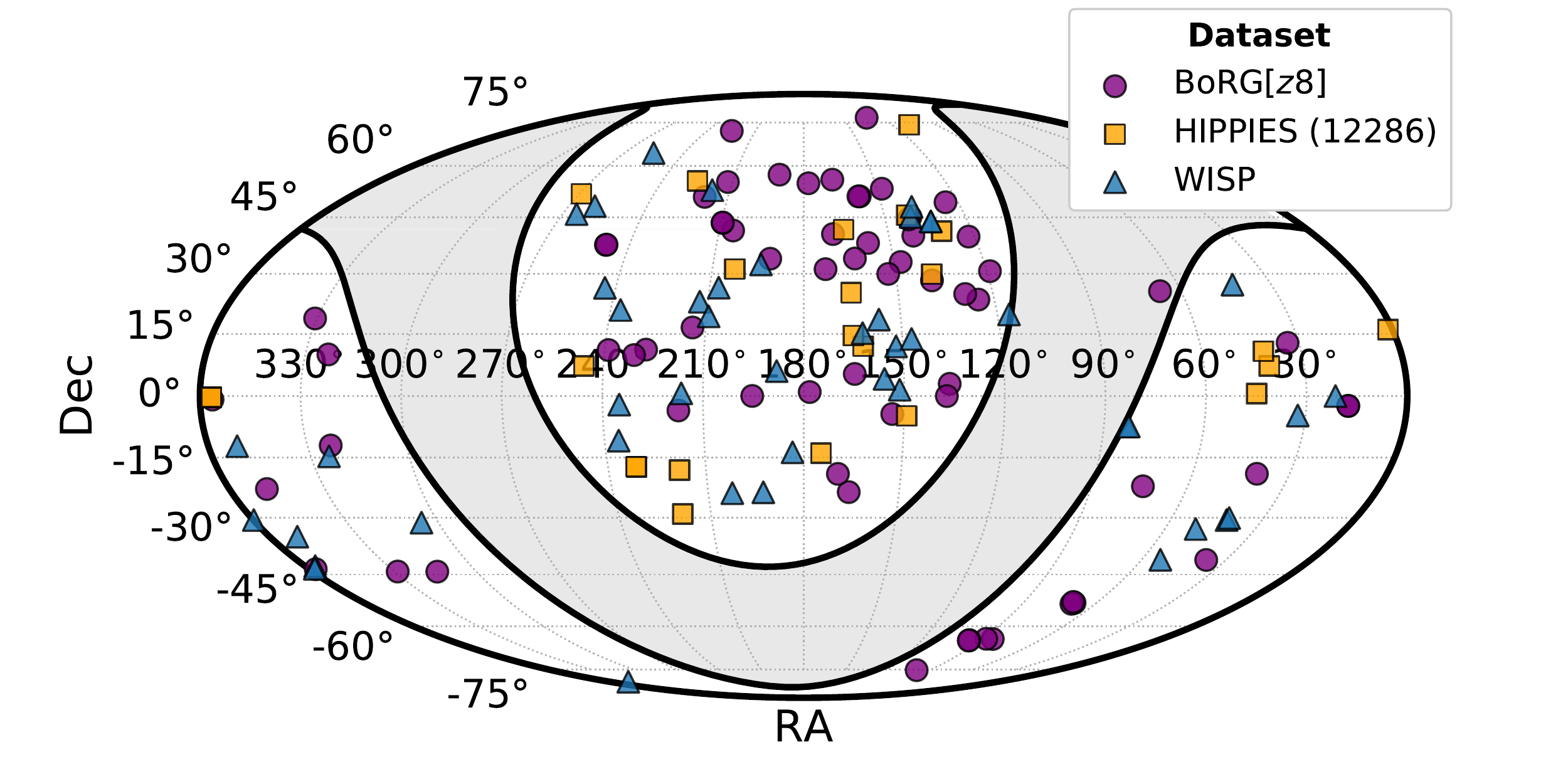}
\caption{The positions of all 132 fields we consider in our analysis. 
Purple circles indicate \borg\ fields, which include 34 fields from the 
BoRG survey (PI: Trenti), 8 HIPPIES fields (PI: Yan,GO 11702), and 22 
parallel fields observed as part of the COS GTO program (PI: Green). 
The 23 HIPPIES fields from GO 12286 are shown as orange squares, and the 
45 WISP fields are shown as blue triangles. (Symbol sizes are not to scale.)
The independent, uncorrelated nature of the fields -- ideal for studies of 
rare, bright galaxies -- is evident from their locations shown here.  
\label{fig:fields}}
\end{figure}

In total, we consider 132 WFC3 parallel fields covering 
$\sim$620 arcmin$^2$: 
64 \borg\ fields (comprised of 34 fields from the BoRG survey, 
8 HIPPIES fields from GO program 11702, 
and 22 parallel pointings from the COS GTO program),
23 additional HIPPIES fields (GO 12286), 
and 45 fields from the WISP survey. 
The locations of all 132 fields are shown in Figure~\ref{fig:fields}, 
where the independent nature of the pointings is clear. 
We provide a summary of these datasets in Table~\ref{tab:surveys} and 
field-specific information about all 132 fields, including filter 
coverage and imaging depths, in Table~\ref{tab:obs}. 
We measure the area covered in each pointing by summing the pixels in the 
F160W weight maps with values greater than 100 counts/sec
($=\mathrm{RMS}<0.1$), as this 
allows us to account for the area lost to cosmic rays, detector artifacts, 
or other bad pixels. 
The imaging depths presented in Table~\ref{tab:obs} are the $5\sigma$ 
limiting magnitudes as measured in circular apertures with radius
$r=0\farcs2$ via the following method. 
We place $\sim$5000$-$10000 uncorrelated apertures randomly 
across the background of each image, avoiding source flux and bad pixels. 
The limiting magnitude quoted in each filter is found by measuring the 
standard deviation of the distribution of fluxes from all apertures, and thus 
represents an average depth across the full image. We note that we have 
not performed an aperture correction to account for the fractional flux 
enclosed by the $r=0\farcs2$ apertures.
The measured depth can 
vary across a field by up to $\pm0.3$ magnitudes from the median, and so
we rely on the noise measured locally around each source 
(see Section~\ref{sec:noise}) to determine detection significances 
that we use when selecting high-redshift candidate galaxies 
(see Section~\ref{sec:selection}).

We briefly explain the characteristics and data reduction of each dataset 
in the following sections. We hereafter refer to the \hst\ filters 
using a letter that specifies the bandpass followed by the three-number 
identification: i.e., \iaband\ for the long pass filter F600LP, 
\hband\ for F160W.

%\startlongtable
\begin{deluxetable*}{c|ccccccccccccc}
\centering
%\tablewidth{0pt}
\tablecaption{Parallel Field Coverage and Depths \label{tab:obs}}
\tabletypesize{\scriptsize}
\tablehead{
\colhead{Field} & \colhead{RA} & \colhead{Dec} & 
\colhead{\vband\tablenotemark{a}} &
\colhead{\iaband} & \colhead{\ibband} & \colhead{\yaband} &
\colhead{\ybband} & \colhead{\jbband} & \colhead{\jband} &
\colhead{\hband} & \colhead{PI} & \colhead{Program} & 
\colhead{Area\tablenotemark{b}}  \\
\colhead{} & \colhead{} & \colhead{} & \colhead{(mag)} &
\colhead{(mag)} & \colhead{(mag)} & \colhead{(mag)} &
\colhead{(mag)} & \colhead{(mag)} & \colhead{(mag)} &
\colhead{(mag)} & \colhead{} & \colhead{} & 
\colhead{(arcmin$^2$)} 
}
\startdata
Par0110$-$0224a\tablenotemark{c} & 01:10:04.4 & $-$02:25:04.6 & 27.54 & 27.09 & \nodata & 27.99 & \nodata & \nodata & 27.74 & 27.42 & Trenti & 11700 & 6.45 \\
Par0110$-$0224b & 01:10:04.4 & $-$02:25:04.6 & 27.51 & \nodata & \nodata & 27.60 & \nodata & \nodata & 27.55 & 27.18 & Trenti & 11700 & 8.22 \\
Par0214+1255 & 02:13:37.8 & $+$12:54:51.6 & \nodata & 26.09 & \nodata & 26.70 & \nodata & \nodata & 26.64 & 26.49 & Yan & 11702 & 4.62 \\
Par0228$-$4102 & 02:27:56.0 & $-$41:01:34.8 & \nodata & 27.16 & \nodata & 27.52 & \nodata & \nodata & 27.39 & 27.16 & Green & 11541 & 4.64 \\
Par0240$-$1857 & 02:40:26.9 & $-$18:57:19.6 & \nodata & 27.05 & \nodata & 27.51 & \nodata & \nodata & 27.38 & 27.13 & Green & 11541 & 4.63 \\
\hline
Par0005+1607 & 00:05:39.5 & $+$16:07:05.3 & \nodata & 27.06 & \nodata & \nodata & 27.17 & \nodata & 27.47 & 27.05 & Yan & 12286 & 4.62 \\
Par0241+0715 & 02:41:48.3 & $+$07:15:35.5 & \nodata & 26.76 & \nodata & \nodata & 27.16 & \nodata & 27.21 & 26.88 & Yan & 12286 & 4.62 \\
Par0245+1051 & 02:45:25.7 & $+$10:51:57.9 & \nodata & 26.80 & \nodata & \nodata & 27.34 & \nodata & 27.29 & 26.91 & Yan & 12286 & 4.62 \\
Par0259+0032 & 02:59:33.4 & $+$00:32:13.4 & \nodata & 26.71 & \nodata & \nodata & 27.08 & \nodata & 27.11 & 26.50 & Yan & 12286 & 4.60 \\
Par0713+7405 & 07:13:22.2 & $+$74:06:03.3 & \nodata & 27.06 & \nodata & \nodata & 27.03 & \nodata & 27.33 & 27.18 & Yan & 12286 & 2.89 \\
\hline
Par96 & 02:09:24.3 & $-$04:43:50.2 & 27.48 & \nodata & 26.88 & \nodata & \nodata & 27.94 & \nodata & 26.97 & Malkan & 12283 & 4.59 \\
Par104 & 10:05:25.1 & $+$01:29:43.5 & 27.02 & \nodata & 26.44 & \nodata & \nodata & 27.78 & \nodata & 26.73 & Malkan & 12283 & 4.58 \\
Par257 & 02:46:37.0 & $-$30:32:05.5 & 26.90 & \nodata & 26.47 & \nodata & \nodata & 27.51 & \nodata & 26.42 & Malkan & 12902 & 4.53 \\
Par296 & 10:27:11.4 & $+$18:35:31.1 & 27.04 & \nodata & 26.63 & \nodata & \nodata & 27.19 & \nodata & 26.40 & Malkan & 12902 & 4.54 \\
Par298 & 09:21:30.5 & $+$45:07:9.5 & 26.97 & \nodata & 26.44 & \nodata & \nodata & 26.99 & \nodata & 26.24 & Malkan & 12902 & 4.26 \\
\enddata
\tablenotetext{a}{The 5$\sigma$ limiting magnitudes are measured for each 
filter in circular apertures with $r=0\farcs2$.}
\tablenotetext{b}{The area in arcmin$^2$ is measured as \hband\ pixels with 
weights of $>100$  (RMS $<0.1$).}
\tablenotetext{c}{For a few of the \borg\ and HIPPIES fields, two 
overlapping \hst\ pointings have been combined to form a single field, 
resulting in varied filter coverage and depths. In these cases, we have 
separated the field into two components, \textit{a} and \textit{b}, and 
consider them separately in our analysis. See the text for details. }
\tablecomments{
Five representative fields from each dataset are presented here.
This table is available in its entirety in machine-readable 
format in the electronic version of the paper.\\
}
\end{deluxetable*}

\begin{figure}
\epsscale{1.2}
\plotone{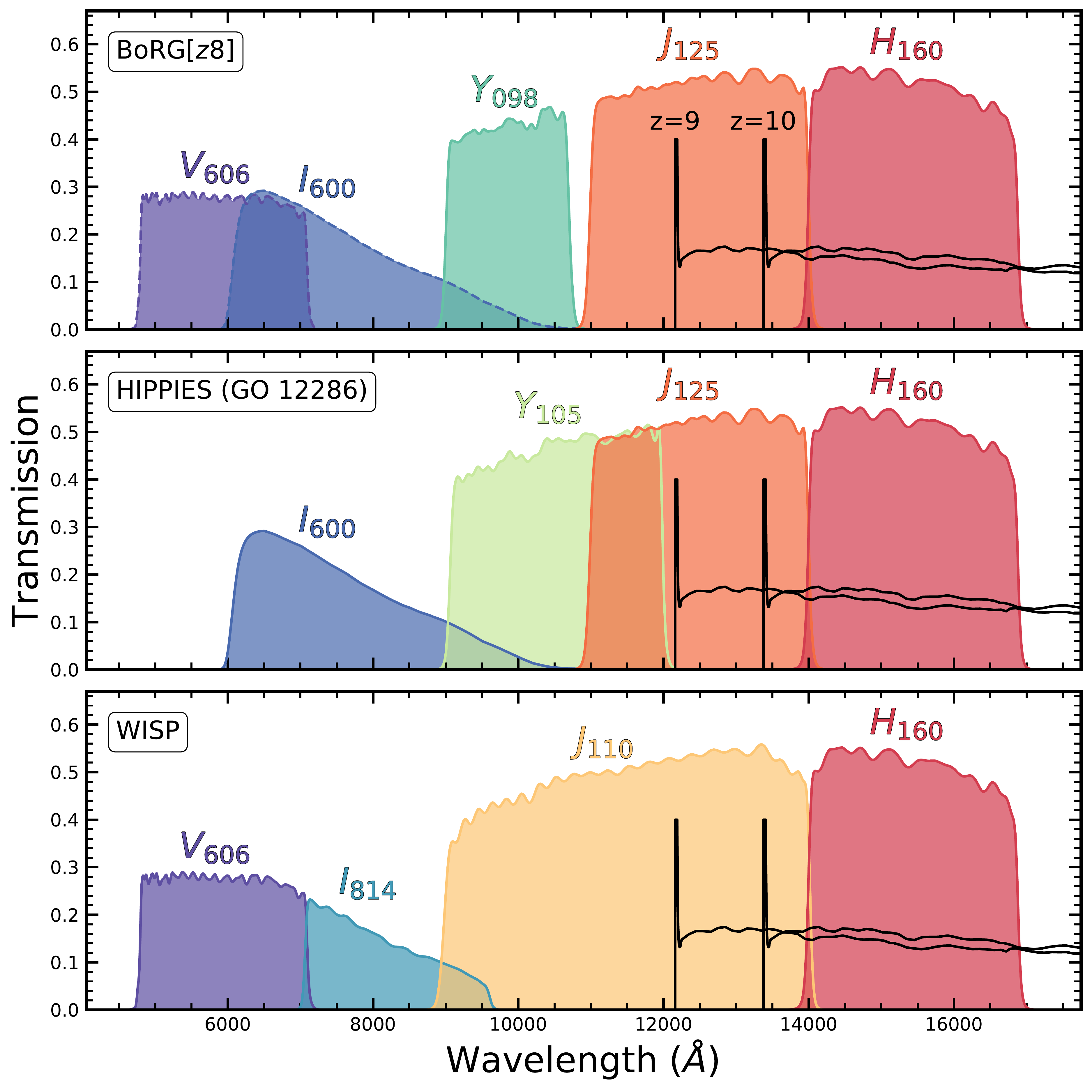}
\caption{The filter coverage of each dataset, showing that of \borg\ (top
panel), HIPPIES GO 12286 (middle), and WISP (bottom). 
The \vband\ and \iaband\ profiles in the top row have dashed outlines to 
indicate that the \borg\ fields are observed with one or the other of these
filters. 
In place of \jband, the WISP survey observed with the broader \jbband,
making WISP fields less sensitive to galaxies at $z\sim9$ than at $z\sim10$.
For reference, we show a model galaxy spectrum at $z=9$ and $z=10$ with 
a weak \lya\ emission line. At $z=9$, the Lyman break is 
near the center of the \jband, while by $z>10$ it has redshifted almost
completely out of the filter, resulting in single-band \hband\ detections.
\label{fig:filters}}
\end{figure}

%%%%%%%%%%%%%%%%%%%%%%%%%%%%%%%%%%%%%%%%
\subsection{\borg} \label{sec:borg}
We include in our analysis data from the first set of fields observed as part
of the BoRG survey \citep[PI: Trenti;][]{trenti2011}. These fields were 
observed in \hst\ Cycles 17 and 19 (GO 11700, 12572) with the \yaband, \jband,
and \hband\ IR filters and the \vband\ UVIS filter (see top row of 
Figure~\ref{fig:filters}).
The BoRG data releases also include data from programs with 
similar observing strategies and filter coverages -- HIPPIES 
\citep[PI: Yan, GO 11702;][]{yan2011} and coordinated parallels from 
the COS GTO team (PI: Green, GO 11528, 11530, 11534, 11541, 12024, 12025) --
though the HIPPIES fields as well as some of the COS parallels were observed
with the \iaband\ filter instead of \vband.
Collectively, these three programs cover 71 fields, with data and results 
described in detail in \citet{trenti2011}, \citet{bradley2012}, and 
\citet{schmidt2014}. These surveys were optimized to detect $z$$\sim$8 
galaxies as \yaband-dropouts with detections in both \jband\ and \hband\ and 
are therefore referred to collectively as \borg. This designation distinguishes
the first set of BoRG observations with those from later \hst\ cycles that 
included the \jhband\ filter.
Imaging with \jhband\ allows for the detection of 
$z\gtrsim9$ galaxies as \jband-dropouts with two-band detections redward of 
the break. These BoRG[$z$910] fields have been used to identify
$z\gtrsim9$ candidates in multiple studies 
\citep{calvi2016,morishita2018,morishita2020a,morishita2021a,bridge2019,rojas-ruiz2020,roberts-borsani2021b}. 
In this paper we focus on the \borg\ fields, and perform an analysis 
similar to that presented in \citet{rojas-ruiz2020}. 

We use the reduced images and weight maps available from the BoRG data 
Delivery 3 from MAST\footnote{\url{https://archive.stsci.edu/prepds/borg}}.
These images are all pixel-aligned and on a scale of 0\farcs08/pixel. 
We convert the weight maps (actually inverse variance maps output by 
the WFC3 pipeline, CALWFC3) to RMS maps for use with source detection via 
$\mathrm{RMS} = 1/\sqrt{\mathrm{WHT}}$. Pixels with zero weight are assigned 
a value of 10$^4$ in the RMS maps.

After visually inspecting all 71 \borg\ fields, we chose to remove seven fields 
from consideration in our analysis. Three of the fields -- Par0540-6409, 
Par0553-6405, and Par1815-3244 -- appear to cover star clusters, resulting 
in a very high density of point sources across the full image. 
Two others -- Par1014-0423 and Par1230+0750 -- are heavily 
contaminated by strong detector persistence, likely resulting from 
observations of bright star clusters immediately preceeding the \borg\ 
exposures. We remove the remaining two -- Par0751+2917 and Par1209+4543 -- 
because they were also observed in Cycle 22 with additional filters (F350LP 
and \jhband) and so are part of the BoRG[$z$910] survey.
One field, Par0110-0224, was covered by both the BoRG and HIPPIES GO 11702 
programs and is therefore partially covered by five filters (including both 
\vband\ and \iaband). We separated this field into two parts -- one with and 
one without \iaband\ -- which are labeled $a$ and $b$, respectively, in 
Table~\ref{tab:obs}. 
We count this as a single field in the total (of the 132), but treat 
them as separate fields in all other aspects of our analysis. Specifically, 
we perform separate completeness simulations to account for the varying filter 
coverage and depth across the field. 
Finally, a few fields are partially affected by significant 
detector issues or other contamination, including scattered light from 
Earth's limb, highly variable background across part of the images, and
extremely strong Dragon's Breath \citep{fowler2017} in the UVIS filters. 
In these cases, we mask out the affected regions in all filters and weight 
maps, therefore accounting for the corresponding loss in total survey area. 
The field areas listed in Table~\ref{tab:obs} reflect the unaffected, 
unmasked portions of these fields.
In total, we include 64 \borg\ fields in our analysis, covering 
$\sim$308 arcmin$^2$. 

%%%%%%%%%%%%%%%%%%%%%%%%%%%%%%%%%%%%%%%%
\subsection{HIPPIES} \label{sec:yan}
The HIPPIES observations from Cycle 17 were reduced and released as part of 
\borg\ (see previous section), due to the similar filter sets and 
observing strategies of the two programs. 
However, HIPPIES observations from Cycle 18 (PI: Yan, GO 12286) were not 
included, and so we count them here as a separate dataset. This program 
used the near-IR filter \ybband\ rather than the medium band \yaband, and also 
used \iaband\ as the bluest filter. 

We downloaded the FLTs and supporting files for this dataset from MAST and 
created reduced mosaics following the same procedure outlined 
in \citet{rojas-ruiz2020}, which we briefly summarize here. 
The final, drizzled images are produced using a pipeline custom built 
for use with HIPPIES data, with the modifications necessary to account for 
the added challenges of undithered, pure parallel observations. 
The pipeline uses an MCMC sampler to align individual images based on 
detected source positions. It then creates mosaic images with a pixel 
scale of 0\farcs1/pixel using the MultiDrizzle software 
package \citep{koekemoer2003}. The output weight maps are then 
scaled by the average amplitude of the correlation measured in blank regions 
of the science images in order to produce RMS maps. 

We rejected a handful of pointings that are severely impacted by features 
such as scattered Earth light, and we masked portions of two others. 
We also separated field Par2346-0021 into two parts due to inconsistent 
filter coverage and imaging depth across the field, similar to that described 
for Par0110-0224 for the \borg\ dataset.  
We include 23 fields ($\sim$108 arcmin$^2$) from the HIPPIES 
GO program 12286.

%%%%%%%%%%%%%%%%%%%%%%%%%%%%%%%%%%%%%%%%
\subsection{WISP} \label{sec:wisp}
Similar to the programs described above, WISP \citep[PI Malkan;][]{atek2010} 
obtained imaging of uncorrelated fields through pure parallel observations. 
However, WISP also acquired slitless spectroscopy using WFC3's two 
near-infrared grisms: G102 and G141. Primarily a spectroscopic program, WISP 
devoted the majority of available integration time to grism observations and 
obtained imaging mainly to aid in extracting and calibrating spectra from the 
slitless grism images. Yet for long parallel opportunities consisting of 
$\geq$5 orbits, WISP pointings were observed with four imaging filters along 
with both grisms. We include 45 of these ``deep'' WISP pointings ($\sim$204 
arcmin$^2$) in our analysis, representing the first search for $z>8$ galaxies 
in WISP observations. 
The WISP pointings are observed with two IR filters (\jbband\ and \hband) and 
the UVIS filter \ibband. Most of the fields are also observed with 
the UVIS filter \vband.

The main difference between the WISP observations we include here and those 
of the other datasets is the use of the \jbband\ filter rather than 
\jband. The considerable width of the \jbband\ filter 
(FWHM$\sim$5000\AA, compared to $\sim$3000\AA\ for \jband) makes it 
challenging to identify sources at $z\sim8-9$ via the dropout technique. 
Galaxies at redshifts of $z\sim7 - 10$ will all be detected in \jbband,
and the resulting photometric redshift probability distributions are too 
broad to reliably distinguish galaxies at the high-redshift end of this 
interval from those at the lower end. 
We account for this selection effect with our field-specific completeness 
simulations. 
Additionally, all 45 WISP fields were observed at 3.6\micron\ with 
\spitzer/IRAC, which helps constrain the photometric redshifts 
(see Section~\ref{sec:irac}).
The detection power for WISP fields returns at $z\sim 10$, when sources 
drop out of the \jbband\ filter and become single-band \hband\ detections.

The WISP data reduction pipeline is presented by \citet{atek2010} and 
Battisti et al. (\textit{in preparation}), and uses the 
WFC3 pipeline CALWF3 with modifications required to address the specific 
challenges of parallel observing. We use the reduced data available through 
the MAST archive\footnote{\url{archive.stsci.edu/prepds/wisp/}} that were
produced with version 6.2 of the WISP pipeline. 
Pixel-aligned images in all filters are available on $0\farcs04$ and 
$0\farcs13$ pixel scales, and we adopt the $0\farcs13$ scale for this 
analysis because it better samples the point spread function -- with the 
undithered parallel data there are not enough exposures to cover the IR 
pixels at the finer resolution.

%%%%%%%%%%%%%%%%%%%%%%%%%%%%%%%%%%%%%%%%
\subsection{Source Extraction and Photometry} \label{sec:photometry}
We detect sources and perform photometry in all three datasets using \se\ 
\citep[v2.5;][]{bertin1996} 
with \hband\ as the detection filter. 
Our source detection parameters are the result of optimization using 5000 
simulated sources inserted randomly into the images from 21 fields. For these 
tests, we chose fields that cover the range of depths, filter coverages, and 
pixel scales present across all three datasets.
For each field, we created 10 realizations of the \hband\ images, adding
200 synthetic sources to each realization. Half of the synthetic sources 
were \sersic\ profiles with half-light radii (\rhalf) and \sersic\ indices 
($n$) pulled randomly from uniform distributions in the ranges 
$0\farcs05<r_{1/2}<0\farcs15$ and $0.5<n<4.5$, respectively. The other 
half were modeled as two-dimensional Gaussian profiles with 
$\sigma_x = \sigma_y$ pulled uniformly from the range 
$0\farcs05<\sigma<0\farcs1$. As we were looking to optimize detection 
parameters, the sources were placed randomly in clean 
areas of the images, i.e., avoiding bad pixels and source flux. 

We then ran \se\ multiple times on each realization, covering a grid of 
source detection parameter values. Specifically, we varied 
the threshold for detection over the range 
$0.3\sigma<$~\texttt{DETECT\_THRESH}~$<3\sigma$
in steps of $0.1\sigma$ and the minimum number of pixels above the threshold
required for detection over the range $3<$~\texttt{DETECT\_MINAREA}~$<10$ 
pixels.
We also inverted each image realization and ran the same grid of parameter 
values to quantify the number of negative pixel groupings that would be 
detected as real sources. In this way, we characterized the contamination 
rates of spurious sources that would be detected as contiguous clusters of 
noise for each set of detection parameters.
We explored the fraction of recovered simulated sources as well as the 
fraction of spurious detections as a function of \hband\ magnitude, choosing 
the parameter values that maximize the recovered fraction ($\geq$80\%)
while minimizing the spurious fraction ($<$5\%) down to $m_{H160} =27$:
\texttt{DETECT\_THRESH}~=~1.1$\sigma$ and \texttt{DETECT\_MINAREA}~=~4 pixels.
Finally, we found that the same set of detection parameters would optimize 
these fractions for all three datasets for $m_{H160}<27$. Any differences 
in the recovery and contamination rates for the datasets occur at 
fainter magnitudes than we will consider in this paper. 
Therefore, although the images are on 
different pixel scales, we use one consistent set of \se\ detection 
parameters for all fields. 

For each field, we create catalogs using \se\ in dual image mode with 
\hband\ as the detection filter and the RMS maps for the weights used in 
source detection and analysis. 
Photometry is performed in both circular and elliptical apertures, with 
each measurement serving a different purpose in our sample selection
(see Section~\ref{sec:selection}). 
We use small circular apertures with radius $r=0\farcs2$ to measure the 
detection significance of sources in all filters, because they capture the 
light in the core of sources while minimizing the contribution of 
off-source noise.
We use small elliptical \citet{kron1980}-like apertures to measure source 
colors, setting the \se\ parameter \texttt{PHOT\_AUTOPARAMS} to (1.2,1.7),
where the first number is the
scaling parameter for the ellipse and the second is the minimum radius below 
which elliptical apertures are converted to circular. 
As shown by \citet{finkelstein2010}, these choices are motivated by the need 
to maximize both the recovered flux and the signal-to-noise (S/N) for faint, 
compact galaxies. 
Additionally, the smaller aperture scaling parameter (here 1.2 compared to the 
\se\ default 2.5) reduces the number of small sources with elliptical 
apertures that are artifically stretched to larger sizes by bright 
neighbors in crowded regions.
In order to similarly determine the optimal parameter
choices for these parallel imaging datasets, we tested the recovered S/N of 
the simulated sources inserted into the 21 fields described above over a 
grid of \texttt{PHOT\_AUTOPARAMS}. The recovered S/N of sources in the \hband\ 
images peaks for parameter values in the range ($1-1.5$,$1.3-2$) with no 
clear winner, and so we adopt (1.2,1.7), following 
\citet{bouwens2010,bouwens2021a} and \citet{finkelstein2010,finkelstein2015a,finkelstein2022a}.
We derive aperture corrections for these small elliptical apertures in 
Section~\ref{sec:photcorrections}.

%%%%%%%%%%%%%%%%%%%%%%%%%%%%%%%%%%%%%%%%
\subsubsection{Photometric Errors} \label{sec:noise}
As the noise in the pixels of drizzled images is partially correlated, 
the total noise in an aperture is larger than the sum in quadrature of 
the noise in individual (uncorrelated) pixels \citep{casertano2000}. 
The flux uncertainties calculated by \se\ are therefore underestimated, 
and the S/N measurements based on these flux uncertainties will be 
artifically high. We therefore perform our own calculation of the noise 
in the images following the process described by 
\citet[][see also \citealt{labbe2003,blanc2008,whitaker2011}]{papovich2016}, 
which we briefly summarize here.

\begin{figure}
\epsscale{1.15}
\plotone{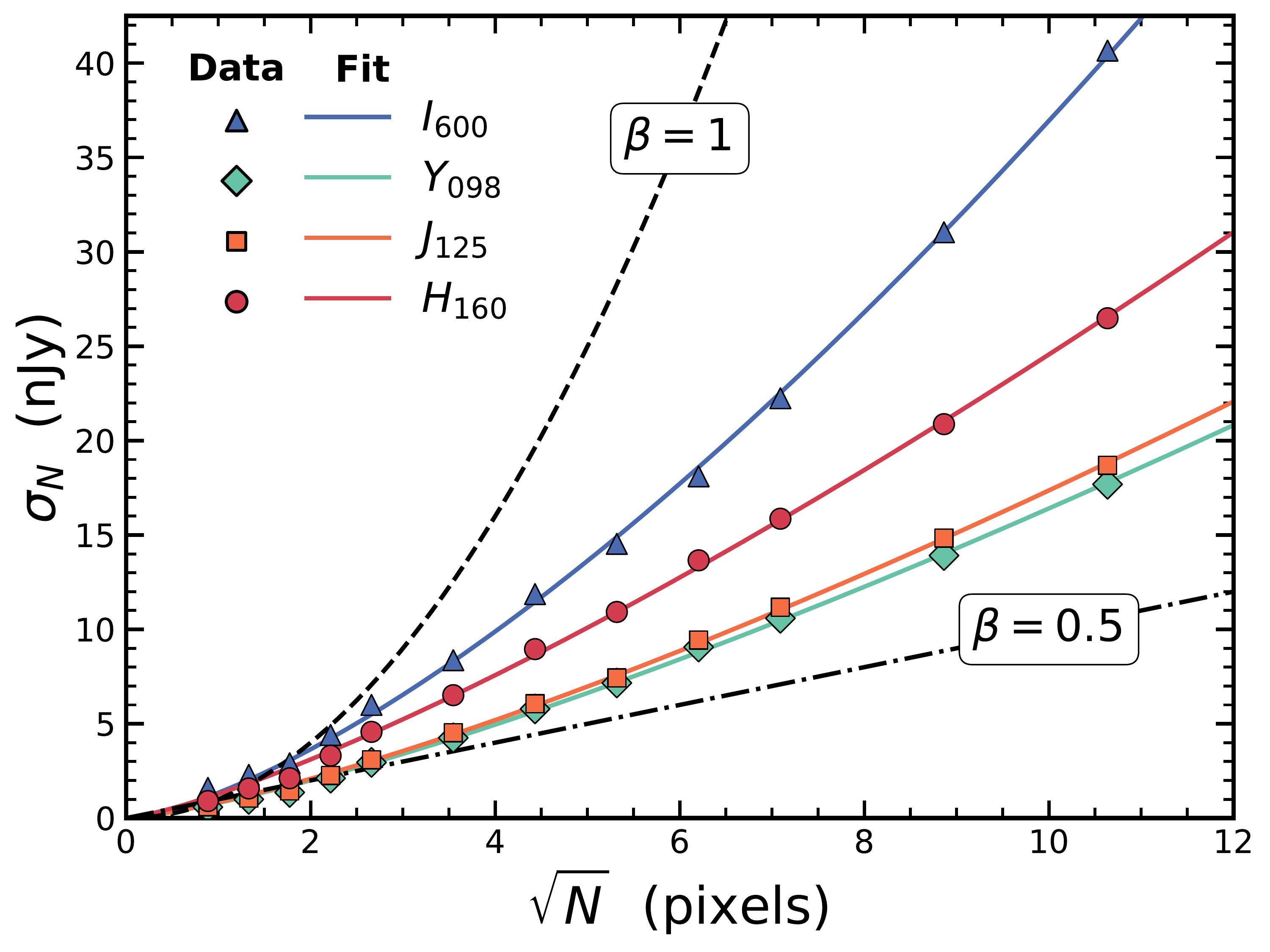}
\caption{An example of the noise calculated for \borg\ field Par0846+7654 as 
a function of aperture size. The noise, $\sigma_N$, is calculated as the 
normalized 
median absolute deviation of background flux measured in uncorrelated circular 
apertures placed on blank regions of the field. We show $\sigma_N$ for the 
filters in this field as blue triangles (\iaband), green diamonds (\yaband),
orange squares (\jband), and red circles (\hband). The functional fits to 
these measurements (equation~\ref{eqn:noise}) are plotted as curves of the 
corresponding colors. The black dashed and dot-dashed lines indicate the 
cases of completely correlated pixels ($\beta=1$) and completely uncorrelated 
pixels ($\beta=0.5$), respectively. The noise in these images is partially 
correlated in all filters. Neglecting to correct for this correlation 
would result in underestimated flux uncertainties and overestimated S/N's 
in our catalog.
\label{fig:noise}}
\end{figure}

In an image with completely uncorrelated pixels, the noise $\sigma_N$ 
measured in an aperture with $N$ pixels is expected to scale as 
$\sigma_N = \sigma_b \times \sqrt{N}$, where $\sigma_b$ is the standard 
deviation of the pixels containing only the sky background. At the other 
extreme, the noise will scale as $\sigma_N = \sigma_b \times N$ for 
completely correlated pixels \citep{quadri2007}. We thus expect the noise to 
scale as $\sigma_N \propto N^{\beta}$, with $0.5 < \beta < 1$ to reflect 
the two limiting cases. For our analysis of the noise as a function of 
aperture size, we place $\sim5000$ non-overlapping circular apertures 
randomly across each image, using the segmentation and RMS maps 
to avoid both detected source flux and bad pixels. The apertures have 
radii ranging from 0.5 pixels to 12 pixels. We use Photutils
\citep[v0.6;][]{bradley2019} to measure the flux in these apertures  
and estimate $\sigma_N$ via the normalized median absolute deviation 
\citep{beers1990} as a robust estimator of the standard deviation in 
the distribution of aperture fluxes.
We then fit the following parameterized function to the relation between 
$\sigma_N$ and $N$:
\begin{equation}
\sigma_N = \sigma_b (\alpha N^{\beta}),
\label{eqn:noise}
\end{equation}
where $\alpha$ and $\beta$ are free parameters. For $\sigma_b$, we again 
use the normalized median absolute deviation to estimate the pixel-to-pixel 
variation in background pixels in the image.
We show the measured $\sigma_N$ and fits for an example field in 
Figure~\ref{fig:noise}.
For each source, filter, and aperture type, we use equation~\ref{eqn:noise} 
to calculate a flux error determined by the number of pixels in the apertures.
Finally, we scale the fit by the value of the RMS map at the position of the 
source divided by the median of the RMS map. This scaling 
maintains any structure present in the RMS map that is indicative of 
location-dependent quality differences. 
We include Poisson photon errors in this scaling parameter, though we note 
that including this shot noise increases the flux uncertainties by only
$\sim3-6$\%.

%%%%%%%%%%%%%%%%%%%%%%%%%%%%%%%%%%%%%%%%
\subsubsection{Photometric Corrections} \label{sec:photcorrections}
We apply three correction factors to the catalog in order to 
account for (1) the missing wings of the PSF in the small Kron 
apertures (i.e., aperture corrections), (2) any unrecovered source flux 
due to the source extraction parameters, and (3) Galactic extinction. 
These corrections are applied in all filters to both the measured fluxes 
and flux uncertainties, thus preserving the optimized S/N in the smaller 
apertures while correcting for any missing or extinguished flux. 
First, we derive aperture corrections in \hband\ as the ratio of the 
fluxes calculated in the small Kron apertures with those using larger 
Kron apertures corresponding to \texttt{PHOT\_APERTURES}~=~(2.5,3.5). 
The median flux ratio in the full catalog is $\sim$0.69 (i.e., source fluxes 
are $\sim$31\% higher when measured in the larger apertures), though the
aperture corrections are calculated and applied on a source-by-source basis.
Next, we use the simulated sources from our completeness simulations 
(see Section~\ref{sec:sims}) to quantify any unrecovered source flux. 
We find that the median recovered \hband\ brightness of sources detected 
with S/N$\geq$5 in all datasets is $\sim$0.14 magnitudes fainter than the 
input brightness after application of the aperture correction.
We apply this correction factor to the total (aperture-corrected) fluxes and 
uncertainties, obtaining a final estimate of the total source flux.
Lastly, we correct for Galactic extinction using the color excess 
$E(B-V)$ for each parallel field taken from the IRSA Galactic Dust 
Reddening and Extinction 
calculator\footnote{\url{https://irsa.ipac.caltech.edu/applications/DUST}}.
This service provides position-based reddening estimates from 
\citet{schlafly2011}. Following \citet{schlafly2011}, we use the 
\cite{fitzpatrick1999} extinction curve for the Milky Way, with 
$R_V = A_V/E(B-V) = 3.1$, where $A_V$ is the extinction in magnitudes in the 
$V$ band, to calculate the extinction at the central wavelength of each 
\hst\ filter. 

Finally, we perform a photometric correction on a small subsample of sources 
in the catalog to account for elliptical apertures that were stretched or 
overly elongated. This aperture stretching can occur for faint 
objects with very close, bright neighboring sources. In these cases, while 
the fluxes measured in the small Kron apertures may be affected, the 
aperture corrections, which rely on the larger Kron aperture, are particularly
unreliable. Following \citet{finkelstein2015a}, we identified sources in the 
catalog that needed this correction as those with a 
$f_{\mathrm{AUTO}}/f_{\mathrm{APER}}$ flux ratio larger than the 95$^{th}$ 
percentile of all sources in their \hband\ magnitude bin. 
Here, $f_{\mathrm{AUTO}}$ refers to the flux measured in the elliptical 
Kron apertures, and $f_{\mathrm{APER}}$ is that 
measured in the circular $r=0\farcs2$ apertures. We also required that the 
sources had a neighbor within 1\arcsec. 
We then determined the median aperture corrections required to go from 
the $r=0\farcs2$ $f_{\mathrm{APER}}$ fluxes to the aperture-corrected 
$f_{\mathrm{AUTO}}$ fluxes in bins of \hband\ magnitude. We applied this 
correction to the $f_{\mathrm{APER}}$ fluxes and uncertainties of the 
affected sources. For these sources only, we use the corrected circular 
aperture fluxes (rather than the Kron fluxes) in measuring source colors.
This correction affected only one of the sources in our sample of high-redshift
candidates (see Section~\ref{sec:results}).

%%%%%%%%%%%%%%%%%%%%%%%%%%%%%%%%%%%%%%%%
\subsection{Stars} \label{sec:psfs}
We next construct a sample of stars in each of the datasets. This sample will 
serve two purposes: to provide an estimate of the half-light 
radius (\rhalf) of unresolved sources in these images for comparison with 
our selected candidates (Section~\ref{sec:spex}), and
to construct an average point spread function (PSF) 
for use in simulating sources for our completeness simulations
(Section~\ref{sec:sims}).
We begin by selecting sources that are $\geq$5\arcsec\ from any neighboring
objects in the catalog and that have measured half-light radii in \hband\
in the range $1 < r_{1/2} < 2.5$ pixels. We also require \hband\ magnitudes 
in the range $16 < m_{H160} < 21$, where the cut at the bright end is 
determined to avoid saturated sources and the cut at the faint end is to 
allow for a clean separation between point sources and extended sources.
Next, we visually inspect the \hband\ imaging of all star candidates, 
rejecting any that lie along an image edge, are heavily affected by bad 
pixels or other detector artifacts, or appear extended or as two 
very close sources that have been extracted as a single source. 
Through this selection, we identify a median of 4 moderately bright, 
isolated stars per field, though the number of stars meeting these 
criteria in a given field ranges from 0 to 15.
In total, we identify 539 stars, 253 in \borg\ fields, 112 in \hippies\ 
fields, and 174 in WISP fields. 

Ideally, we would create a composite PSF for each filter in each \hst\ 
pointing, constructed from stars that are located across the full field. 
In this way, we would account for any variations in the telescope focus 
during each set of observations that could result in field-to-field variations 
in the PSF. By including stars across the full field, we would also account 
for the detector location dependence of the PSF.
However, as these \hst\ surveys were designed as extragalactic programs, 
they attempted to avoid stars as much as possible.  
Many of the \hst\ pointings contain only a handful of isolated stars in the 
desired magnitude range, and some fields contain none. As there are not 
enough stars to construct a PSF for each individual \hst\ pointing, we 
instead use a single PSF for all fields of a given dataset. To validate 
this approach, we explored 
the field dependence of the PSF by creating individual stacks of stars for 
each field containing at least three point sources.
We constructed a radial profile of each combined PSF by computing the
azimuthally-averaged flux in circular apertures of increasing radii and used
it to measure the full width at half maximum (FWHM).
We found that while the FWHM of the PSF varies from field to field, it does 
so by an acceptably small amount. For example, the standard deviation of 
\hband\ FWHMs measured for individual fields is $\leq$0\farcs04 for all three 
datasets.
We therefore create a combined PSF for each dataset/filter combination
that we use for all fields in that dataset.

\begin{figure}
\epsscale{1.2}
\plotone{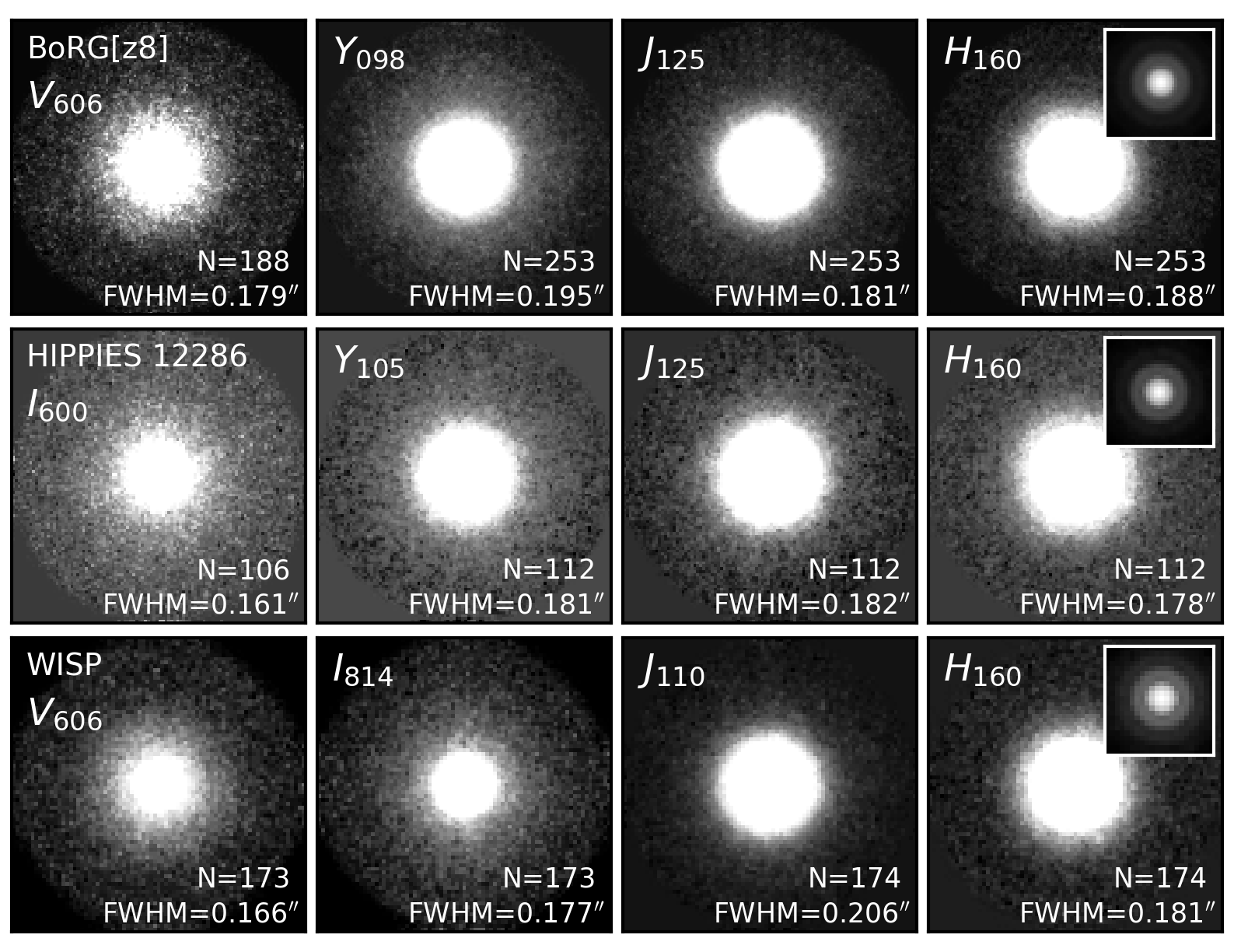}
\caption{
Median-combined PSFs for each filter and dataset. The image stamps are 
10\arcsec$\times$10\arcsec, displayed on a zscale interval and linear stretch
with no image smoothing or pixel interpolation. We also show the central 
3\arcsec$\times$3\arcsec\ of the \hband\ PSFs in the inset panels to zoom in 
on the core of the PSF. The insets are displayed with a min-max interval and 
a log stretch. In each panel, we list the measured FWHM of the PSF profile 
as well as the number of stars that were combined in the stack. 
We display the \vband\ PSF rather than the \iaband\ PSF for the 
\borg\ dataset, as $\sim$2/3 of the \borg\ fields are covered by \vband. The 
\iaband\ FHWM is included in Table~\ref{tab:psfs}.
\label{fig:psfs}}
\end{figure}

For each dataset/filter, we create a combined PSF in the following way. 
The individual image postage stamps of each identified star are resampled 
by a factor of 10, registered to a common centroid using subpixel shifts, 
and then resampled back to their original pixel scale.
We normalize each image by the flux measured in a circular aperture of radius 
$r=0\farcs2$. Because the fields are observed at a range of roll angles, 
we rotate each stamp by a randomly-assigned angle in 
order to randomize the location of the diffraction spikes. Finally, we 
median combine all image stamps. Combining multiple 
stars more fully samples the undersampled \hst/WFC3 PSF, as together they 
provide images of point sources with a range of subpixel centroids.
We perform this stacking on each of the three datasets separately, to 
preserve the pixel scales and drizzling patterns. 
The combined PSFs for each filter and dataset are displayed in 
Figure~\ref{fig:psfs}, where we show the full stamps for all filters and 
the central cores of the PSFs in the \hband\ images as insets. 
We calculate the radial profile and FWHM as described above.
Each PSF panel in  
Figure~\ref{fig:psfs} is labeled with the measured FWHM and the number of 
stars that are included in the stack. We also list the measured FWHMs in 
Table~\ref{tab:psfs}.
\begin{deluxetable}{ccccc}
\centering
\tablecaption{Empirical PSFs and Point Source \rhalf \label{tab:psfs}}
%\tabletypesize{\scriptsize}
\tablehead{
\colhead{Filter} & \colhead{$N_{\mathrm{stars}}$} &
\colhead{Measured FWHM} & \colhead{\hband\ \rhalf}
& \colhead{$\sigma_{r1/2}$} \\
\colhead{} & \colhead{} & \colhead{(arcsec)} & \colhead{(pixels)}
& \colhead{(pixels)}}
\startdata
\multicolumn{5}{l}{\borg} \\ 
\vband\ & 188 & 0\farcs179 & \nodata & \nodata \\ 
\iaband\ & 77 & 0\farcs157 & \nodata & \nodata \\
\yaband\ & 253 & 0\farcs195 & \nodata & \nodata \\
\jband\ & 253 & 0\farcs181 & \nodata & \nodata \\
\hband\ & 253 & 0\farcs188 & 1.69 & 0.20 \\
\hline
\multicolumn{5}{l}{\hippies} \\ 
\iaband\ & 106 & 0\farcs161 & \nodata & \nodata \\
\yaband\ & 112 & 0\farcs181 & \nodata & \nodata \\
\jband\ & 112 & 0\farcs182 & \nodata & \nodata \\
\hband\ & 112 & 0\farcs178 & 1.39 & 0.14 \\
\hline
\multicolumn{5}{l}{WISP} \\ 
\vband\ & 173 & 0\farcs166 & \nodata & \nodata \\
\ibband\ & 173 & 0\farcs177 & \nodata & \nodata \\
\jbband\ & 174 & 0\farcs206 & \nodata & \nodata \\
\hband\ & 174 & 0\farcs181 & 1.37 & 0.26 \\
\enddata
\tablecomments{Our measured FWHMs of the empirical, pixelated 
PSFs are larger than the pre-pixelated values reported in the WFC3 Instrument
Handbook (\url{hst-docs.stsci.edu/wfc3ihb}).}
\end{deluxetable}

Finally, we use this sample of visually-vetted stars to measure a typical 
\rhalf\ for stars in each dataset. 
While \se\ provides a measurement of stellarity that can be useful in 
separating point and extended sources (\texttt{CLASS\_STAR}), we find that 
it is most reliable for bright stars, $m_{H160} \lesssim 17$. 
The stars in this sample with \hband\ magnitudes fainter than 
$\sim$17 have \texttt{CLASS\_STAR} values that range from 1.0 (as expected 
for point sources) all the way down to $\sim$0.2 (closer to that expected for 
extended sources). We therefore choose to use the \rhalf\ as measured in 
\hband\ with \se\ to classify unresolved sources. 
In Table~\ref{tab:psfs}, we provide the median and standard deviation of 
\rhalf\ for the stars in each dataset, values we use in Section~\ref{sec:spex}
to idenfity potential stars in our sample of high-redshift candidates.

%%%%%%%%%%%%%%%%%%%%%%%%%%%%%%%%%%%%%%%%%%%%%%%%%%%%%%%%%%%%%%%%%%%%%%%
\section{Sample Selection} \label{sec:sample}
We select candidate \zsamp\ galaxies through a combination of criteria 
related to detection significance, photometric redshift fitting, and visual 
inspection. Whereas many have selected high-redshift galaxies through strict 
color cuts \citep[e.g.,][]{oesch2013,bouwens2015a,bouwens2016a,trenti2011,bradley2012,schmidt2014,bernard2016},
we use a more holistic approach by 
including all available \hst\ and \spitzer\ imaging in our 
photometric redshift selection \citep[as in, e.g.,][]{mclure2010,finkelstein2015a,finkelstein2022a,bouwens2019a,bouwens2021a,bowler2020,rojas-ruiz2020}. As the Lyman break is the main spectral 
feature within the \hst\ filter coverage for galaxies at $z\gtrsim8$, 
the photometric redshift fitting is akin to a Lyman break selection. 
However, this approach allows for the selection of sources that might lie 
right on the edge or even outside of a selection window in color-color 
space. Yet in constructing a potentially more complete sample, we must take 
extra care with contaminants. We therefore aim to be conservative and as 
objective as possible in our process for source vetting, removing subjective 
steps entirely where possible and quantifying our subjectivity where not. 
We describe the aspects of our high-redshift galaxy selection and source 
vetting in the following sections.

%%%%%%%%%%%%%%%%%%%%%%%%%%%%%%%%%%%%%%%%
\subsection{Photometric Redshifts} \label{sec:photz}
We use the Easy and Accurate $z_{\mathrm{phot}}$ from Yale
\citep[\eazy\, version 2015-05-08;][]{brammer2008} software to calculate 
photometric redshifts for each source, real and simulated. 
The \eazy\ code fits all photometric measurements to a synthetic spectral
energy distribution (SED) template via a chi-squared minimization
process. In deriving a best-fitting template, it can consider linear 
combinations of templates from a user-supplied set, and the resulting fits 
are therefore less dependent on the choice of input templates. 
We use the 12 \texttt{tweak\_fsps\_QSF\_12\_v3} \eazy\ templates  
that are based on the Flexible Stellar Population Synthesis (FSPS) 
models \citep{conroy2009,conroy2010} and account for systematic 
differences between observed galaxy colors and the models. Following 
\citet{finkelstein2022a},
we also add the spectrum of Q2343-BX418, a young ($<$100 Myr) galaxy at 
$z=2.3$ with high equivalent width nebular emission lines \citep{erb2010}. 
We include this low mass ($M* \sim 10^9 M_{\odot}$), 
low metallicity ($Z \sim 1/6 Z_{\odot}$), 
blue (UV continuum slope $\beta=-2.1$) template as galaxies with blue colors
are expected at high redshift \citep[e.g.,][]{bouwens2009,finkelstein2009}.
We also add a version of this spectrum with all \lya\ emission 
removed to approximate attenuation by a neutral intergalactic medium (IGM) 
while still providing 
a template with strong optical emission lines that can affect 
\spitzer/IRAC colors (e.g., \oiii\ and \hb\ for galaxies at $z\sim8-9$).
We allow \eazy\ to construct best-fitting templates through linear combination 
pulling from all 14 of these input templates. 

In fitting templates over a grid of redshifts, \eazy\ compares the input 
source fluxes and flux errors with the synthetic fluxes of the templates 
convolved through each filter. We use the total fluxes computed 
in the small elliptical Kron apertures as the input source fluxes
(see Section~\ref{sec:photometry}, \ref{sec:photcorrections}), with flux 
uncertainties computed as described in Section~\ref{sec:noise}. 
We also add a minimum fractional error of 0.05 to the flux uncertainties 
in all filters (\eazy\ parameter \texttt{SYS\_ERR}) to allow for 
systematic uncertainty in our flux measurements. 
Template fitting is performed in the redshift range $0.01 \leq z \leq 12$ 
in steps of $\Delta z = 0.01$. 
At each redshift step, \eazy\ applies IGM attenuation following 
\citet{madau1995} and includes the absorption by the \lya\ forest 
and damped \lya\ systems as prescribed by \citet{inoue2014}.
As the population of galaxies at redshifts $8-10$ is currently not 
well-understood, we assume a flat luminosity prior.
This approach has the benefit of not biasing us against potential true 
high-redshift sources, yet also means we are treating sources at all 
redshifts the same, i.e., a galaxy at $z\sim2$ is equally likely as one 
at $z\sim10$. We must therefore take extra care in considering the 
contamination fraction of lower-redshift galaxies in our sample, which
we discuss in Section~\ref{sec:contam}.

%%%%%%%%%%%%%%%%%%%%%%%%%%%%%%%%%%%%%%%%
\subsection{Selection Criteria} \label{sec:selection}
We begin by imposing an initial set of detection criteria on the full 
\se\ catalog using the S/N as measured in circular apertures of radius 
$r=0\farcs2$. These small apertures capture the light at the central core 
of the source positions as detected in the \hband\ image and therefore 
maximize the measured S/N. 
First, we require that sources be detected in the \hband\ band with a 
S/N$_H > 5$. As we are aiming to select $z\gtrsim9$ galaxies, we expect 
the sources to have ``dropped out'' of the $V$, $I$, and $Y$ filters. We
therefore only consider sources with a S/N~$<2$ in all available filters 
blueward of the $J$ band (see Figure~\ref{fig:filters}), 
permitting a S/N~$>1.5$ in at most one of these filters.
This criterion allows for a detection up to S/N~$=2$ in one, but not both, of 
the bluer filters, such that a coincidental $<2\sigma$ noise fluctuation in an 
optical filter or a partial detection in, e.g., \ybband\ for 
$8.5\lesssim z \lesssim 9$ will not disqualify a source.

We next select sources based on their \hband\ magnitude, requiring 
$22 < H_{160} \leq 26.5$. 
The cut at the bright end helps remove contaminants such as stars and 
lower-redshift galaxies, as $H_{160}\sim22$ corresponds to an absolute 
UV magnitude of $M_{UV} \sim -25$ at $z\sim9$. 
We choose the magnitude limit at the faint end 
in an attempt to remove spurious sources. As our analysis involves 
relatively shallow \hst\ imaging (as compared to the Hubble Ultra Deep 
Field or the Hubble Frontier Fields, for example), we focus on bright
high-redshift candidates. The majority of the fields in these 
datasets\footnote{We note that $\sim$30 of the fields we include in our 
analysis have $5\sigma$ depths brighter than 26.5 as presented in 
Table~\ref{tab:obs}. However, these depths are averaged over the entire 
field and are not always indicative of the depth in all areas of the field.}
have $5\sigma$ depths $\gtrsim$26.5, and so the likelihood of detecting 
noise as spurious sources increases for magnitudes fainter than 26.5.
We use the total fluxes measured in the elliptical apertures 
for these magnitude cuts.

As an additional check against including spurious sources in our 
selection, we impose a minimum size criterion. Given the similarities in 
the depth and noise properties of the parallel imaging we consider in this 
paper and that of \citet{rojas-ruiz2020}, we similarly expect a half-light 
radius criterion to remove the majority of hot pixels and cosmic 
rays from our sample.  
Following the example of \citet{rojas-ruiz2020}, we used a random forest 
algorithm to obtain a quantitative size cut for each dataset. 
In order to construct a training set, we performed an initial visual 
inspection of 2500 randomly-selected sources 
from each dataset that satisfied the detection significance criteria in the 
\hband\ and all optical filters. 
Based on our visual inspection of the imaging in all available filters, 
we classified each source as `real' or `spurious'. These classifications 
became the target values or class labels input to the random forest. 
To this sample, we added an equivalent number of randomly-selected 
simulated sources that satisfied the S/N criteria, each of which were 
classified as `real'. For each dataset, we thus had a 
sample of 5000 sources that we fed to a Random Forest Classifier 
using the Python Scikit-learn package \citep{scikit-learn}. Though we included 
information such as the \hband\ magnitude, source isophotal area, elongation, 
and surface brightness, we found that the half-light radius was the most 
discerning parameter. Through this analysis, we found that the majority of 
sources classified as `spurious' could be removed by imposing a 
half-light radius cut at \rhalf$>$1.3 pixels for \borg\ fields 
and $>$1.1 pixels for the \hippies\ and WISP datasets.
This difference in \rhalf\ values is expected due to the different pixel 
scales of each dataset and corresponds to $r_{1/2} > 0\farcs1$ for the 
\borg\ (0\farcs08/pixel) and HIPPIES GO 12286 (0\farcs10/pixel) fields and 
$r_{1/2} > 0\farcs14$ for the WISP (0\farcs13) fields. The \rhalf\ values 
are measured by \se\ in the \hband. 
We do not include a maximum effective radius criterion in our selection, as 
it has been noted that such a criterion can remove real sources along with 
lower-redshift interlopers \citep[e.g.,][]{holwerda2020}.

Finally, 
we use the redshift probability distribution functions (PDFs, which we 
denote $p(z)$),
that \eazy\ calculates for each source to select a sample of candidates with
preferred high-redshift solutions. Specifically, we require that $>$70\% 
of the integrated redshift probability be at $z>8$. This criterion ensures 
that the majority of the redshift probability resides at high redshift with 
less than 30\% contained in lower-redshift solutions. We do not enforce 
any additional criteria related to the photometric redshift fits at this 
stage, as the PDFs derived from measurements in only four filters 
provide minimal constraining power. 
For example, the redshift probabilty distribution functions are flat for 
the majority of our candidates at $z\gtrsim10$, where the \lya\ break 
has redshifted almost entirely out of the $J$ band and the source is detected
in only a single filter.
While we do not select sources based on the $\chi^2$ of the best-fitting 
template, we explore the $\Delta \chi^2$ of the best-fitting 
templates at high and low redshift in Section~\ref{sec:contam}.

In summary, in order for sources to be considered \zsamp\ candidates,
we require that they:
\begin{itemize}
  \item are detected in \hband\ with (S/N)$_H>5$;
  \item have \hband\ magnitudes in the range $22 \leq H_{160} \leq 26.5$; 
  \item are undetected at the $2\sigma$ level in $V$, $I$, and $Y$, allowing 
    a S/N$>$1.5 in at most one of these filters;
  \item have a half light radius of \rhalf$>$1.3 (1.1) pixels for 
    \borg\ (HIPPIES GO 12286 and WISP)
  \item have at least 70\% of their redshift probability distribution 
    function at $z>8$.
\end{itemize}

We find 193 sources that pass these selection criteria: 116 from 
\borg\ fields, 48 from HIPPIES GO 12286, and 29 from WISP 
fields. In the following section we screen this sample 
for cases of detector persistence from previously observed bright 
targets, which could masquerade as high-redshift galaxies.

%%%%%%%%%%%%%%%%%%%%%%%%%%%%%%%%%%%%%%%%
\subsection{Persistence} \label{sec:persistence}
Image persistence
occurs when a bright source saturates the detector and leaves a residual
charge that appears as a ghost image in subsequent exposures. This situation
is particularly problematic in fields for which the \hband-band is observed
before the $J$-band, as persistence fades with time and so can mimic a
$J$-dropout. The BoRG observing strategy was designed to mitigate image
persistence by observing with the $Y$ band followed by \jband\ and
\hband\ in every orbit \citep{trenti2011}. Persistence is therefore
expected to impact both the dropout and detection filters such that the
residual ghost images will not be selected as dropout candidates.
However, the other datasets we include in our analysis did not always
perform their observations in this order. Even some of the HIPPIES
observations from GO 11702 that are included in the \borg\ dataset were not
scheduled such that the $Y$-band images could shield the detection filters.
The WISP observing strategy is particularly prone to self persistence,
which occurs when the offending bright targets are observed in the same
visit as the affected image.
As a slitless spectroscopic survey, WISP observed fields in direct
imaging--grism pairs, with the \jband\ and \hband\ images used for
source extraction and wavelength calibration of the G102 and G141 grism
observations. The zeroth orders of bright stars are then common
causes of persistence in the direct images that follow grism exposures.

We therefore performed a detailed persistence check for all candidates in 
the following way. 
For each candidate, we searched the MAST archive for observations
taken in the 24 hours prior to each of the individual exposures (FLTs) that
went into the \hband\
mosaics. If any of those observations had a count $>$100,000 e$^-$ within
a 10$\times$10 pixel box centered on the position of a given candidate, the
candidate was removed from our sample. We note that the use of a 10 pixel box
is likely overly conservative, and that the pixel with the highest counts
need not be the central pixel. However, considering a box rather than a few
central pixels allows for offsets in the drizzle solution, as the world 
coordinate solution in the undrizzled image (that has not been distortion
corrected) could be off by a few pixels at the position of the candidates.
We identify 10 candidates as heavily affected by persistence and
remove them from our sample.

\begin{figure}
\epsscale{1.2}
\plotone{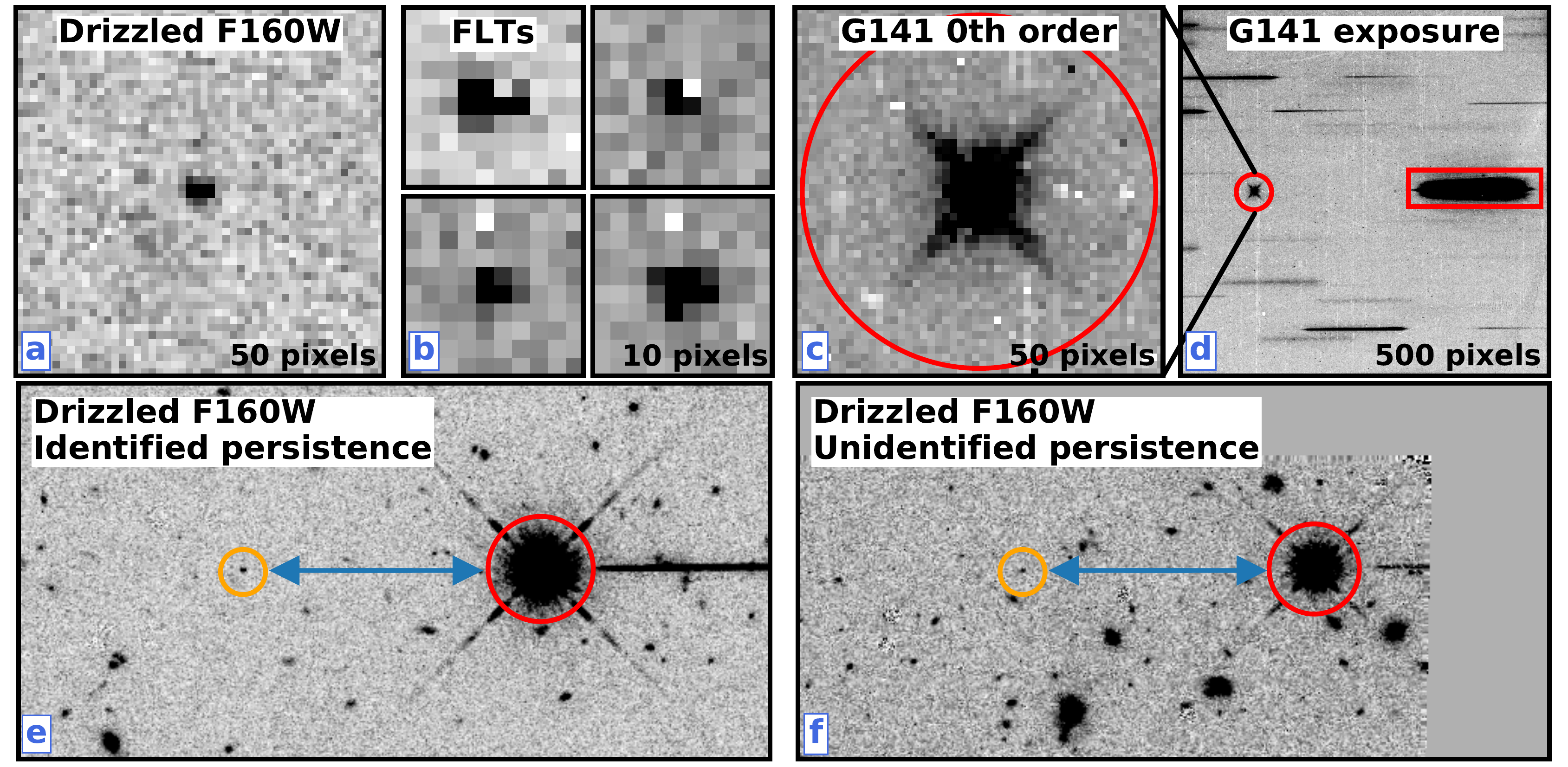}
\caption{Two examples of WISP sources identified as self persistence and
removed from our sample. First, in panel (a) we show a 50$\times$50 pixel
stamp of a candidate identified as persistence due to the high detector counts
in FLTs observed before the \hband\ observations. In panel (b), we
show 10$\times$10 pixel stamps of the four affected FLTs that contributed to
the \hband\ mosaic. The cause of the persistence was the
zeroth order of a bright star observed with the G141 grism just prior to the
\hband\ exposures. We show a zoom-in
of this zeroth order in panel (c) and a larger portion of the G141 image
in panel (d) with the zeroth and first orders indicated.
As can be seen in panel (e), the position of self persistence resulting from
this specific observing strategy will be offset by $\sim$180-195 pixels from the
offending bright source. We have identified and rejected two additional WISP
sources in our sample based on their position relative to a bright star, one
of which is shown in panel (f). These candidates were not identified as
persistence based on counts from previous observations at the candidate
position on the detector, indicating that even this thorough
check can miss cases of persistence.
\label{fig:persistence}}
\end{figure}

However, given the WISP observing strategy, we find that even this aggressive
approach is likely to have missed some cases of self persistence in WISP 
images -- where zeroth orders appear as \jbband-dropout candidates.
For each source, the grism dispersion determines the location of
all spectral orders in the grism exposures with respect to the position of
the source in the direct image. We can therefore expect persistence from
G141 zeroth orders to exist in the \hband\ image $\sim$180-195 pixels 
(23\farcs4-25\farcs4) to the left of a bright star, where the exact offset 
is location dependent. 
We illustrate this situation in
Figure~\ref{fig:persistence}. First, in panel~\ref{fig:persistence}a, we
show an example of a WISP candidate that was identified as persistence based
on the counts in previously-observed FLTs as described above.
This persistence is caused by a zeroth order in an earlier G141 observation
that is shown in panels~\ref{fig:persistence}c and \ref{fig:persistence}d.
We show the \hband\ image of this candidate in panel~\ref{fig:persistence}e,
where it can be seen that this candidate is at the expected distance to the
left of a bright star. In Figure~\ref{fig:persistence}f,
we show another WISP candidate that was \textit{not} identified by our
persistence check described above (the maximum counts in all FLTs was
$<$8,000 e$^-$), yet lies at the expected position.
We have removed 2 such WISP sources from our sample based on their
positions relative to a very bright source. We note that we are only able to
identify these additional cases of persistence because of the specifics of the
WISP observing strategy, and that similar unidentified cases of persistence may
still contaminate the rest of our sample. Detecting candidates in a second
imaging band is the only way to confirm such single-band detections are real
sources. We show all 12 sources that are rejected as persistence in
Figure~\ref{fig:rejects}d in the appendix, and indicate those that were 
identified based on their distance from a bright star.

After removing the 12 cases of persistence, there are 181 remaining
candidates. We further explore these sources in Section~\ref{sec:inspection}
with a visual inspection process to
remove spurious or misclassified sources.

%%%%%%%%%%%%%%%%%%%%%%%%%%%%%%%%%%%%%%%%
\subsection{Visual Inspection} \label{sec:inspection}
Thus far, we have selected candidate \zsamp\ galaxies using only 
quantitative criteria. However, we find that the majority of sources 
in our sample at this stage are spurious detections, 
where artifacts such as diffraction spikes and partially-removed satellite 
trails are 
selected as high-redshift candidates. Given the wavelength dependence of the 
PSF, diffraction spikes will be largest in the \hband\ images and can therefore
appear as $J$-band dropouts. Similarly, a satellite trail present in the 
\hband\ image can also cause false dropout candidates, and many of the 
fields were observed with only two or three \hband\ exposures, too few to 
fully remove a satellite trail through outlier rejection during mosaic 
creation. Spurious sources are therefore detected in the increased noise 
of these satellite remnants. 
Hot pixels and cosmic rays that impacted the detector in a compact area 
can also masquerade as high-redshift sources. The drizzling 
process and kernel smoothing that is applied by \se\ can both act to spread 
out the light into nearby pixels, giving cosmic rays a more extended 
appearance like that of real sources.
As these pure parallel datasets are undithered, such features
will not all be removed by image stacking. While bad pixels on the IR 
detector should appear in the same location in all IR filters, cosmic 
rays in the \hband\ images can appear as $J$-band dropouts in our selection.
A careful visual inspection is necessary to identify and remove such 
spurious candidates and other sample contamination.

However, visual inspection is an inherently subjective process, and the 
effect that visually rejected candidates have on the sample incompleteness is 
typically not accounted for with completeness calculations.
We therefore use simulated sources (see Section~\ref{sec:sims}) to 
quantify any biases that may enter our sample selection through the 
visual inspection process.
We compile a set of 362 sources, consisting of all selected 
real candidates and an equal number of simulated candidates selected in the 
same fields and via the same criteria. 
We randomize this set of real and simulated candidates and 
carefully inspect each one, accepting or rejecting them without knowing 
which is simulated. We discuss using the simulated sources to quantify 
the effect of the visual inspection on sample incompleteness in 
Section~\ref{sec:sims_inspection}.
For the rest of this section we refer to the 
inspection and rejection of the real candidates. 
We create a set of categories for rejection, ensuring that we use a 
standardized classification scheme in all visual inspections. The 
classification categories we consider are as follows. 

First, we consider whether the source in question is a spurious detection. 
As mentioned above, diffraction spikes and satellite remnants are common 
examples of spurious detections. 
Additionally, we reject sources located immediately along image 
edges where the noise levels are significantly higher, because these sources 
at best have unreliable photometry and at worst are only partially extracted. 
We also expect spurious sources may be identified on the edges of 
bright sources, where the \se\ deblending can be too aggressive. In these 
oversplit regions, the light from a bright source gets separated and 
attributed to multiple sources. The PSF is smaller in bluer filters, and
so any oversplit regions are closer to the bright object core.
Oversplit regions in $J$ and $H$ can therefore appear as 
high-redshift galaxies that have dropped out of the bluer filters.
We consider this possibility during our inspection, but do not identify any 
sources in this category.
As part of our visual inspection, we also inspect the RMS maps at the 
position of each candidate and reject sources with clusters of bad pixels 
identified within the $r=0\farcs2$ circular apertures used to determine 
detection significance. 
Our final category of spurious detections includes hot pixels, cosmic rays, 
and other artifacts that are not identified as bad pixels in the RMS maps. 
We identify these features visually as sources appearing as single pixels, 
disjointed clusters of pixels that may have been smoothed together during 
source detection, or otherwise strange morphologies. 

We then consider two additional categories for candidate rejection, 
cases in which a source is identified as real yet was incorrectly selected 
due to problems related to photometric measurement or redshift fitting.
As described in Section~\ref{sec:photcorrections}, the \se\ Kron 
apertures of small sources in close proximity to a large, bright neighbor 
can be stretched by the light from the neighbor. We have attemped to 
identify sources with affected apertures and correct the corresponding 
fluxes (see Section~\ref{sec:photcorrections}), and one of the 181 
selected candidates required this correction (Par0456-2203\_473, see 
Section~\ref{sec:cand0456_473}). Here, we 
use our visual inspection process to check that the elliptical apertures
(or, in the case of the one corrected source, the circular aperture that 
replaced the elliptical aperture) 
closely match the source morphology for all candidates.
We also inspect all available $V$, $I$, and $Y$ band images at each source 
position to ensure that there is no significant flux that was missed in the 
catalog measurements. 

In summary, we consider the following categories of spurious sources or 
unreliable measurements in our visual inspection (and indicate in parentheses
the number of real sources rejected in each category):
\begin{enumerate}
  \item Is the candidate a real source or a spurious detection (166 rejected,
    see Appendix~\ref{app:rejections})? I.e., is the source:
    \begin{itemize}
      \item[1a.] a diffraction spike or satellite remnant (48 rejected, see 
        Figure~\ref{fig:rejects}a);
      \item[1b.] along an image edge or detected in an area of increased 
        noise near an image edge (12, Figure~\ref{fig:rejects}b);
      \item[1c.] a bad pixel in the weight map (25, Figure~\ref{fig:rejects}c);
      \item[1d.] the oversplit region of a bright neighboring source (0); or
      \item[1e.] a hot pixel, cosmic ray, or source with a strange 
        morphology (81, Figure~\ref{fig:reject_1d1e})?
    \end{itemize}
  \item Is the aperture drawn correctly, or was the Kron ellipse stretched 
    by a neighboring source and not subsequently corrected? (0)
  \item Is there significant optical flux that was not measured correctly in 
    the catalog? (0)
\end{enumerate}

As can be seen, all 166 candidates we rejected through visual inspection 
were identified as some type of spurious detection or false source. While 
85 of these rejections are objectively motivated (diffraction spikes, 
satellite trails, image edges, and bad pixel clusters), 81 are based on 
subjective opinion during the inspection (category 1e). 
In Section~\ref{sec:sims_inspection}, we use the classifications of 
simulated sources performed as part of the ``anonymous'' visual inspection 
to explore these 81 classifications.

Following this visual inspection, there are 15 candidates that have 
satisified all of our selection criteria. We further explore these sources
in Sections~\ref{sec:irac}, \ref{sec:spex}, and \ref{sec:contam}.

%%%%%%%%%%%%%%%%%%%%%%%%%%%%%%%%%%%%%%%%
\subsection{IRAC}  \label{sec:irac}
With 15 candidates remaining after rejecting sources from our visual 
inspection and persistence checks, we now turn to incorporating  
\spitzer/IRAC \citep{fazio2004} imaging to our analysis. 
Imaging at these redder wavelengths is a crucial tool for disentangling the 
spectral energy distributions of true high-redshift galaxies and 
sources that are common contaminants in high-redshift samples
\citep[e.g.,]{finkelstein2022a}.
The $J-H$ colors of both passive and dusty star-forming galaxies at 
$z\sim2-3$ can be indistinguishable from those at $z>8$, and these 
contaminants will likely by undetected in bluer filters given 
the shallow imaging of these parallel datasets. Additionally, as the 
majority of galaxies at $z>8$ are compact or unresolved in 
the \hst\ NIR imaging, M, L, and T dwarf stars can similarly contaminate 
\hst-selected high-redshift samples. 
However, the spectral energy distributions of all three types of contaminants 
are expected to diverge from those of high-redshift galaxes at 
$\lambda \gtrsim 2$\micron, and even shallow IRAC imaging can help 
distinguish between them.

\begin{deluxetable*}{c|cccc}
\centering
\tablecaption{IRAC Coverage \label{tab:irac}}
%\tabletypesize{\scriptsize}
\tablehead{
\colhead{Field} & \colhead{Program} & \colhead{PI} & 
\colhead{Ch1 AORKEY} & \colhead{Ch2 AORKEY}
}
\startdata
Par335 & 10041 & Colbert & 51773440 & \nodata \\
Par0440-5244 & 14253 & Stefanon & 69070080 & 69070336 \\
Par0456-2203 & 12058 & Bouwens & 59360512, 59360768 & \nodata \\
Par0456-2203 & 14253 & Stefanon & \nodata & 69071360 \\
Par0713+7405 & 11121 & Finkelstein & 53188864 & \nodata \\
Par0713+7405 & 12058 & Bouwens & 58126848, 58127104 & \nodata \\
Par0713+7405 & 14253 & Stefanon & 69082368 & 69075200, 69082624 \\
Par0756+3043 & 12058 & Bouwens & 58126336, 58126592 & \nodata \\
Par0756+3043 & 14253 & Stefanon & \nodata & 69078016 \\
Par0843+4114 & 14253 & Stefanon & 69082368 & 69082624 \\
Par0926+4000 & 14304 & Stefanon & 69982720 & \nodata \\
Par0926+4536 & 11121 & Finkelstein & 53189120, 53189376 & \nodata \\
Par0926+4536 & 12058 & Bouwens & 58125568, 58125824 & \nodata \\
Par0926+4536 & 14253 & Stefanon & \nodata & 69091328 \\
Par0956-0450 & 11121 & Finkelstein & 53189632 & \nodata \\ 
Par0956-0450 & 14253 & Stefanon & \nodata & 69096704 \\
Par1033+5051 & 12058 & Bouwens & 58125056, 58125312 & \nodata \\
Par1033+5051 & 14253 & Stefanon & 69105152, 69105408 & \nodata \\
Par1301+0000 & 80134 & Colbert & 42568960 & \nodata \\
Par1301+0000 & 14253 & Stefanon & 69136896 & 69137152 \\
Par2346-0021 & 90045 & Richards & 46957568, 46961152 & 46957568, 46961152 \\
\enddata
%\tablecomments{
%}
\end{deluxetable*}

We searched the Spitzer Heritage Archive hosted by 
IRSA\footnote{\url{sha.ipac.caltech.edu/applications/Spitzer/SHA}, 
\spitzer\ Level 2 data: \url{www.ipac.caltech.edu/doi/irsa/10.26131/IRSA413}}
for all available IRAC imaging at the positions of each candidate and 
downloaded the coresponding Level 2 (``Post Basic Calibrated Data''; PBCD) 
mosaic images. These mosaics 
are processed by the IRAC pipeline version 
S19.2\footnote{except for Program 12058 coverage of Par0926+4536 and 
Par1033+5051, which used version S.19.1. See  
\url{irsa.ipac.caltech.edu/data/SPITZER/docs/irac/iracinstrumenthandbook} 
for information about the IRAC pipeline.}
and are on a 0\farcs6 pixel scale. 
\spitzer/IRAC 3.6\micron\ imaging exists for 11 of the \hst\ parallel 
fields in which we identify candidates (nine of which also have 4.5\micron\
imaging), amounting to IRAC coverage of all but two of the 15
candidates. The imaging was obtained as part of seven unique programs, many 
of which were designed with the goal of following up  
high-redshift candidates previously identified in these fields.
Table~\ref{tab:irac} lists the program information for each IRAC 
dataset that we include in our analysis. 

We measure IRAC photometry using a method similar to that described in 
\citet{rojas-ruiz2020}. 
For each candidate, we create background-subtracted IRAC stamps that are 
30\farcs6$\times$30\farcs6 (51$\times$51 pixels). We consider each 
IRAC observation separately, 
such that a source observed at multiple position angles or through multiple 
programs may have multiple stamps per channel. This approach results in 
36 stamps (24 of which are at 3.6\micron) for the 13 sources 
with IRAC coverage.
In our photometric analysis, we treat each stamp as an independent measurement. 
We model all sources in each stamp with the 
\galfit\footnote{\url{users.obs.carnegiescience.edu/peng/work/galfit/galfit.html}} image-fitting software \citep[v3.0][]{peng2010} as demonstrated by
\citet{finkelsteink2015}.
Given the lower resolution of IRAC images, this modeling approach is 
necessary to deblend the light from the candidate and any neighboring sources.
We use source positions and magnitudes from the \hband\ catalog as inputs 
to \galfit, including everything in the catalog down to $H_{160}=25$ and 
making the 3.6\micron\ and 4.5\micron\ magnitude initial guesses
one magnitude brighter than the \hband\ value.
We constrain source positions in the \galfit\ model to be within 
$\pm$1.5 pixels (0.9\arcsec) of the input values and model magnitudes to 
be brighter than 40. We can safely assume that sources that hit this lower 
magnitude constraint are 
undetected in the IRAC imaging, and so we iteratively remove them from the 
model. 
Extended sources, defined as those with a semi-major axis in the \hband\
\se\ catalog that is $>$2$\times$ larger then the FWHM of the IRAC 
point response function, are modelled as \sersic\ profiles. However, the 
majority of sources in the IRAC stamps -- all high-redshift candidates and 
almost all of their neightbors -- are modelled as point sources. 
\begin{figure}
\plotone{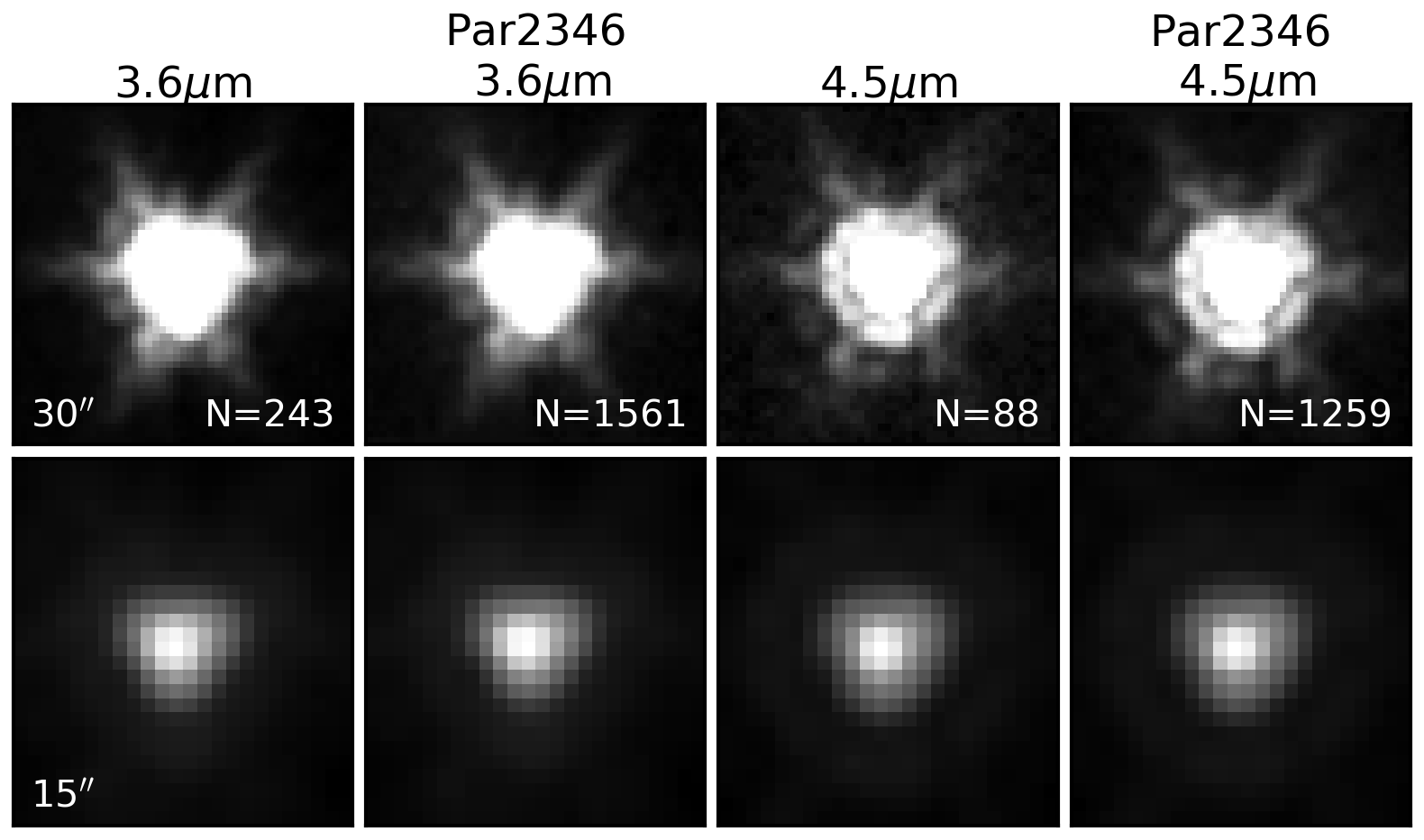}
\caption{The median-combined PSFs created from IRAC 3.6\micron\ point sources
(left two columns) and 4.5\micron\ point sources (right two columns). 
The separate PSFs for Par2346-0021 are labeled `Par2346'. The top row 
displays the full stamps ($30\arcsec \times 30\arcsec$) on a zscale interval 
and linear stretch, and the number of individual point sources that were 
combined to create each PSF are indicated by $N$. 
The bottom row zooms in on the central 15\arcsec\
and uses a min-max interval with a square root stretch to highlight the PSF 
core.
\label{fig:iracpsfs}}
\end{figure}

\begin{figure*}
\epsscale{1.1}
\plotone{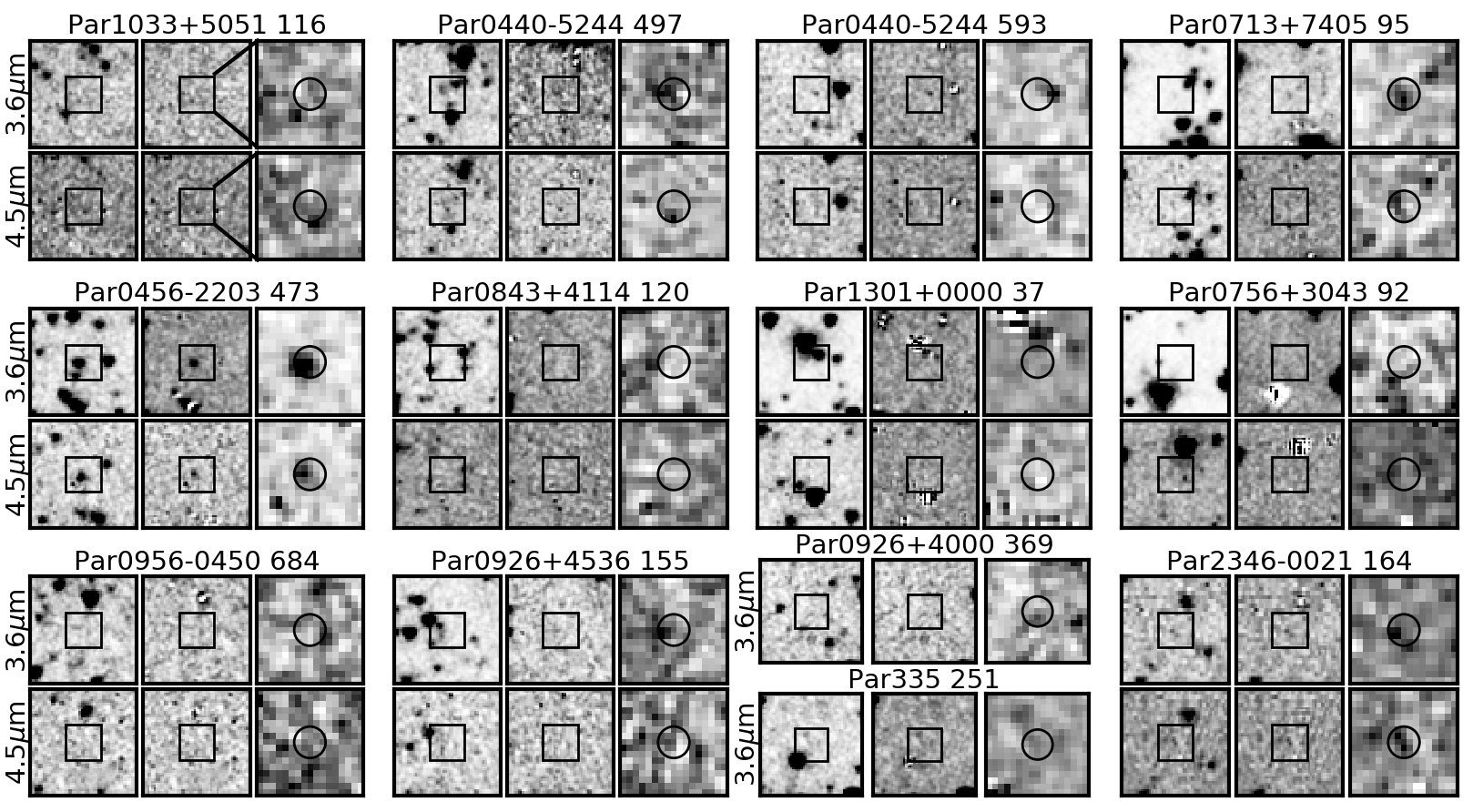}
\caption{\spitzer/IRAC imaging of the 13 candidates. For each candidate 
we show a 30\farcs6 IRAC image centered at the candidate's position (left), 
the 30\farcs6 residual map with all source flux 
removed except that of the candidate (middle), and a 10\farcs2 zoom-in
of the residual map to highlight the structure close to the candidate 
position (right). In each set of stamps, the top and bottom rows display the 
3.6\micron\ and 4.5\micron\ (where available) images, respectively. 
The black squares indicate the size of the zoom-in stamp, and the circles
have radius 2.5 pixels, the size used for aperture photometry as described 
in the text.
For fields observed multiple times, we show a single observation 
as an example. Additional images and residual maps are similar to those 
shown here. The image stamps are displayed with no image smoothing and 
on zscale intervals calculated individually for each stamp and zoom.
The deblending at the target positions is mostly clean with the exception 
of Par0456-2203\_473, which is discussed in Section~\ref{sec:cand0456_473}.
\label{fig:galfit}}
\end{figure*}
For the point source models, we create median PSFs for each IRAC channel 
in the same manner as described in Section~\ref{sec:psfs}. 
Specifically, we run \se\ on the IRAC images and identify point sources 
as bright ($14 < M_{\mathrm{IRAC}} < 18$), isolated (no 
neighbors within 15\arcsec) sources with a measured half light radius between 
$0.9\arcsec < r_{1/2,\mathrm{IRAC}} < 1.35\arcsec$. We visually inspect all 
sources to remove any with bad pixels or undetected neighbors in the \se\ 
catalog, resample all image stamps by a factor of 10 to allow for sub-pixel 
centroiding, and median combine the resulting stack. We do not randomly 
rotate each stamp, as all the IRAC mosaics are aligned by detector 
coordinates. In total, we combine 243 (3.6\micron) and 88 (4.5\micron) sources
to create the PSFs, which are shown in Figure~\ref{fig:iracpsfs}.
In Figure~\ref{fig:galfit}, we show the results of running \galfit\ on 
the 13 candidates with IRAC coverage. 

We use a two-fold approach to measuring the IRAC photometry of the sources 
in our sample. First, we adopt the \galfit\ model magnitudes for all IRAC 
observations in which \galfit\ successfully modeled the target source, as 
was the case for 23 out of the 36 separate stamps. 
For the remaining 13, \galfit\ did not measure a significant 
flux at the source position, i.e., the \galfit\ model magnitude hit the 
constraint at 40 mag and the source was removed from the model. 
In these cases, we measure the photometry of the target in the \galfit\ 
residual maps with all neighboring source flux removed. We use circular 
apertures of radius $r=2.5$ pixels, corresponding to a diameter of 3\arcsec\
($\sim$1.5$\times$ the FWHM of the 3.6\micron\ warm mission point 
response function\footnote{\url{https://irsa.ipac.caltech.edu/data/SPITZER/docs/irac/iracinstrumenthandbook/5}}).
The black circles in the zoomed-in residual maps in Figure~\ref{fig:galfit}
indicate the aperture sizes used to measure the candidate fluxes.
We apply aperture corrections of 1.69 (3.6\micron) and 1.70 
(4.5\micron), obtained by measuring the flux of our custom PSFs in circular 
apertures of increasing radii.

We estimate the flux uncertainty in these measurements in the follow way.
We place $r=2.5$ pixel circular apertures randomly across the full, 
background-subtracted IRAC images, using \se\ segmentation maps to
avoid source flux. With Photutils, we measure the flux in these apertures, 
fit a Gaussian to the flux distribution, and take the standard deviation as 
the flux uncertainty in the apertures.
We then treat this $1\sigma$ uncertainty as a noise floor, adopting it as 
the minimum uncertainty for both the \galfit\ model fluxes and those 
measured through aperture photometry. Indeed, the majority of the 
\galfit\ uncertainties are larger than the background-measured standard 
deviation, indicating that the \galfit\ photometry is also accounting for 
uncertainties related to the source flux deblending. 

Finally, we rerun \eazy\ for each high-redshift candidate,
incorporating all available \hst\ and \spitzer/IRAC photometry.
As mentioned previously, we fit each IRAC observation
with \galfit\ separately, leading to multiple photometric measurements 
for some candidates.
We adopt a weighted mean of the available measurements for each IRAC
channel in our \eazy\ runs. 
We note that the \eazy\ results -- including the
best-fitting SED templates and the redshift probabiltiy distributions -- are
nearly identical if we instead treat each IRAC flux as an independent 
measurement, i.e., with multiple instances of the IRAC $3.6\mu$m and 
4.5\micron\ filters in the \eazy\ setup files. The best-fitting redshifts 
from each method (weighted mean versus multiple photometric measurements)
differ by at most $\Delta z < 0.05$. 

With the exception of one candidate (Par2346-0021\_164, discussed in 
Section~\ref{sec:par2346}), the \hst\ and \spitzer/IRAC photometry 
continue to prefer a high-redshift solution. 
For many of these candidates, the addition of the IRAC photometry reduces 
the fraction of the redshift PDF that lies at $z<6$, decreasing this
fraction from $\sim$0.1-0.2 in most cases to $<$0.01. The most extreme
improvements were for candidates Par0926+4000\_369 
($p(z$$<$$6)_{\mathrm{HST}}=0.295$ to $p(z$$<$$6)_{\mathrm{HST+IRAC}} <0.001$) and
Par0843+4114\_120 
($p(z$$<$$6)_{\mathrm{HST}} = 0.236$ to $p(z$$<$$6)_{\mathrm{HST+IRAC}} =0.009$).
This rejection of the lower-redshift
solutions typically occurs when the upper limits measured in one or both of the 
IRAC channels fall below the flux density expected for lower-redshift, red 
and dusty galaxies. 
For some candidates, the \eazy\ run with IRAC photometry even helped narrow 
the redshift probability peak at $z>8$, thereby tightening the photometric 
redshift constraints. The 95\% interval of the redshift PDF at $z>8$ decreases
by $z\sim0.36$ on average and as much as 0.9 for candidate 
Par0926+4000\_369. In these cases the IRAC measurements, as either 
detections or upper limits, help constrain the expected location of the 
4000\AA\ break. 
We show the best-fitting SED templates and photometric 
redshift PDFs incorporating  \hst\ and IRAC photometry for our high-redshift 
candidates in Figures~\ref{fig:sample1} and \ref{fig:sample1b}, 
and discuss the IRAC results of individual candidates in 
Section~\ref{sec:cands}.

\subsubsection{Field Par2346-0021} \label{sec:par2346}
While almost all of the IRAC observations listed in Table~\ref{tab:irac} 
covered a single pointing with relatively small dithers,  
those covering HIPPIES field Par2346-0021 (PI: Richards; PID 90045) are a 
little different. 
This program aimed to cover a large enough volume to study the clustering 
and luminosity function of quasars at $z>3$, and so it created wide-area maps 
along the Sloan Digital Sky Survey (SDSS) `Stripe 82'.
The PBCD maps at the position of our high-redshift candidate cover
$\sim$$2.2^{\circ} \times 0.65^{\circ}$ and are significantly shallower than 
the IRAC imaging available for the rest of our sample. The 1$\sigma$ flux 
uncertainties in these large maps, measured in $r=2.5$ pixel circular 
apertures as described above, are $\sim$1545 nJy and 
$\sim$1407 nJy at 3.6\micron\ and 4.5\micron, respectively, while the 
median uncertainties for the rest of the fields are 302 nJy and 737 nJy.
This difference corresponds to limiting magnitudes that are $\sim$1.8 
(3.6\micron) and 0.7 (4.5\micron) magnitudes brighter in this field than 
the others we consider in Section~\ref{sec:irac}.
The much larger imaging area also provides far more point sources for
the creation of a custom PSF.

We therefore measure IRAC photometry for the candidate in Par2346-0021 
separately, following the same steps described in Section~\ref{sec:irac} but 
with field-specific custom PSFs. 
We create the PSFs for this field using only sources identified 
in the Par2346-0021 IRAC maps, though we use a slightly 
more conservative magnitude cut when identifying point sources because there 
are so many to work with: $14 < M_{\mathrm{IRAC}} < 17.5$.
The PSFs we create for Par2346-0021 are comprised of 1561 (3.6\micron) and 
1259 (4.5\micron) sources and are also displayed in Figure~\ref{fig:iracpsfs}
in columns titled `Par2346'.
The aperture corrections we measure using the 3.6\micron\ PSF (1.67) is 
similar to that measured for the other IRAC observations, and the 4.5\micron\ 
aperture correction (1.98) is slightly larger. 
We then proceed as before with \galfit\ and measuring source photometry and
flux uncertainties. 

\begin{figure}
\gridline{\fig{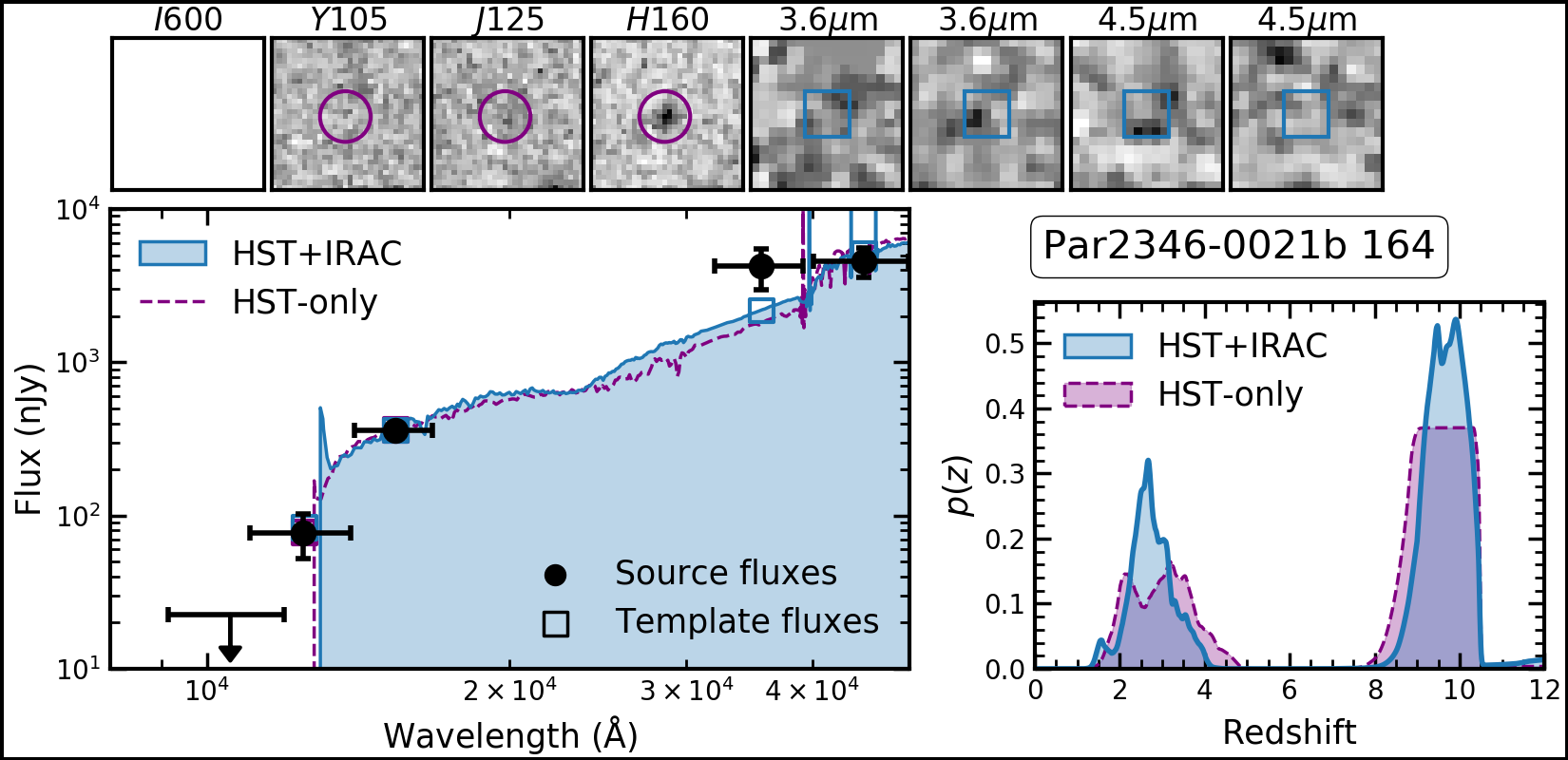}{0.45\textwidth}{}}%{(a)}
\vspace{-6mm}
\gridline{\fig{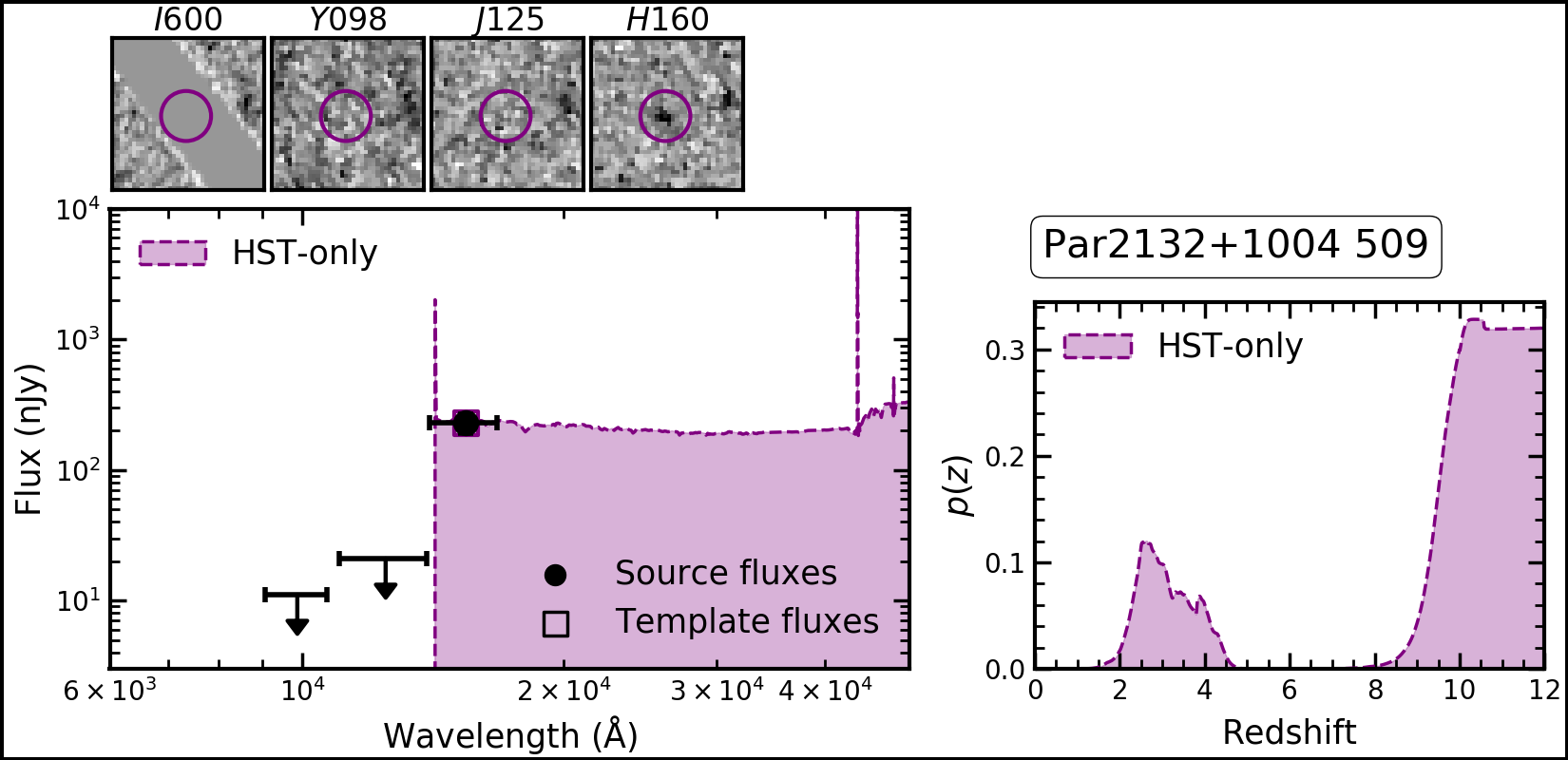}{0.45\textwidth}{}}%{(a)}
\caption{The image stamps, photometry, best-fitting SEDs and redshift PDFs
for two candidates we have removed from our sample. For each candidate we
show 3\arcsec\ \hst\ stamps and, where available, 10\arcsec\ \galfit\ 
residual maps for the IRAC imaging from each individual AOR with all modeled 
flux removed except that attributed to the candidate.
The purple $r=0\farcs5$ circles identify the position of the candidate, and
the blue 3\arcsec\ squares indicate the size of the \hst\ stamps on the larger
IRAC images. 
Beneath the image stamps we show the SEDs in the
left panel and the photometric redshift PDFs in the right, with the fit to
\hst+IRAC photometry (solid blue curve) and \hst-only photometry (dashed purple
curve).  
\textit{Top:} We remove Par2346-0021\_164 due to an increase in its redshift 
PDF at $z\sim2-4$ after including IRAC photometry. 
\textit{Bottom:} We remove Par2132+1004\_509 from our sample primarily due to 
its low $\Delta \chi^2 = 2.53$, but also because it is only observed in 3 
\hst\ bands and is not covered by \spitzer/IRAC. 
\label{fig:par2346}}
\end{figure}

The \galfit\ model brightnesses calculated for Par2346-0021\_164, 
when combined with the \hst\ photometry, reduce our confidence in this 
candidate.
As shown in Figure~\ref{fig:par2346}, this source has a very red $H-[3.6]$ 
color that is consistent with a dusty, red SED.  
The photometric redshift PDF obtained with the \hst+IRAC photometry 
(blue, solid line) has an increased peak around $z\sim3$ when compared with 
that from the \hst\ photometry alone (purple, dashed). Due to this added 
probability at lower redshifts, only 62\% of the integrated PDF is at 
$z>8$, rather than the 70\% threshold we require as part of our selection
criteria.

While we may opt to keep an \hst-selected source in our sample if we find 
that we cannot trust the IRAC photometry, that is not the case for this 
candidate. It lies outside of the existing \iaband\ imaging in this field,
and so is only covered by three \hst\ filters, and the \hst-only photometric
redshift solution is not well-constrained (the 68\% redshift interval 
ranges from $z=3.07$ to 10.06). Additionally, the 
candidate does not appear to have any close neighbors in the \hband\ image,
and so we consider the IRAC photometry reliable. Although the IRAC imaging 
is shallow, the \galfit\ model magnitudes for the candidate are significant. 
We therefore remove Par2346-0021\_164 from our sample as a possible $z<7$
contaminant, but note that it could be a viable $z\sim9$ candidate.

%%%%%%%%%%%%%%%%%%%%%%%%%%%%%%%%%%%%%%%%
\subsection{Photometric Redshift Goodness-of-Fit}
While we did not include a minimum $\chi^2$ threshold on the \eazy\ templates
as part of our selection criteria, we now explore the goodness-of-fit of
these photometric redshifts, including IRAC where available.
Specifically, we consider the difference between the $\chi^2$ of the
best-fitting template and the minimum $\chi^2$ at $z<6$.
This $\Delta \chi^2$ provides an estimate of the quality of the photometric
redshift fit at high-redshift compared to the best-possible fit at lower
redshift. In all cases, the $\chi^2$ of the best-fitting template
is lower than that at $z<6$, with a median
$\Delta \chi^2 = \chi^2_{z<6} - \chi^2_{\mathrm{best}} = 16.8$.
However, the $\Delta \chi^2$ for candidate Par2132+1004\_509 is 2.53,
lower than the value corresponding to the 95\% confidence interval
($\chi^2 > 4$). This candidate also has a broad PDF (the redshift interval
containing 68\% of the PDF ranges from $z=4.01$ to 11.51),
is not covered by IRAC imaging, and falls on the UVIS chip gap
in the \iaband\ imaging (see Figure~\ref{fig:par2346}).
We therefore consider this source as an unreliable high-redshift
candidate and remove it from our sample.

%%%%%%%%%%%%%%%%%%%%%%%%%%%%%%%%%%%%%%%%
\subsection{Comparison with Stellar Colors} \label{sec:spex}

Finally, we explore whether any of the $z\gtrsim9$ candidates are likely to 
in fact be stars. 
Low-mass stars and brown dwarfs can have NIR colors that are very similar to 
those of high-redshift galaxies, and so are common sources of contamination 
in Lyman break samples.
The case for potential contamination is made worse by the fact that 
all but one of the candidates in our sample are unresolved in the 
\hst\ imaging. 
Eight of the candidates have half-light radii (\rhalf) as measured in 
\hband\ that are smaller than the median \rhalf\ of the stars in 
their respective datasets (see Section~\ref{sec:psfs}). The \rhalf\ of another
four candidates are within $1\sigma$ of the median values. 
The one exception is Par0456-2203\_473, which has a larger \rhalf\
and is discussed further in Section~\ref{sec:magnification}. 
The very compact sizes measured for the majority of our sample 
are not necessarily concerning. These sources have fairly low 
S/N in undithered data, and so are not expected to show significant 
extended morphologies. However, it does make a comparison with stellar 
colors crucial. 

We therefore compare the colors of the 13 remaining candidates with 
those of M, L, and T dwarfs. The stellar colors are calculated from 
spectra in the range $\sim$0.63$-$2.5\micron\ taken with the medium-resolution ($R \sim 2000$) spectrograph 
SpeX \citep{rayner2003} at the NASA Infrared Telescope Facility (IRTF). 
We downloaded from the IRTF Spectral 
Library\footnote{\url{irtfweb.ifa.hawaii.edu/~spex/IRTF_Spectral_Library}} 
\citep{burgasser2014}
582 observations of 449 unique stars (132 M dwarfs, 210 L dwarfs,
and 107 T dwarfs), covering the full range of temperature subclasses and 
including $J$, $H$, and $Ks$ photometry from the Two Micron All Sky 
Survey (2MASS). 

\begin{figure}
\epsscale{1.1}
\plotone{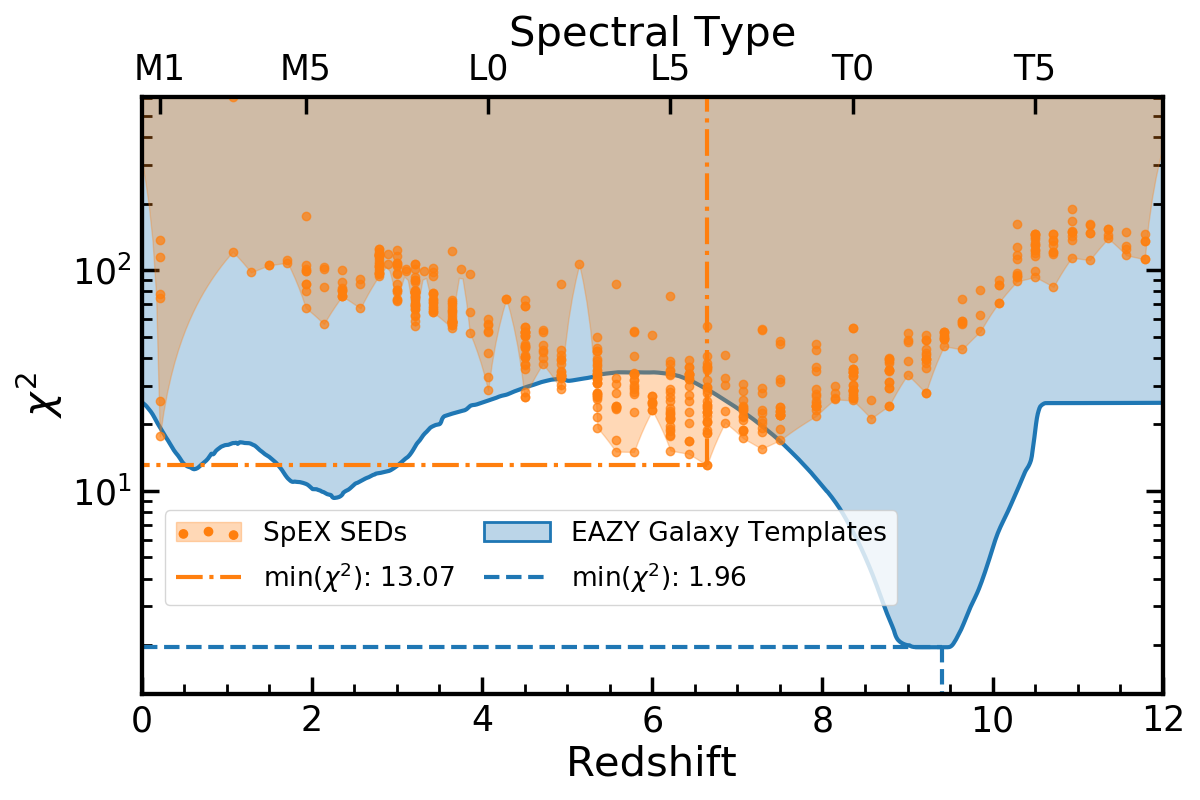}
\caption{The SpeX-derived goodness-of-fit $\chi^2$ values as a function of 
spectral type (orange, top axis) compared with that determined 
by \eazy\ as a function of redshift (blue, bottom axis) for high-redshift 
candidate Par0713$+$7405 ID 95. The $\chi^2$ calculated for each SpeX spectrum
is identified by a separate point, indicating the variation measured for a 
given spectral type. The minimum galactic and stellar $\chi^2$ values are 
identified by horizontal dashed and dashed-dotted lines, respectively. 
The difference in $\chi^2$ values for this source is 
$\Delta \chi^2 = \chi^2_{\mathrm{stellar}} - \chi^2_{\mathrm{galactic}} = 11.11$, which is the smallest $\Delta \chi^2$ of all 
candidates in our sample. All candidates are better fit by galactic templates
than by MLT dwarf spectral templates.
\label{fig:spex}}
\end{figure}

We calculate broadband \hst\ fluxes by integrating each SpeX spectrum through 
the \hst\ filter profiles shown in Figure~\ref{fig:filters}. The filters 
are interpolated onto the observed wavelength array of each spectrum such that 
only the portion of the filter profile that overlaps the spectrum is used in 
calculating broadband fluxes. For each SpeX spectrum, we remove any fluxes 
that result from a $<$25\% wavelength overlap between the filter and the 
spectrum. Following \citet{finkelstein2022a}, we then 
assign IRAC magnitudes to each stellar spectrum using the median 
$Ks-[3.6]$ and $[3.6]-[4.5]$ presented by \citet{patten2006}. We match the 
SpeX stars to the colors in Tables 1 and 3 from \citet{patten2006} by 
spectral type, using $\Delta\mathrm{type}<1$ where possible and 
$\Delta\mathrm{type}<3$ for a few sources with no closer spectral matches. 
We convert all photometry -- 2MASS $J$, $H$ and $Ks$ from the SpeX 
Library, and 3.6\micron\ and 4.5\micron\ obtained by matching to 
\citet{patten2006} -- from Vega to AB magnitudes using 
$m_{\mathrm{Vega}} + 0.91$, 1.39, 1.85, 2.79 and 3.26 for 2MASS $J$,
$H$, $Ks$, and \spitzer/IRAC 3.6\micron\ and 4.5\micron, respectively. 

Finally, with a full suite of photometry for each SpeX spectrum, we calculate 
a $\chi^2$ goodness of fit for the photometry of each high-redshift candidate.
We then compare this SpeX-derived $\chi^2$ with that from the best-fitting 
galactic template determined by \eazy\, and find that 
$\chi^2_{\mathrm{EAZY}} < \chi^2_{\mathrm{SpeX}}$ for all candidates.
As an example, in Figure~\ref{fig:spex}, we show the $\chi^2$ measured for each 
SpeX specturm given the photometry of Par0713$+$7405\_95. Of all 
candidates in our sample, this source has both the lowest minimum 
SpeX-derived $\chi^2$ (=13.07) and the smallest 
$\Delta \chi^2 (\mathrm{SpeX} - EAZY)=11.1$. 

Therefore, although almost all of the 13 candidates are unresolved, 
none of their \hst/WFC3 and \spitzer/IRAC (when available) colors are better 
fit by these 500$+$ M, L, and T dwarf spectra. 
We conclude that all 13 
candidates are better fit by galactic spectral templates and do not remove 
any from the sample due to stellar contamination.
Future observations with \jwst/NIRCam imaging, for example, will vastly
improve on the stellar contamination analysis currently possible in 
high-redshift Lyman break samples by providing both higher-resolution imaging 
in the NIR and imaging in 5$+$ filters redward of the expected position of 
the Lyman break.

\begin{deluxetable*}{cccccccDDccc}
%\tabletypesize{\small}
%\tablenum{4}
\tablehead{
\colhead{Par} & \colhead{ID} & \colhead{Dataset} & \colhead{RA} 
& \colhead{Dec} & \colhead{$m_{160}$} & \colhead{$r_{1/2}$} 
& \multicolumn2c{$z_{\mathrm{phot}}$} & 
\multicolumn2c{$p(z>8)$} & \colhead{68\%} & \colhead{$\Delta \chi^2$} 
& \colhead{$M_{\mathrm{UV}}$}}
\decimalcolnumbers
\tablecaption{Sample of $z\sim9-10$ Candidates} \label{tab:sample}
\startdata
Par1033+5051 & 116 & \borg\ & 10:32:44.7 & $+$50:50:30.1 & 25.99 & 0\farcs11 & 8.28 & 81.29\% & 7.86$-$8.68 & 21.25 & $-21.20^{+0.15}_{-0.14}$ \\
Par0440-5244 & 497 & \borg\ & 04:39:47.0 & $-$52:43:55.2 & 25.47 & 0\farcs15 & 8.53 & 89.06\% & 7.98$-$9.07 & 23.59 & $-21.76^{+0.17}_{-0.15}$  \\
Par0713+7405 & 95\tablenotemark{c} & HIPPIES 12286 & 07:13:13.2 & $+$74:05:04.5 & 25.74 & 0\farcs14 & 9.23 & 98.16\% & 8.79$-$9.65  & 7.33 & $-21.32^{+0.18}_{-0.19}$ \\
Par0456-2203 & 473\tablenotemark{c} & \borg\ & 04:55:51.6 & $-$22:01:36.1 & 24.52 & 0\farcs33 & 9.39 & 93.45\% & 8.81$-$9.80  & 4.70 & $-19.87^{+0.56}_{-0.36}$ \\
Par0259+0032\tablenotemark{a} & 194\tablenotemark{c} & HIPPIES 12286 & 02:59:37.6 & $+$00:33:01.8 & 24.79 & 0\farcs13 & 9.36\tablenotemark{b} & 98.71\% & 8.77$-$9.95 & 8.81 & $-21.13^{+1.06}_{-0.59}$ \\
Par0926+4000 & 369 & \borg\ & 09:25:33.6 & $+$40:00:44.9 & 24.68 & 0\farcs13 & 10.49 & $>$99.99\% & 10.10$-$11.38 &  28.66 & $-22.88^{+0.11}_{-0.12}$ \\
Par335 & 251 & WISP & 15:47:43.4 & $+$20:58:23.3 & 24.74 & 0\farcs19 & 10.53 & $>$99.99\% & 10.24$-$11.61 & 20.41 & $-22.85^{+0.12}_{-0.12}$ \\
Par0956-0450 & 684 & HIPPIES 12286 & 09:56:50.7 & $-$04:49:38.8 & 24.88 & 0\farcs12 & 10.54 & $>$99.99\% & 10.49$-$11.63  & 69.05 & $-22.74^{+0.08}_{-0.09}$ \\
Par0843+4114 & 120 & HIPPIES 12286 & 08:43:25.9 & $+$41:14:00.3 & 26.39 & 0\farcs13 & 10.98 & 99.07\% & 10.30$-$11.68 & 7.79 & $-21.23^{+0.24}_{-0.20}$ \\
Par1301+0000 & 37 & \borg\ & 13:01:13.6 & $-$00:01:26.6 & 25.42 & 0\farcs13 & 11.02 & 99.99\% & 10.46$-$11.61 & 16.49 & $-22.20^{+0.17}_{-0.15}$ \\
Par0756+3043 & 92 & \borg\ & 07:55:54.4 & $+$30:42:22.1 & 26.03 & 0\farcs14 & 11.03 & 99.83\% & 10.39$-$11.71 & 10.83 & $-21.59^{+0.22}_{-0.20}$ \\
Par0440-5244 & 593 & \borg\ & 04:39:55.2 & $-$52:43:49.6 & 26.46 & 0\farcs11 & 11.07 & $>$99.99\% & 10.46$-$11.72 & 17.99 & $-21.16^{+0.15}_{-0.14}$ \\
Par0926+4536 & 155 & HIPPIES 12286 & 09:26:26.2 & $+$45:37:26.2 & 26.37 & 0\farcs13 & 11.15 & 99.99\% & 10.59$-$11.74 & 16.78 & $-21.26^{+0.23}_{-0.20}$ \\
\enddata
\tablenotetext{a}{This field does not have \spitzer/IRAC coverage.}
\tablenotetext{b}{This redshift is based on an upper limit estimated for a 
contaminated \jband\ flux. See text for details.}
\tablenotetext{c}{These candidates were found to experience intermediate-to-strong
lensing, and so the $M_{\mathrm{UV}}$ values have been corrected for magnification
as described in Section~\ref{sec:magnification}.}
\tablecomments{
Columns are: (1-2) field and candidate IDs; (3) Dataset containing field; 
(4-5) candidate right ascension and declination; 
(6) \hband\ apparent magnitude;
(7) half-light radius as measured in \hband; 
(8) best-fitting redshift from \eazy, selected as the peak of the redshift PDF;
(9) percentage of the integrated PDF that is at $z>8$; 
(10) 68\% interval of the redshift PDF; 
(11) difference in the $\chi^2$ of the best-fitting \eazy\ template and the 
minimum $\chi^2$ at $z<7$: 
$\Delta \chi^2 = \chi^2_{z<7} - \chi^2_{\mathrm{best}}$;
(12) median absolute magnitude and 68\% interval of magnitude distributions 
as described in Section~\ref{sec:magnification}.
}
\end{deluxetable*}

%%%%%%%%%%%%%%%%%%%%%%%%%%%%%%%%%%%%%%%%%%%%%%%%%%%%%%%%%%%%%%%%%%%%%%%
\section{Results} \label{sec:results}

%%%%%%%%%%%%%%%%%%%%%%%%%%%%%%%%%%%%%%%%
\subsection{High-redshift Candidates} \label{sec:cands}
After applying all selection criteria, visual inspection, persistence 
screening and stellar color comparison, we have a sample of 13 
high-redshift candidate galaxies spanning the range 
$8.3 \lesssim z \lesssim 11$. 
In Table~\ref{tab:sample}, we list the candidates, their positions, \hband\
magnitudes, half-light radii measured in \hband, and their photometric 
redshifts. The photometric redshift fits all include IRAC where available. 
We also provide the percentage of the integrated photometric 
redshift PDF that lies at $z>8$ (recall that we require $p(z>8) > 70\%$ for 
sample selection) and the redshift interval containing 68\% of the redshift 
PDF in columns (9) and (10), respectively.
The lower bound of this interval is above $z=8$ for all but two of
the candidates, and these both have $p(z>8) > 80\%$. Column (9) also 
demonstrates that while the redshift PDFs are relatively well-constrained to 
be at high redshifts, they are still quite broad with 68\% $p(z)$ intervals
ranging from $\sim$0.8 to $\sim$1.4. Column (11) lists the difference between 
the $\chi^2$ of the best-fitting template and the minimum $\chi^2$ at $z<7$ 
for the same \eazy\ run (see discussion in Section~\ref{sec:irac}).

\begin{figure*}
\gridline{\fig{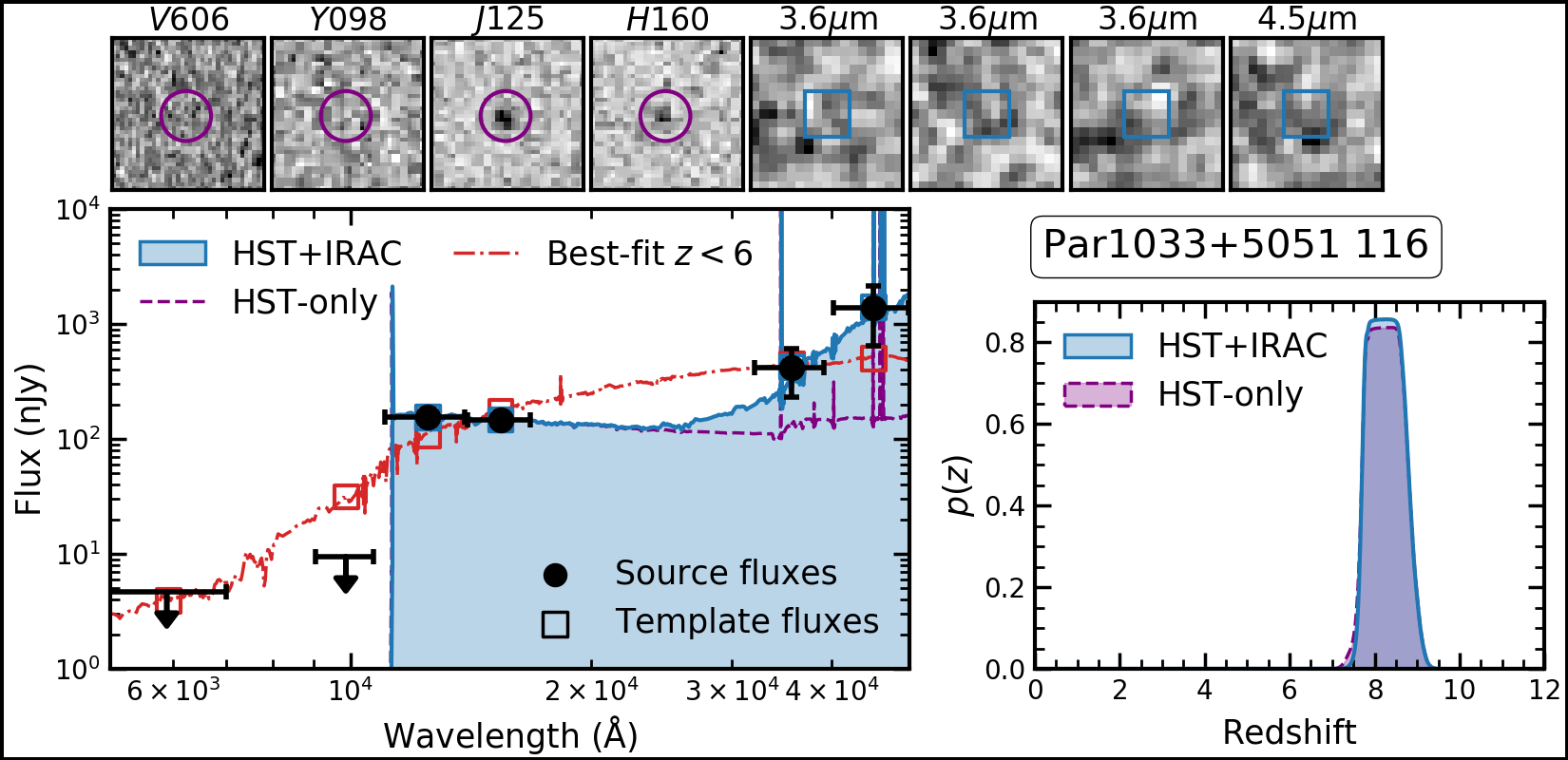}{0.45\textwidth}{}%{(a)}
          \fig{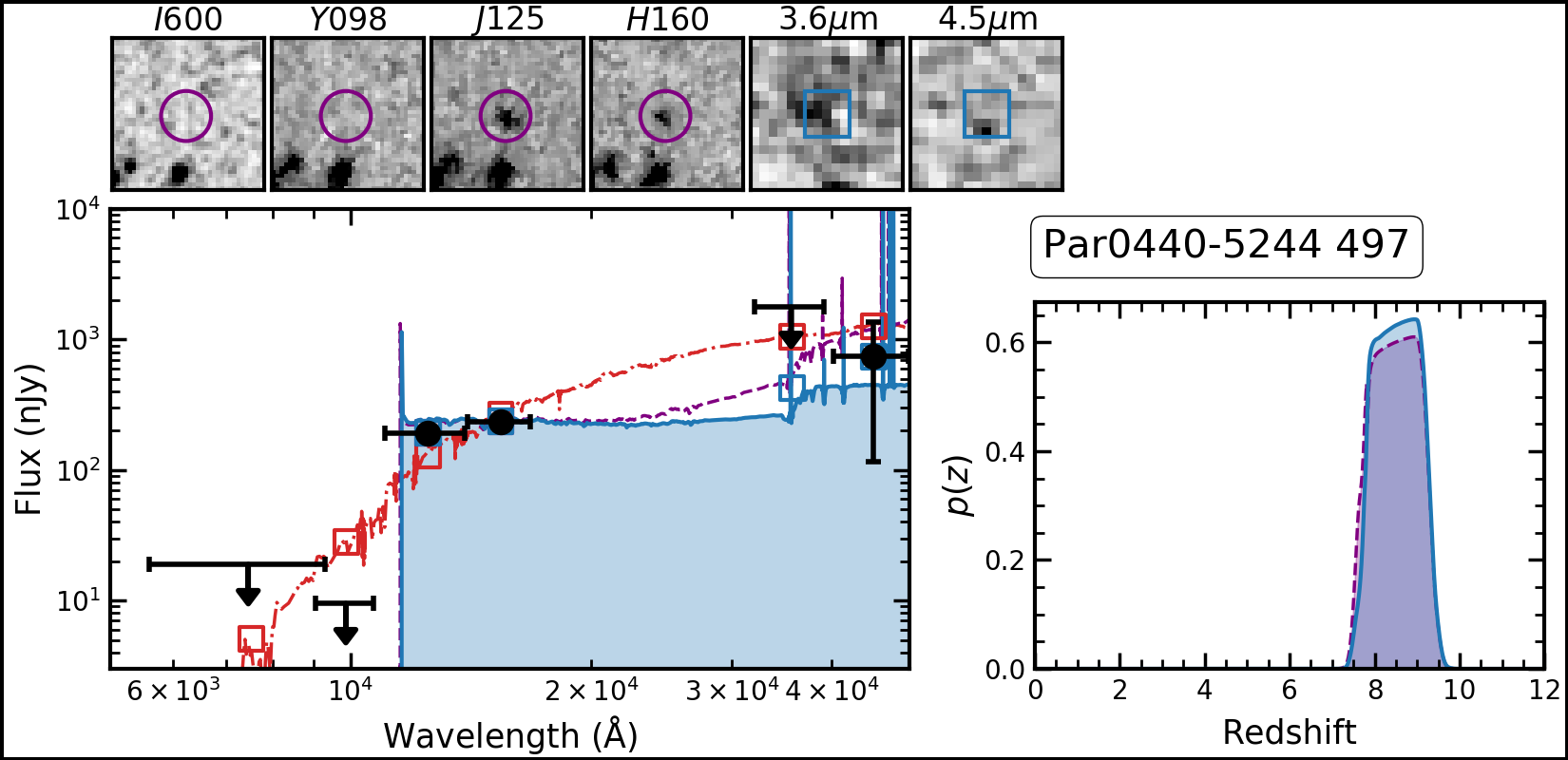}{0.45\textwidth}{}}%{(b)}}
\vspace{-6mm}
\gridline{\fig{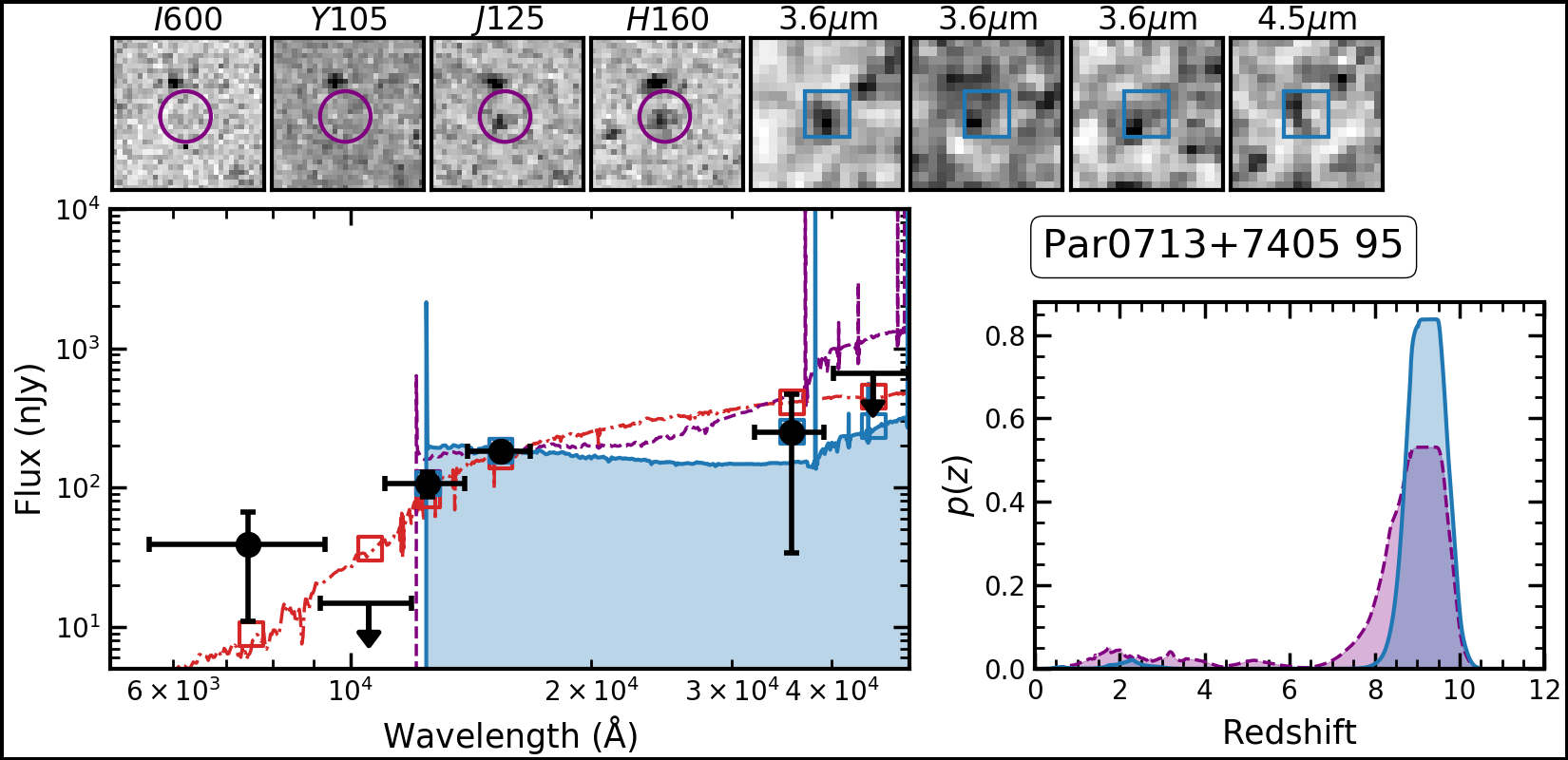}{0.45\textwidth}{}%{(c)} 
          \fig{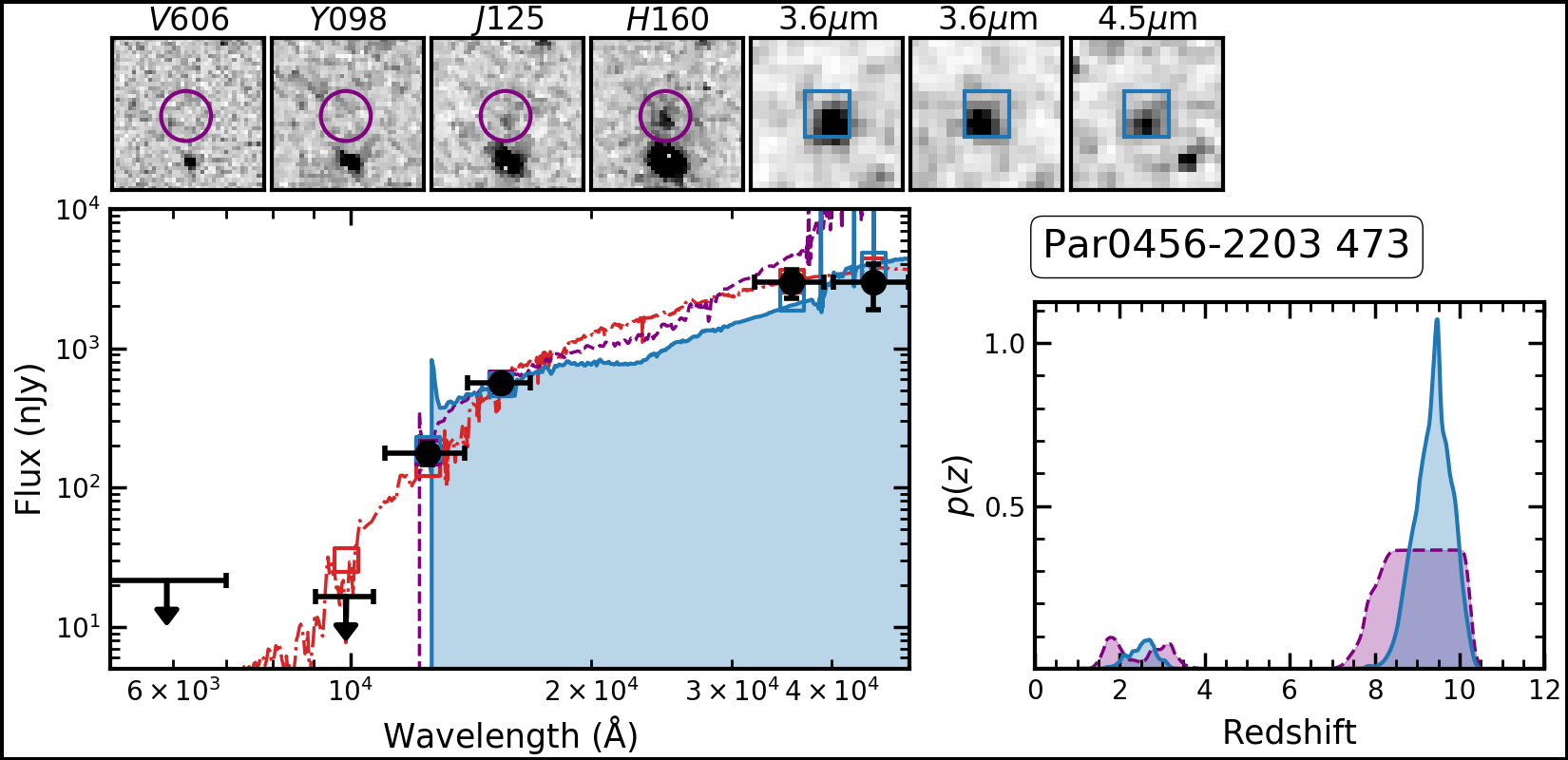}{0.45\textwidth}{}}%{(d)}} 
\vspace{-6mm}
\gridline{\fig{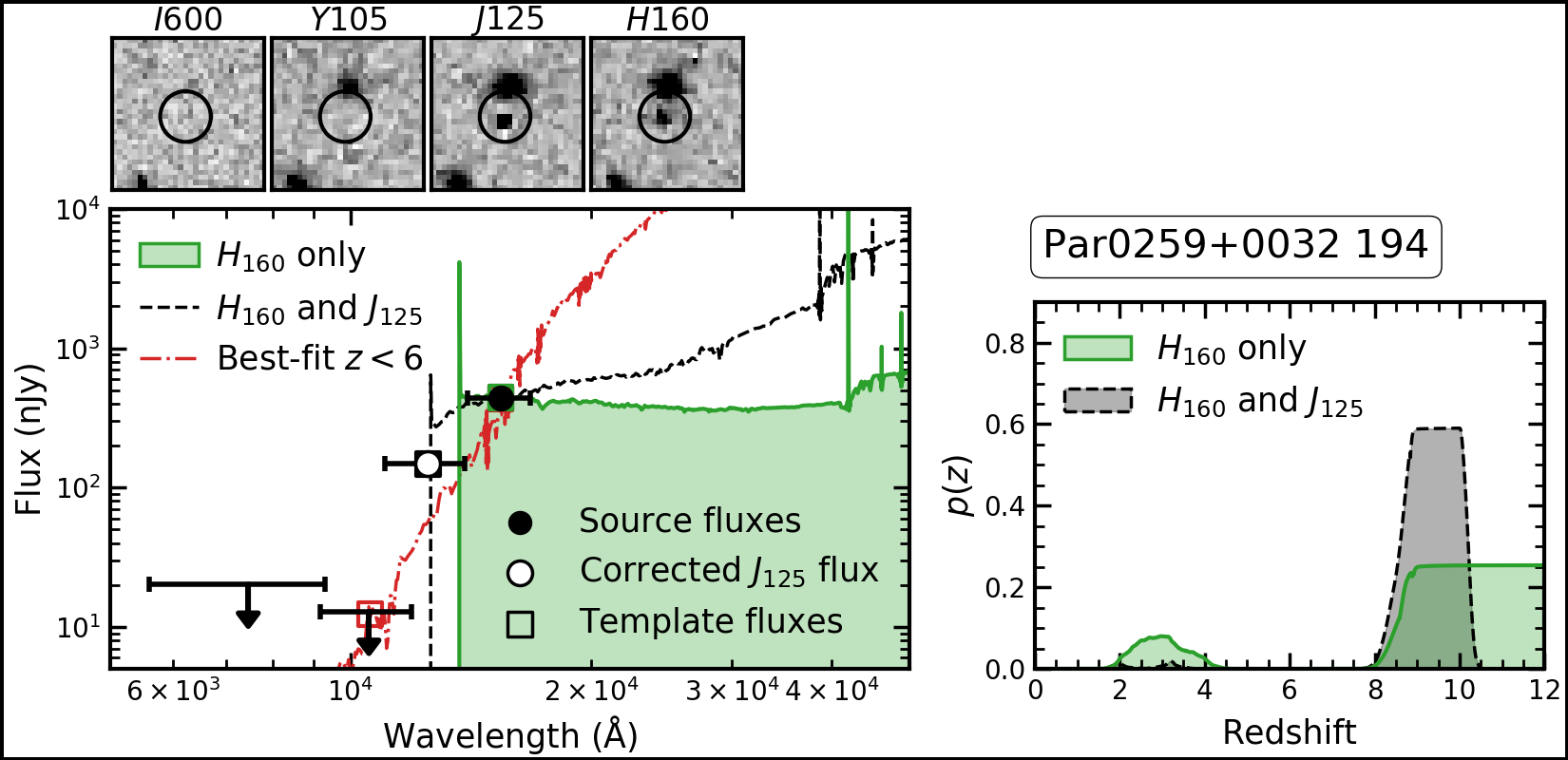}{0.45\textwidth}{}%{(e)} 
          \fig{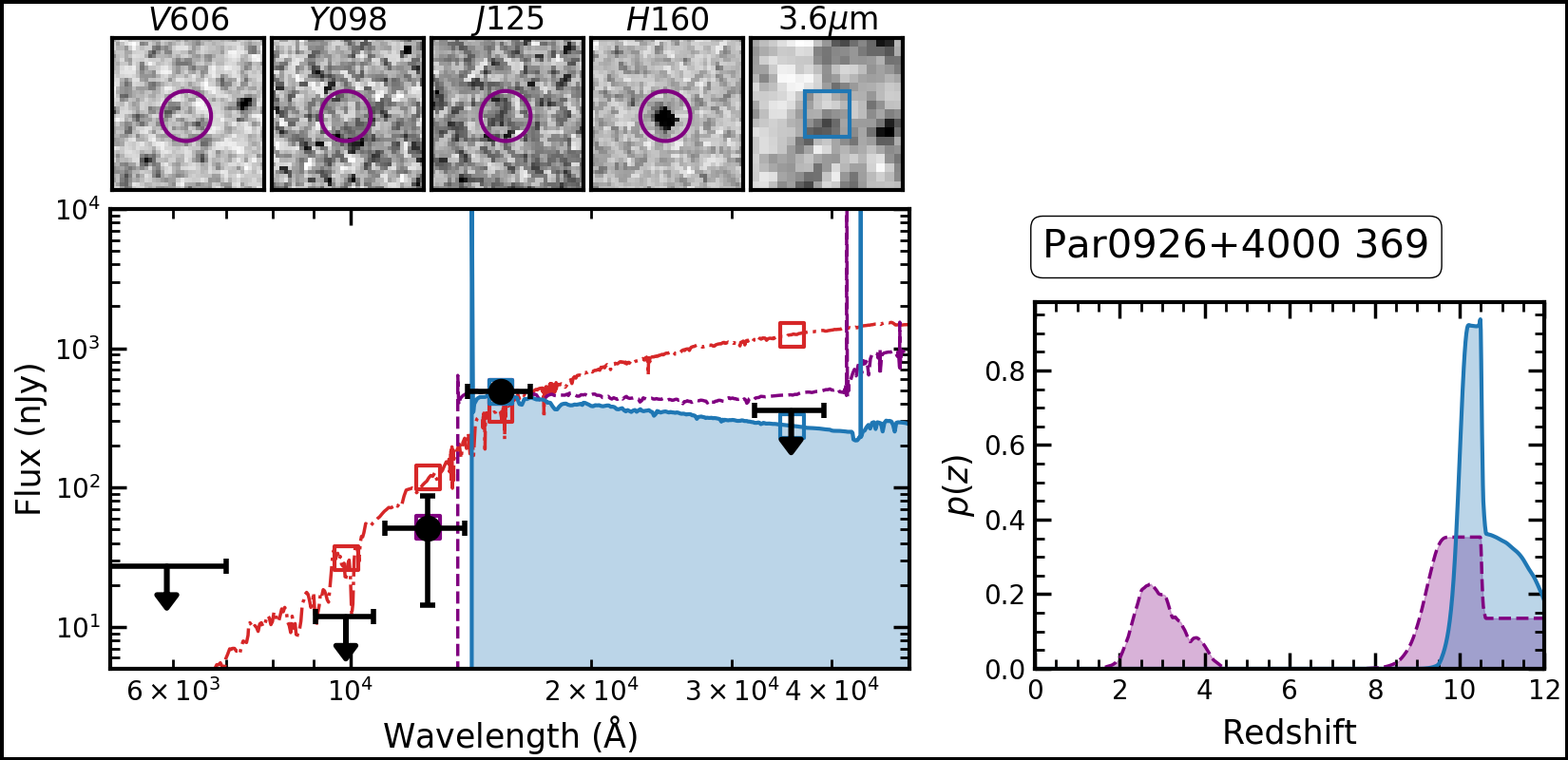}{0.45\textwidth}{}}%{(f)}} 
\caption{The image stamps, photometry, best-fitting SEDs and redshift PDFs
for the first 6 candidates in our sample, sorted by photometric redshift. 
For each source, we show all available imaging including 3\arcsec\ \hst\ 
stamps and 10\arcsec\ \galfit\ residual maps for the IRAC imaging from 
individual AORs with all modeled flux removed except that attributed to 
the candidate. The purple $r=0\farcs5$ circles identify the position of the 
candidate in the \hst\ images, and the blue 3\arcsec\ squares indicate the 
size of the \hst\ stamps in the larger IRAC images. 
Beneath the image stamps we show the SEDs in the 
left panel and the photometric redshift PDFs in the right.
We show the measured photometry as filled circles 
and the predicted fluxes for each template as open squares. The blue solid 
curve and blue shaded regions indicate the best-fitting template and PDF 
including all \hst+IRAC photometry. The dashed purple curves indicate the 
best-fitting template/PDF considering only \hst\ photometry. The inclusion 
of the IRAC photometry almost always tightens the width of the PDF at high 
redshifts while reducing the portion of the PDF present at lower redshifts. The 
red dot-dashed curve shows the best-fitting template/PDF at low redshift,
when \eazy\ is restricted to $z<6$. 
These candidates are all better fit with a high-redshift galactic spectral 
template than that of a lower-redshift galaxy.
We use a different color scheme with Par0259+0032\_194, and show the 
photometric redshift fits when the \jband\ photometry
is excluded (solid green curve) and when a lower limit is asssumed as described
in the text (dashed black curve). 
\label{fig:sample1}}
\end{figure*}

\begin{figure*}
\gridline{\fig{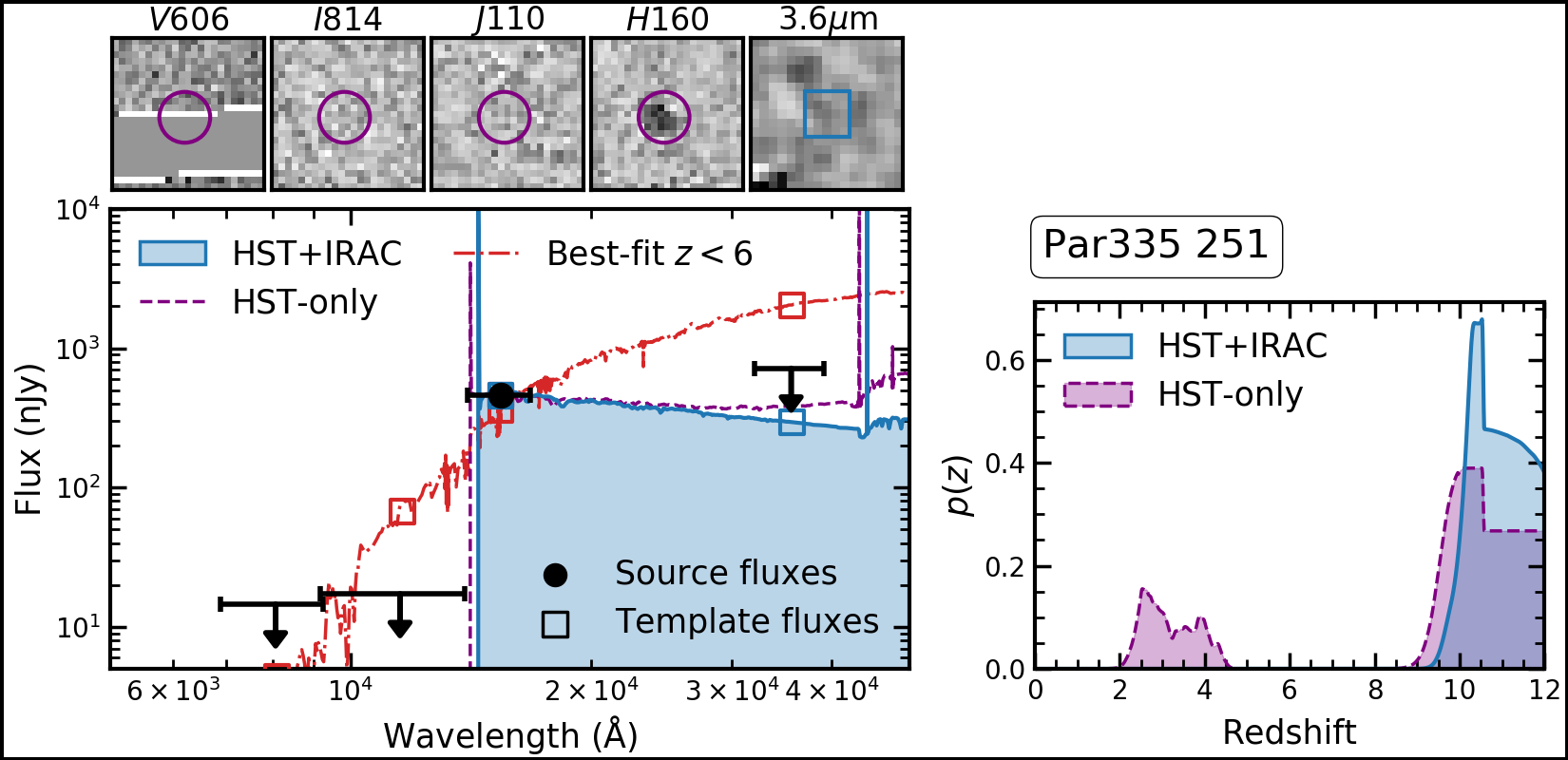}{0.45\textwidth}{}%{(g)} 
          \fig{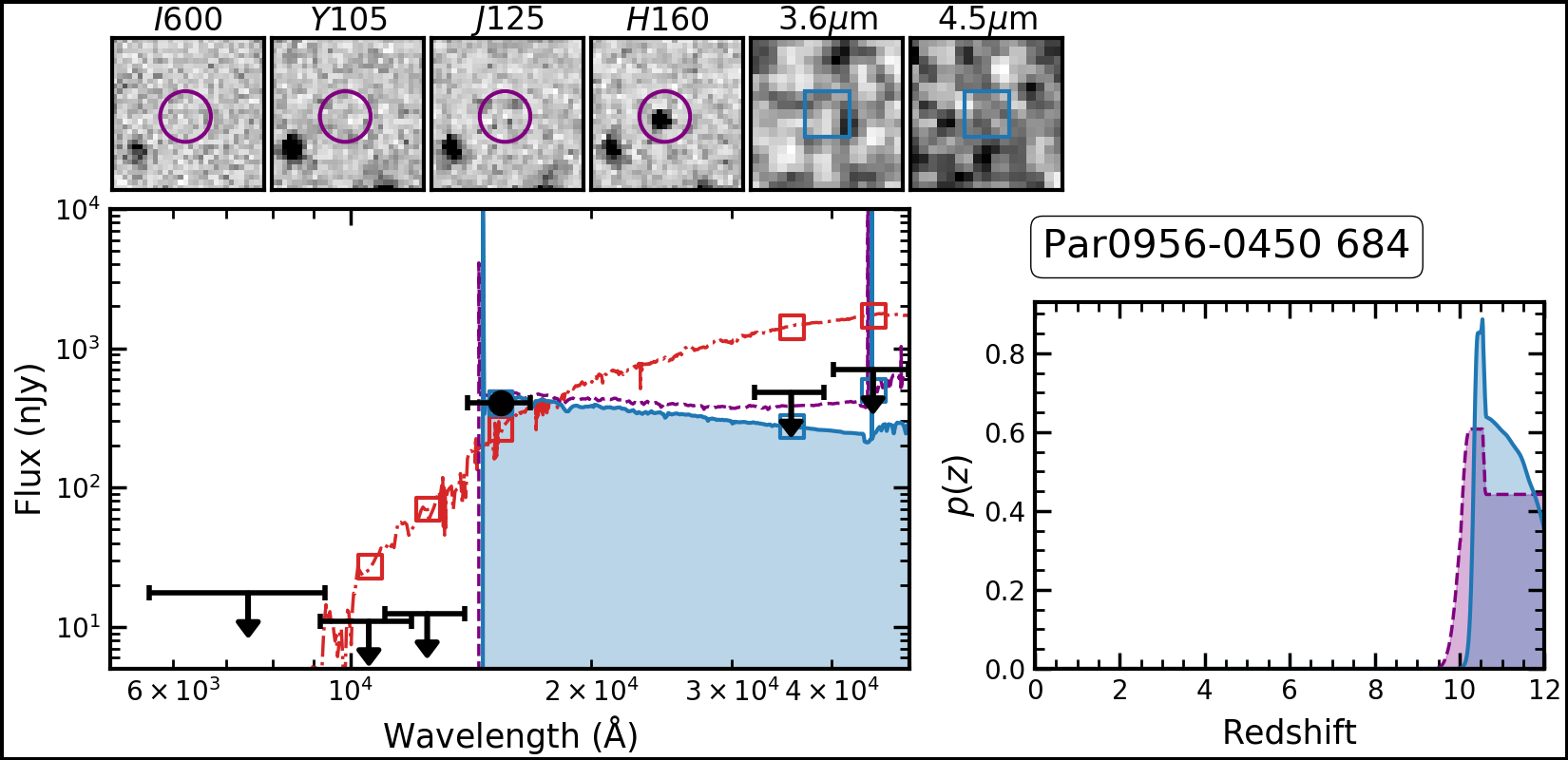}{0.45\textwidth}{}}%{(h)}} 
\vspace{-6mm}
\gridline{\fig{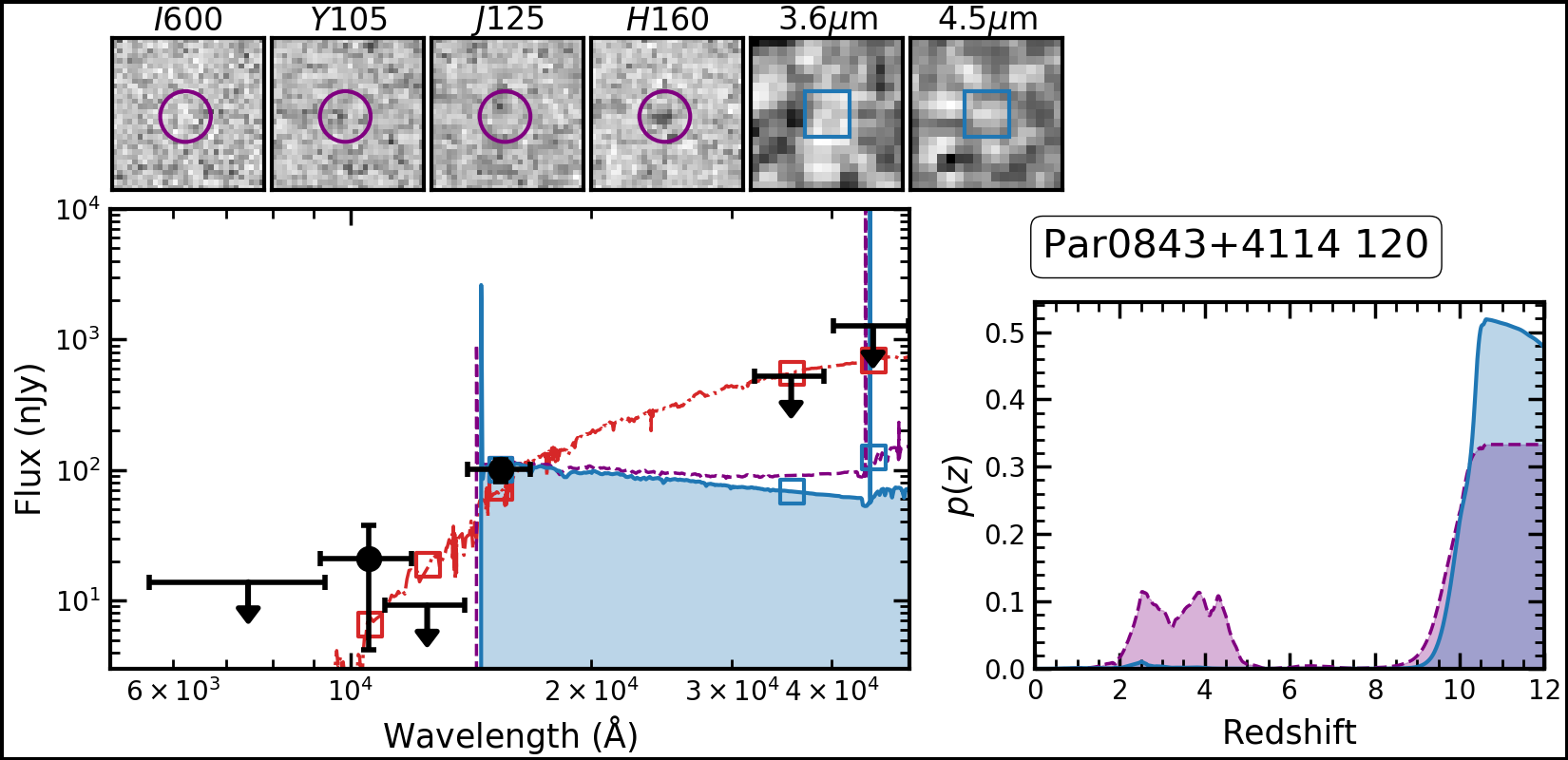}{0.45\textwidth}{}%{(i)} 
          \fig{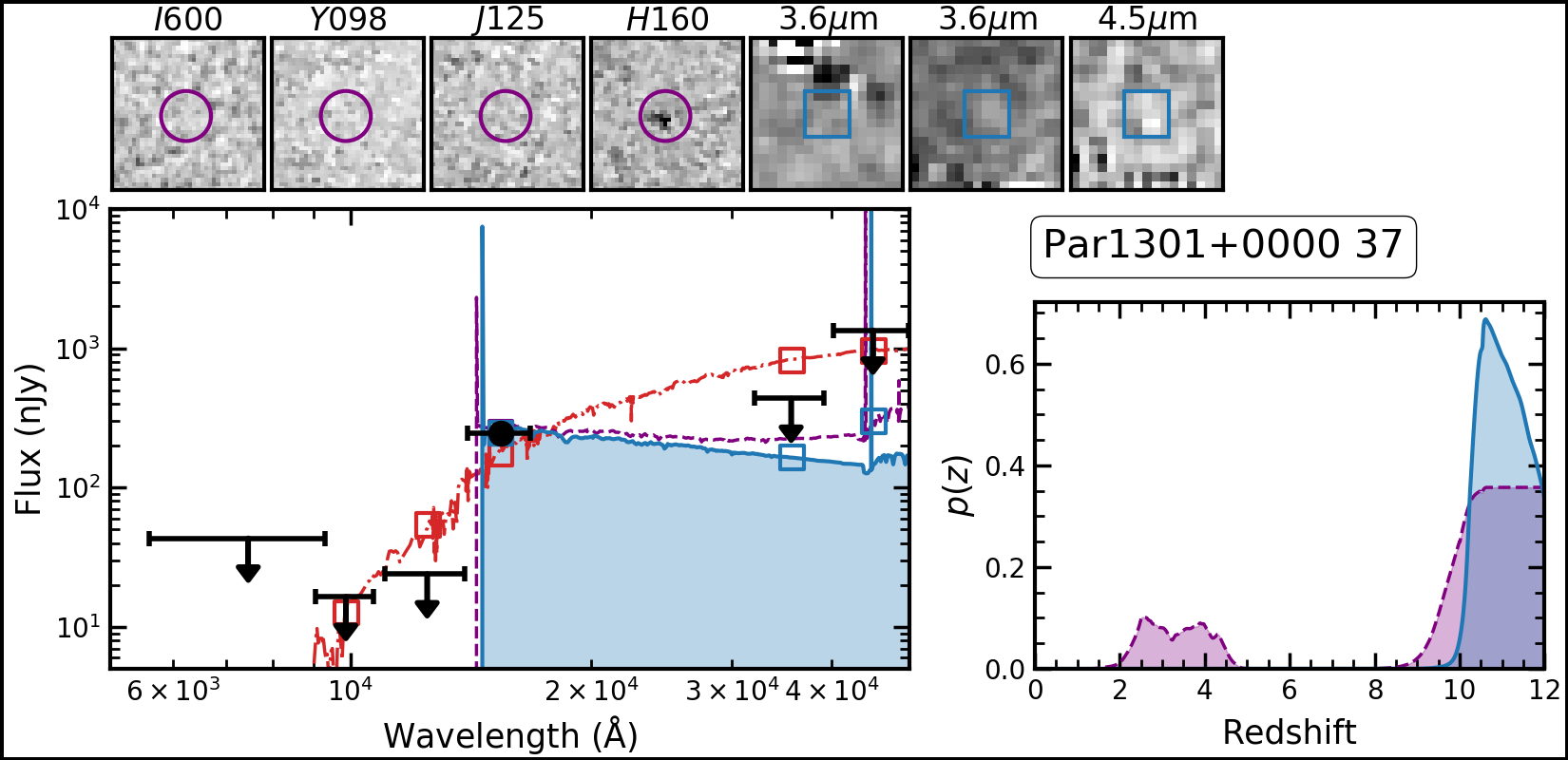}{0.45\textwidth}{}}%{(j)}} 
\vspace{-6mm}
\gridline{\fig{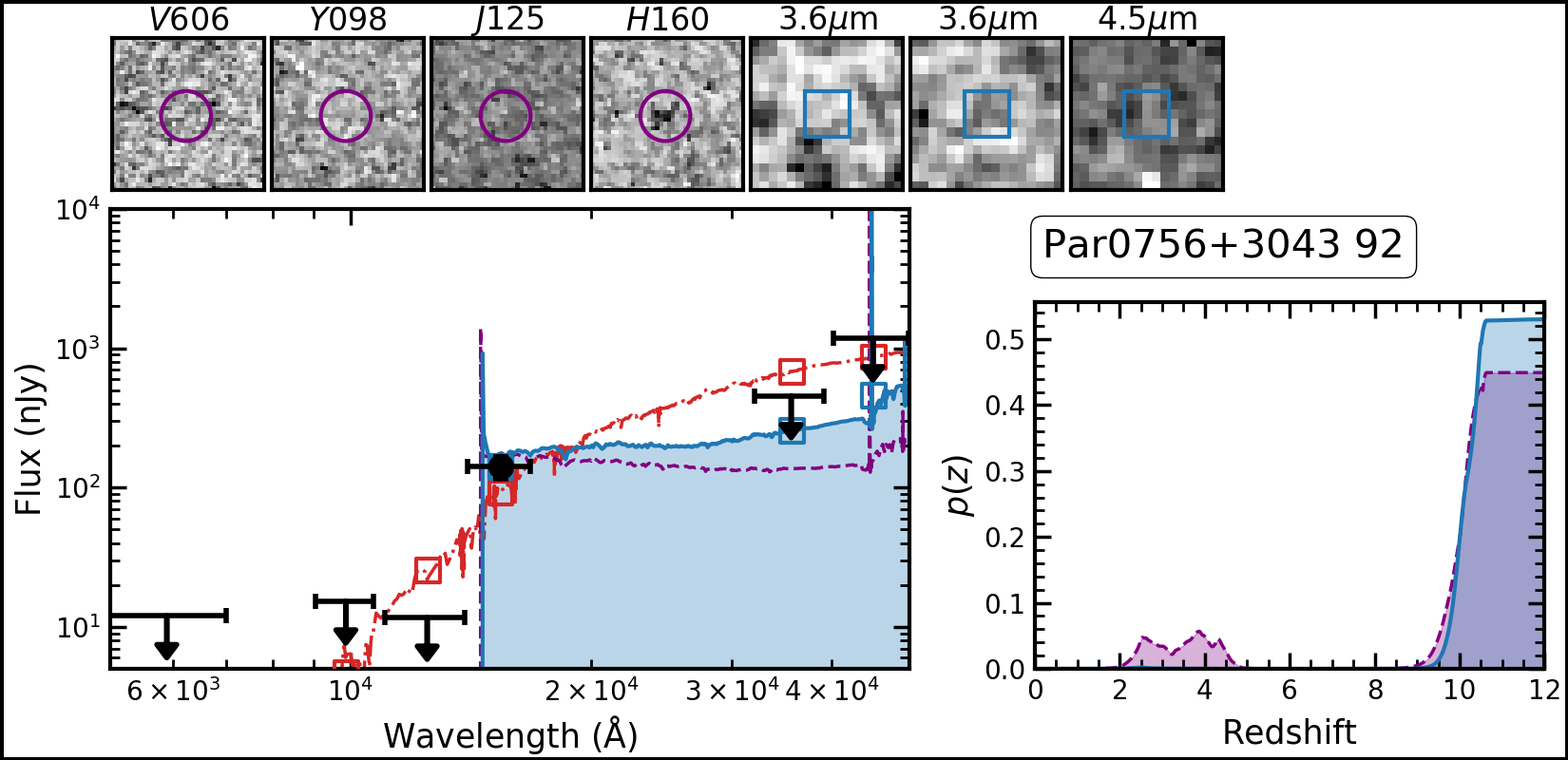}{0.45\textwidth}{}%{(k)} 
          \fig{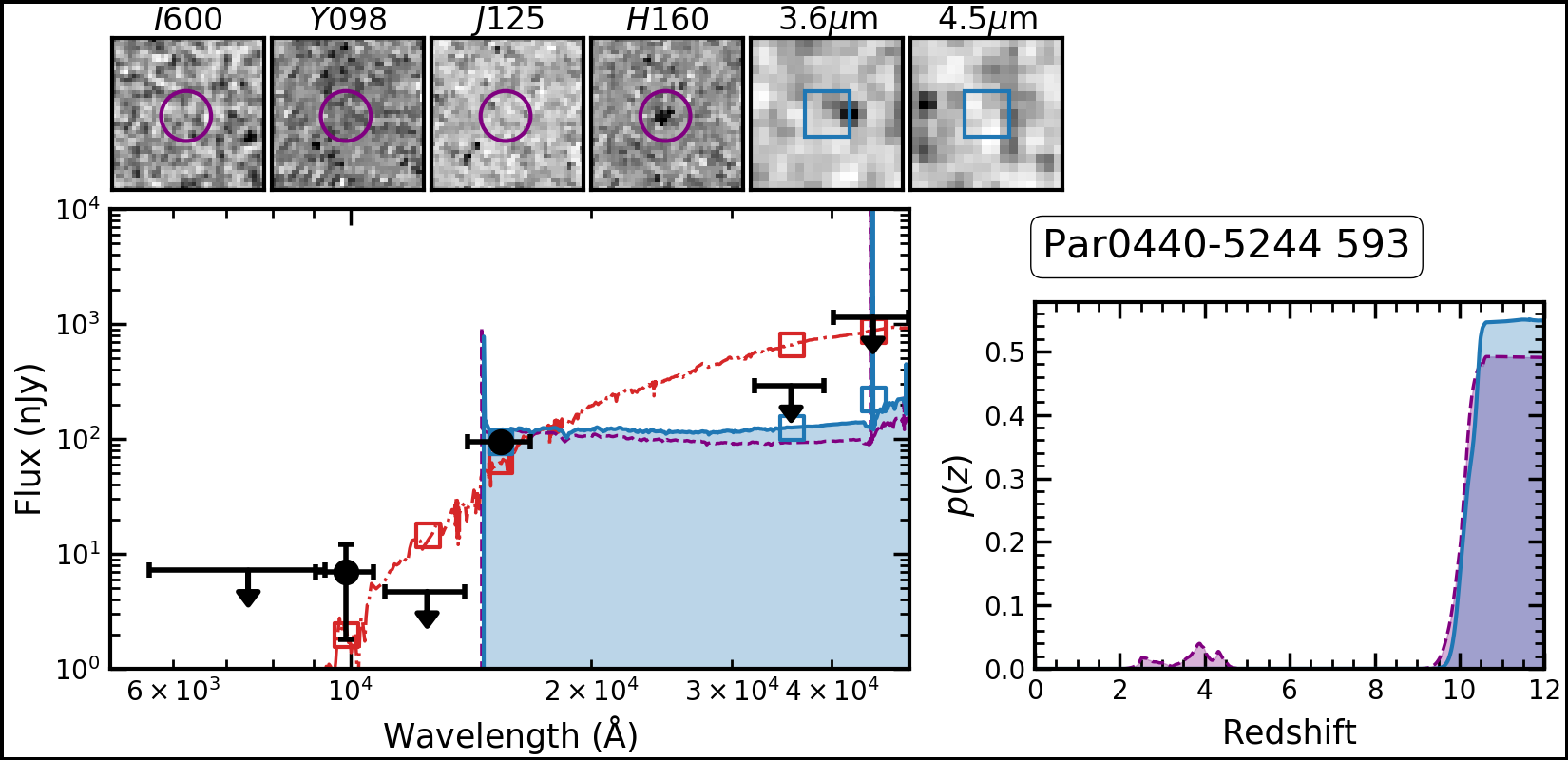}{0.45\textwidth}{}}%{(l)}} 
\vspace{-6mm}
\gridline{\fig{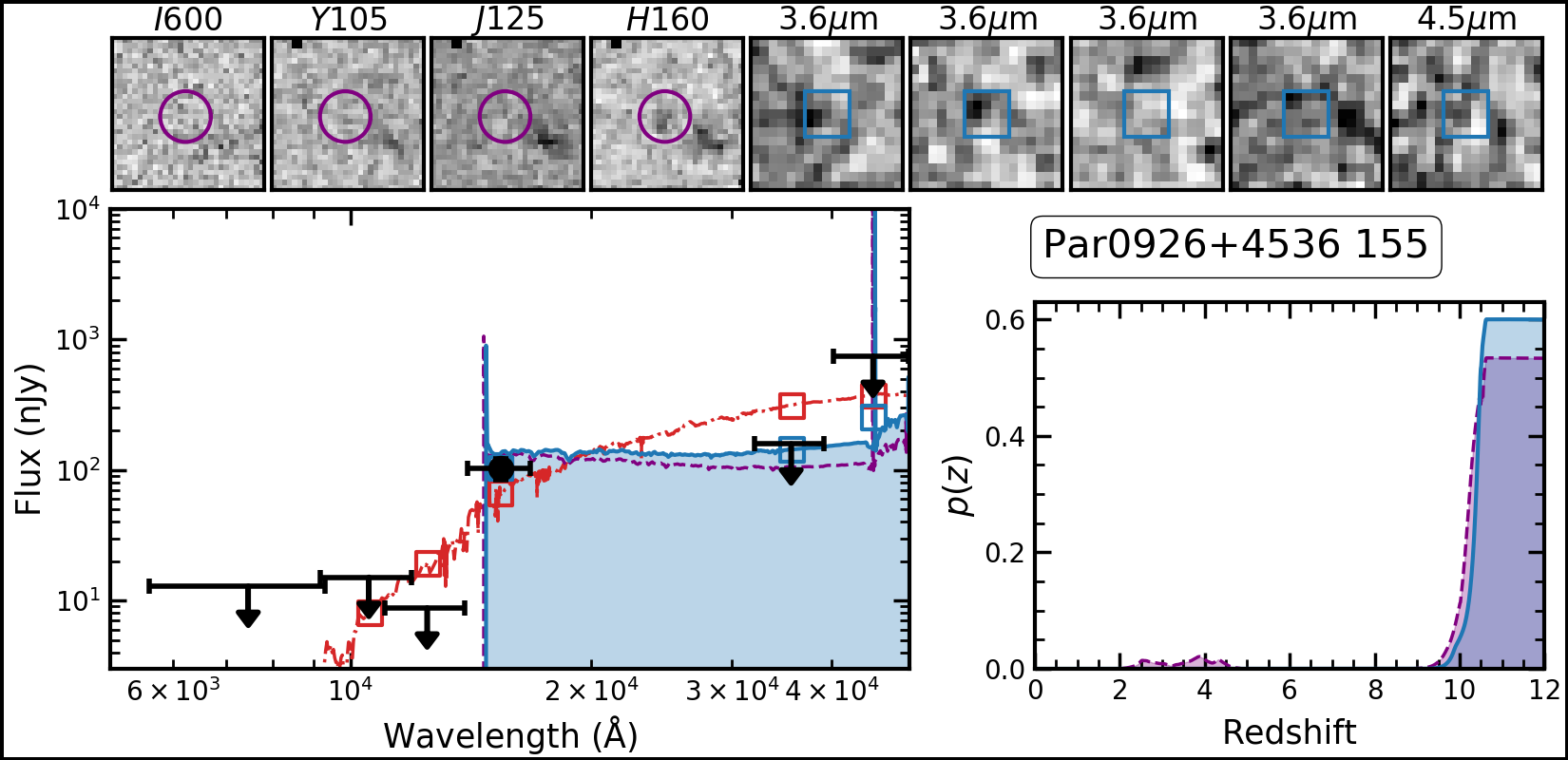}{0.45\textwidth}{}%{(m)} 
        }
\caption{The candidate plots for the remaining 7 candidates. Plots are 
displayed as in Figure~\ref{fig:sample1}. 
\label{fig:sample1b}}
\end{figure*}

In Figures~\ref{fig:sample1} and \ref{fig:sample1b} we show the best-fitting 
SED and photometric redshift PDFs 
for the \hst\ photometry (purple dashed) and the \hst+IRAC photometry when 
available (blue solid). We also show the best-fitting template obtained 
by limiting \eazy\ to $z<7$ as a red dash-dotted curve.
The lower-redshift ($z<7$) templates are ruled out by the upper limits measured 
blueward of the expected Lyman break, especially those in the $Y$-bands for 
candidates at $z<10.5$ and in the $J$-bands for candidates at $z>10.5$. The 
addition of the \spitzer/IRAC photometry also significantly helps, 
especially the upper limits measured in the 3.6\micron\ images for 
candidates at $z>10$. The IRAC non-detections rule out dusty galaxies
with redder $H$-[3.6\micron] colors in favor of higher-redshift sources with 
a flat spectrum or bluer color.
Postage stamps of all available filters are shown along the top, where we 
display the \galfit\ residual maps for each individual IRAC observation.
All modeled fluxes are removed from the residual maps except that attributed
to the candidate. 

In the following subsections, we briefly discuss each candidate.

%%%%%%%%%%%%%%%%%%%%%%%%%%%%%%%%
\subsubsection{Par1033+5051\_116} \label{sec:cand1033_116}
Candidate Par1033+5051\_116 ($m_{160}=25.99$) is detected in both \jband\ 
and \hband. As a two-band detection, we consider this candidate to be 
higher-confidence than the single-band detections in the sample, both 
because it is less likely to be a spurious source and because its 
photometric redshift is better constrained. With a photometric 
redshift of $z_{\mathrm{phot}}=8.28$, this candidate is at the low end of 
our sample, yet the 68\% interval extends out to $z\sim8.7$. 

%%%%%%%%%%%%%%%%%%%%%%%%%%%%%%%%
\subsubsection{Par0440-5244\_497 \& 593} \label{sec:cands0440}
Candidate Par0440-5244\_497 ($m_{160}=25.47$) is also a two-band detection, 
with a $z_{\mathrm{phot}}=8.53$. A second candidate in our sample, 
Par0440-5244\_593 ($m_{160}=26.46$) is detected in the same \hst\ pointing 
at a distance of $\sim$1\farcm25 from source 497, corresponding to 
$\sim$0.35 pMpc at $z=8.5$. 
However, as Par0440-5244\_593 is a single-band detection with a photometric 
redshift of $z_{\mathrm{phot}}=11.07$, the proper distance between 
the two sources is $\sim$58 pMpc. Here we have adopted each of their 
photometric redshifts and used the average $z_{\mathrm{phot}}$ 
to convert their comoving distance to a proper distance.  
Given the redshift separation, we do not expect these sources to be 
associated with the same overdensity 
(see discussion in Section~\ref{sec:brightgals}).

%%%%%%%%%%%%%%%%%%%%%%%%%%%%%%%%
\subsubsection{Par0713+7405\_95} \label{sec:cand0713_95}
The third candidate in Table~\ref{tab:sample}, Par0713+7405\_95 
($m_{160}=25.74$), is also a two band detection. While the photometry in 
Figure~\ref{fig:sample1}
shows a detection in \iaband, we note that it is not significant with 
S/N~$<1.5$. The inclusion of the IRAC photometry helps to narrow the width of 
the primary redshift PDF peak, placing $z_{\mathrm{phot}}=9.23$.
However, the rejection of the lower-redshift solution is driven by the 
non-detection in \ybband.

%%%%%%%%%%%%%%%%%%%%%%%%%%%%%%%%
\subsubsection{Par0456-2203\_473} \label{sec:cand0456_473}
The IRAC deblending for Par0456-2203\_473 is uncertain due to its close,
bright neighbor. The flux remaining in the IRAC residual maps has been 
attributed to the candidate, but it is not fully centered in the blue box that 
indicates the size of the \hst\ stamp. This flux may therefore be partially 
or fully coming from the neighboring source. With such a close neighbor, 
any measured fluxes at IRAC wavelengths will be uncertain until the 
much higher resolution \jwst/NIRCam can be used to 
separate the light from these sources. However, the \hst-only 
photometry yields a high-redshift solution that is still preferred even 
when these relatively bright IRAC fluxes are included.
We note that Par0456-2203\_473 has a bright \hband\ magnitude 
($m_{H160} = 24.52$), which could make it one of the brightest candidates 
detected at $z>8$. We calculate an absolute UV magnitude for this source 
of $M_{1500} = -22.86$, $\sim$0.2 brighter than the $z\sim10$ candidate 
presented by \citet{morishita2018} in the BoRG[$z$910] fields.
However, as discussed in Section~\ref{sec:magnification}, this source is 
strongly lensed by its very close neighbor, and is likely much fainter 
intrinsically. This source is also the only one in our sample that received
the aperture correction for stretched apertures described in 
Section~\ref{sec:photcorrections}. The \hband\ magnitude in 
Table~\ref{tab:sample} is an aperture-corrected flux calculated in a circular 
aperture, rather then the elliptical Kron apertures used for the other 
candidates. We note that we measure a large half-light radius for this 
candidate, $r_{1/2} = 0\farcs33$. This \rhalf\ is consistent with the 
radius measured by \citet{holwerda2015} for low-redshift interlopers, yet
is not inconsistent with a high redshift nature \citep{holwerda2020}. Also,
as a strongly lensed source, this 0\farcs33 radius may not represent the 
intrinsic size of this candidate.

%%%%%%%%%%%%%%%%%%%%%%%%%%%%%%%%
\subsubsection{Par0259+0032\_194} \label{sec:cand0259_194}
A fifth candidate, Par0259+0032\_194 ($m_{160}=24.79$), is potentially another two-band 
detection, though there is a cluster of bad pixels in the \jband\ image 
at the position of the source (there is no such cluster affecting \hband). 
The \jband\ flux of this source was masked out in our photometric catalog, and 
the source is treated as having no \jband\ coverage in our analysis
(sample selection, photometric redshift fitting, comparison with SpeX spectra).
However, the removal of \jband\ leads to a poorly-constrained redshift solution
with the source placed at a higher redshift than it likely should, considering
it may very well be detected in \jband. 
We re-measure the \jband\ flux of Par0259+0032\_194 after masking out the 
bad pixels (using \se\ in dual image mode with \hband\ as before). 
We use this \jband\
flux to measure a new photometric redshift, $z_{\mathrm{phot}}=9.36$, 
which is quoted in Table~\ref{tab:sample}.
We adopt this redshift PDF (grey ``\hband\ and \jband'' curve in 
Figure~\ref{fig:sample1}) in estimating the candidate's
magnification and absolute magnitude posterior (Section~\ref{sec:magnification})
and constructing the luminosity function (Section~\ref{sec:densities}). 
We note that this \jband\ 
flux is likely a lower limit, and a brighter \jband\ flux 
would indicate a stronger spectral break and therefore more strongly reject the 
smoother slope of a lower-redshift, red galaxy. In Figure~\ref{fig:sample1},
we show the original SED and redshift PDF for Par0259+0032\_194 (i.e., 
excluding the \jband\ flux) in green and those including the estimated 
\jband\ lower limit in black, using a different color scheme as the rest of the 
figure to signify this special case.

%%%%%%%%%%%%%%%%%%%%%%%%%%%%%%%%
\subsubsection{Par335\_251} \label{sec:cand335_251}
Included in our sample is the first $z>8$ candidate from the WISP survey,
Par335\_251 ($m_{160}=24.74$). 
It has a photometric redshift of $z_{\mathrm{phot}}=10.53$ and 
is detected only in the \hband, as expected for WISP fields (see discussion in 
Section~\ref{sec:veff}). As can be seen in Figure~\ref{fig:sample1}, it 
lies on the UVIS chip gap in \vband, and so we treat the source as having 
no \vband\ coverage. 
As Par335 has IRAC 3.6\micron\ coverage, we 
have this additional constraint for the WISP source, and so we include it 
in our sample.
Additionally, as a slitless spectroscopic survey, WISP observed Par335 with 
both of WFC3's IR grisms: G102 ($8.8-1.1$ \micron, $R\sim201$) and G141
($1.07-1.7$ \micron, $R\sim130$). The integration times were tuned to 
achieve approximately uniform sensitivity at all wavelengths for a line of 
a given flux, and for this field are $\sim$8800 seconds in G102 and  
$\sim$3800 seconds in G141.

\begin{figure}
\epsscale{1.2}
\plotone{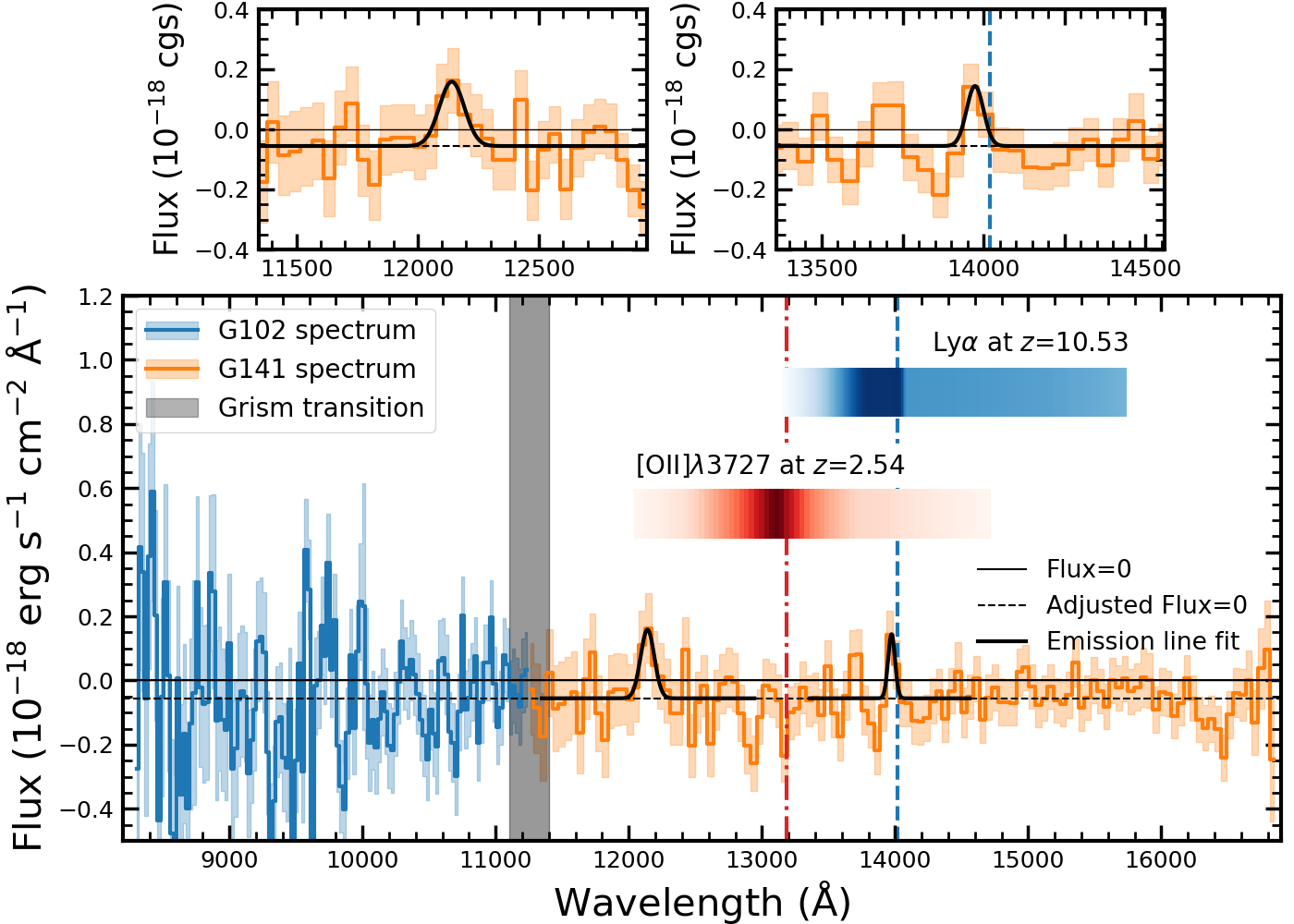}
\caption{The one-dimensional WISP spectra in the WFC3 grisms G102 (blue) and 
G141 (orange) of candidate Par335\_251. The light shaded regions show the 
1$\sigma$ uncertainties.
The best-fitting high-redshift solution ($z_{\mathrm{phot}}=10.53$) would 
place \lya\ at $\lambda_{\mathrm{obs}}\sim1.4$\micron\ (blue dashed vertical 
line). The best-fitting low-redshift solution would place \oii\ at 
$\lambda_{\mathrm{obs}}\sim1.32$\micron\ (red dash-dotted line).
We show a horizontal rectangle for each emission line weighted by the
redshift PDFs and extending to their 95\% intervals. 
There are potential emission lines at 
$\sim$ 12200\AA\ (S/N=2.3) and 14000\AA\ (S/N=1.5) that are broadly consistent
with the $z$$\sim$2 redshift PDF, though this low-redshift solution is 
disfavored with a $\Delta \chi^2=20.41$. The fits to each line are 
described in the text and displayed in the top panels.
The 14000\AA\ feature would be consistent with \lya\ at $z=10.53$, 
though a \lya\ detection at this redshift is unlikely due to the relatively 
short integration times. 
While there is no strong indication that this candidate is at a lower 
redshift, its true nature is not clear from this spectrum.
\label{fig:wispspec}}
\end{figure}

While these spectra will be 
far too shallow to detect an emission line or continuum for a source at 
$z\sim10.5$, it is instructive to explore them for any spectral features 
that may indicate the source is at a lower redshift. 
We downloaded the extracted, calibrated spectra from the WISP page on 
MAST (see footnote 4) and plot them in Figure~\ref{fig:wispspec}.
Assuming the best-fitting redshifts for the high- and low-redshift solutions, 
we show the expected location of \lya\ at $z=10.53$ and \oii$\lambda$3727 
at $z=2.54$. 
There are two possible emission lines at 
$\lambda_{\mathrm{obs}} \sim 12200$\AA\ and
$\lambda_{\mathrm{obs}} \sim 14000$\AA\ that lie
on either edge of the expected range for \oii\ given the low redshift 
solution. 

We measure these lines using a Monte Carlo analysis, assigning fluxes at 
each wavelength step that are pulled randomly from a normal distribution 
centered at the flux value and with $\sigma$ equal to the 
flux uncertainty. We then fit the fluxes with a Gaussian 
profile using least-squares minimization to determine the best fit, and 
repeat this process 1000 times. For each line, the median of all 1000 fits 
is displayed in the top panels of Figure~\ref{fig:wispspec}. We define the 
S/N of the line as the flux of the median fit divided by the standard 
deviation of the fits. The potential emission lines at
12200\AA\ and 14000\AA\ have S/N=2.3 and 1.5, respectively. 
We note that the sky background has been over-subtracted in the spectrum, 
and so we fit a new continuum (black horizontal dashed line) from which 
we measure the strength of each line. If fit instead with a continuum 
at flux=0, the S/N of each line would decrease.

Both lines have S/N$<$3, and lie far from the expected observed wavelength 
for \oii$\lambda$3727 given the low redshift solution's narrow PDF. 
Additionally, the $\Delta \chi^2$ between the best-fit solution at 
$z\sim10.5$ and the minimum $\chi^2$ at lower redshift is 20.41, indicating
that the low-redshift solution is disfavored. 
The potential line at $\lambda_{\mathrm{obs}} \sim 14000$\AA\ 
would be very consistent with \lya\ at a redshift of $z=10.53$. However, 
such a detection would require hours more integration time, as in 
\citet{zitrin2015} and \citet{larson2022}.
While the WISP spectrum does not exhibit clear signs of being at a lower 
redshift, it cannot help constrain the nature of Par335\_251.

%%%%%%%%%%%%%%%%%%%%%%%%%%%%%%%%
\subsubsection{Remaining Single-Band Detections}\label{sec:cands_single}
The remaining five candidates:
\begin{itemize}
\item Par0956-0450\_684 ($m_{160}=24.88$, $z_{\mathrm{phot}}=10.54$)
\item Par0843+4114\_120 ($m_{160}=26.39$, $z_{\mathrm{phot}}=10.98$)
\item Par1301+0000\_37 ($m_{160}=25.42$, $z_{\mathrm{phot}}=11.02$)
\item Par0756+3043\_92 ($m_{160}=26.03$, $z_{\mathrm{phot}}=11.03$) 
\item Par0926+4536\_155 ($m_{160}=26.37$, $z_{\mathrm{phot}}=11.15$)
\end{itemize}
are all single-band detections in the \hst\ imaging. All five have
IRAC coverage at both 3.6\micron\ and 4.6\micron, and are non-detections 
in both bands. While the non-detections in the bluer \hst\ filters 
as well as both IRAC channels help to rule out lower-redshift contaminants,
there is an increased concern that these single-band detections are 
spurious sources (e.g., hot pixels or cases of persistence).
Three of these candidates are from the \hippies\ dataset, which has been 
noted to suffer from increased levels of persistence and contamination
\citep[e.g.,][]{morishita2021a}. We have done our best to identify and remove 
spurious sources (see Sections~\ref{sec:inspection} and \ref{sec:persistence}),
yet the possibility remains that some or all of these candidates are 
contaminants in our sample. Spectroscopic confirmation or imaging in more 
filters with NIRCam is needed to confidently evaluate the high-redshift 
nature of these sources.

%%%%%%%%%%%%%%%%%%%%%%%%%%%%%%%%%%%%%%%%
\subsection{Comparison with Previous Studies in the Same Fields} \label{sec:others}
The datasets we include in our analysis are from surveys that were designed 
to detect high-redshift galaxies. The BoRG survey in particular was optimized 
for searches for $z\gtrsim8$ sources, and several teams have previously 
reported results for galaxies in these fields. In this section, we discuss 
results from the previous studies that were performed in the same \hst\ 
fields that we consider in this paper. As this is the first search for $z>8$ 
candidates in \hippies\ or WISP fields, we focus here on results in the 
\borg\ fields.

We begin by discussing the \borg\ results aimed at searching for galaxies 
at $z\sim8$ as \yaband\ dropouts \citep{trenti2011,bradley2012,schmidt2014}. 
These studies require a significant 
detection in \jband\ and use \hband\ as a second detection band to 
rule out spurious sources. 
We are focusing our search at higher redshifts and allow for single-band 
detections, and so do not expect to recover
many of the candidates presented in these papers. However, we do find 
two candidates in common, which is unsurprising because the redshift PDFs of 
some sources in our sample extend to $z\sim8$. Our two lowest-redshift
candidates, Par1033+5051\_116 and Par0440-5244\_497,
were previously published by \citet{bradley2012}. For both 
candidates, we measure slightly brighter \hband\ magnitudes:
$m_{H160}=25.99$ (Par1033+5051\_116) and 25.47 (Par0440-5244\_497) compared 
with 26.2 and 25.9, respectively, based on the \jband-\hband\ colors 
reported in \citet{bradley2012}. 
While \citet{bradley2012} do not calculate photometric redshifts for their 
candidates, the completeness they present for a representative field peaks 
at $z\sim7.8-8$, which is consistent with the 68\% intervals we measure for 
our redshift PDFs. 
We do not recover the remainder of the candidates presented in these three 
papers because our selection function 
is more sensitive to galaxies at higher redshifts than those identified 
as \yaband-dropouts. We also focus on a slightly brighter magnitude 
range than these \borg\ studies, requiring $m_{H160} \leq 26.5$ while 
many of the candidates presented in these papers are in the range 
$26.7 \lesssim m_{H160} \lesssim 27.2$.

Next, \citet{morishita2020a} perform a search for quasars at $z\sim8$ in 
the \borg\ fields (among others), 
focusing on the identification of Lyman break candidates 
that have point-source morphologies rather than resolved shapes. 
\citet{morishita2020a} also employ a combination of color cuts and 
detection significance criteria in their sample selection, but additionally 
require that the $\geq$70\% of the source redshift PDFs be at $z>6.5$. 
\citet{morishita2020a} are again focusing on a slightly lower redshift range
than we are, and so it is not surprising that our samples do not overlap. 
We do not select their one candidate in a BoRG[$z8$] field because it is 
detected in \ybband\ with a S/N$>2$.
We do find one candidate in the same field as they do (Par0926+4536\_155), 
but our candidate is $\sim$2\arcmin\ away from the one they identify. 

While the \borg\ survey was designed to detect galaxies at $z\sim8$ as 
\yaband-dropouts, \citet[][hereafter B16]{bernard2016} performed a search 
comparable to ours for 
$z\sim9-10$ galaxies in these fields. They focused on the 62 fields with 
an exposure time in \hband\ of at least 900 seconds, a survey area 
covering 293 arcmin$^2$. (We include 66 pointings and make no exposure time 
cut.) \citetalias{bernard2016} identified three $z\sim10$ 
candidates. 
They perform their search using the Lyman break technique to identify \jband\
dropouts, implementing a $J_{125} - H_{160} > 1.5$ color selection that 
will select a strong \jband$-$\hband\ break while still allowing for \jband\ 
detections. This color selection is coupled with a detection requirement of 
S/N$_{H160} \geq 8$ and non-detections required in the $V$ and \yaband\ 
filters (S/N$_V$ and S/N$_{Y098} < 1.5$).
They also implemented a threshold of \texttt{CLASS\_STAR}$ < 0.95$ and 
performed a visual inspection of their candidates.
This source selection method is in contrast to our process, which we base 
on photometric redshifts rather than color-cuts (Section~\ref{sec:sample}).
As discussed in Section~\ref{sec:psfs}, we also find that the accuracy of 
the \texttt{CLASS\_STAR} flag decreases with magnitude, and so 
we did not incorporate it into our selection but instead compare the 
colors of our sources to identify stellar contaminants.
We also note that \citetalias{bernard2016} did not have access to \spitzer/IRAC 
imaging of their candidates, but applied for IRAC follow-up which we are 
able to include in our analysis here (\spitzer\ PID 12058, PI Bouwens). 

\citetalias{bernard2016} identified a further three candidates as 
contaminants based on source size. 
Following \citet{holwerda2015}, the authors adopt 
0\farcs3 as an upper limit for the half-light radius of their candidates and 
reject larger sources as low-redshift interlopers. 
On the other hand, \citet{holwerda2020} find that while such a size 
criterion could remove up to 75\% of low redshift interlopers, it may also 
reject $\sim$20\% of real, bright galaxies in the targeted redshift range.
Source size and morphology is another area where the higher resolution of 
the \jwst/NIRCam imaging will significantly improve our understanding of the 
size-luminosity relation out to $z\gtrsim10$. 
We do not apply an upper limit size cut to our sample selection, though 
our candidates are fairly compact with (non-PSF-corrected) half-light 
radii in the range 0\farcs11$-$0\farcs19 (with the exception of 
Par0456-2203\_473, see Section~\ref{sec:cand0456_473}).

We do not recover any of the three candidates presented in \citetalias{bernard2016}.
Two of their candidates (borg\_0240-1857\_25 and borg\_0456-2203\_1091) 
satisfied our selection criteria but were rejected during visual inspection 
in category 1e (a hot pixel, cosmic ray, or source with a strange
morphology). We identify these two sources in Figure~\ref{fig:reject_1d1e}.
Our rejection of these two candidates illustrates the 
necessarily subjective nature of high-redshift candidate searches with \hst, 
where potentially real candidates may be removed in an effort to clean 
samples of spurious sources. 
Interestingly, we identify a different candidate in one of the same fields,
Par0456-2203\_473. Our candidate is located $\sim$56\arcsec away from 
borg\_0456-2203\_1091, yet was not selected by \citetalias{bernard2016}. 
We did not detect the third candidate presented in \citetalias{bernard2016}
(borg\_1153+0056\_514) with enough significance in \hband\ 
to meet our S/N$_{H160}>5$ criterion (S/N=4.78) 
and so this source was not selected in our sample. 
The \se\ parameters used 
to detect objects in an image, the choice of aperture size and shape, and 
the way in which the noise is calculated can all affect the detection 
significance. 
These discrepencies between the sample we present in Section~\ref{sec:cands} 
and that of \citetalias{bernard2016} highlight how dependent $z>8$ candidate 
samples are on the sample selection methods used.

%%%%%%%%%%%%%%%%%%%%%%%%%%%%%%%%%%%%%%%%%%%%%%%%%%%%%%%%%%%%%%%%%%%%%%%
\section{Low-Redshift Contamination} \label{sec:contam}
The datasets we consider in this paper are relatively shallow
compared to those of the well-studied fields such as CANDELS or the Hubble
Frontier Fields that are the best current options for detecting
$\leq$$L*$ high-redshift galaxies.
While these parallel datasets are particularly well-suited to detecting
rare bright galaxies needed to explore the bright end of the luminosity
function, they are likely not deep enough to detect the far more common
faint galaxies at lower-redshifts.
The possibility remains that our selection criteria are in fact identifying
faint, low redshift galaxies. Faint, red galaxies at $z\sim2-3$ are of 
particular concern. Whether passive or star-forming and dusty, their 
steepening red SEDs may only rise above our detection thresholds
in the reddest filters ($J$ and $H$).
Extreme emission line galaxies at lower redshift are another possible source 
of contamination, where high equivalent width lines such as \oii, \oiii, 
or \ha\ could boost the \hband\ flux at $z\sim3$, $z\sim2$, or $z\sim1.3$, 
respectively \citep[e.g.,][]{atek2011,faisst2016a}.
In both cases, photometry in IRAC bands can help identify these sources 
as their SEDs likely increase past 2\micron. However, a faint contaminant
with an emission-line-boosted \hband\ flux may fall below the IRAC detection
limit. Additionally, one of our candidates (Par0259+0032\_194) does not have 
IRAC coverage, and we consider the IRAC deblending for another candidate 
(Par0456-2203\_473) to be unreliable. 
While \spitzer/IRAC imaging should help break these redshift degeneracies, 
the shallow depths of the IRAC imaging as well as the uncertainties in the 
IRAC deblending complicate the picture. 
Our conservative selection criteria and the imaging currently
available at $\sim3-5$\micron\ are not sufficient to conclusively rule out 
such contaminants.
In the following sections, we therefore explore the low-redshift contamination 
fraction in our sample from faint galaxies falling below our detection 
thresholds.

%%%%%%%%%%%%%%%%%%%%%%%%%%%%%%%%%%%%%%%%
\subsection{Estimate from Stacking} 
\begin{figure}
\plotone{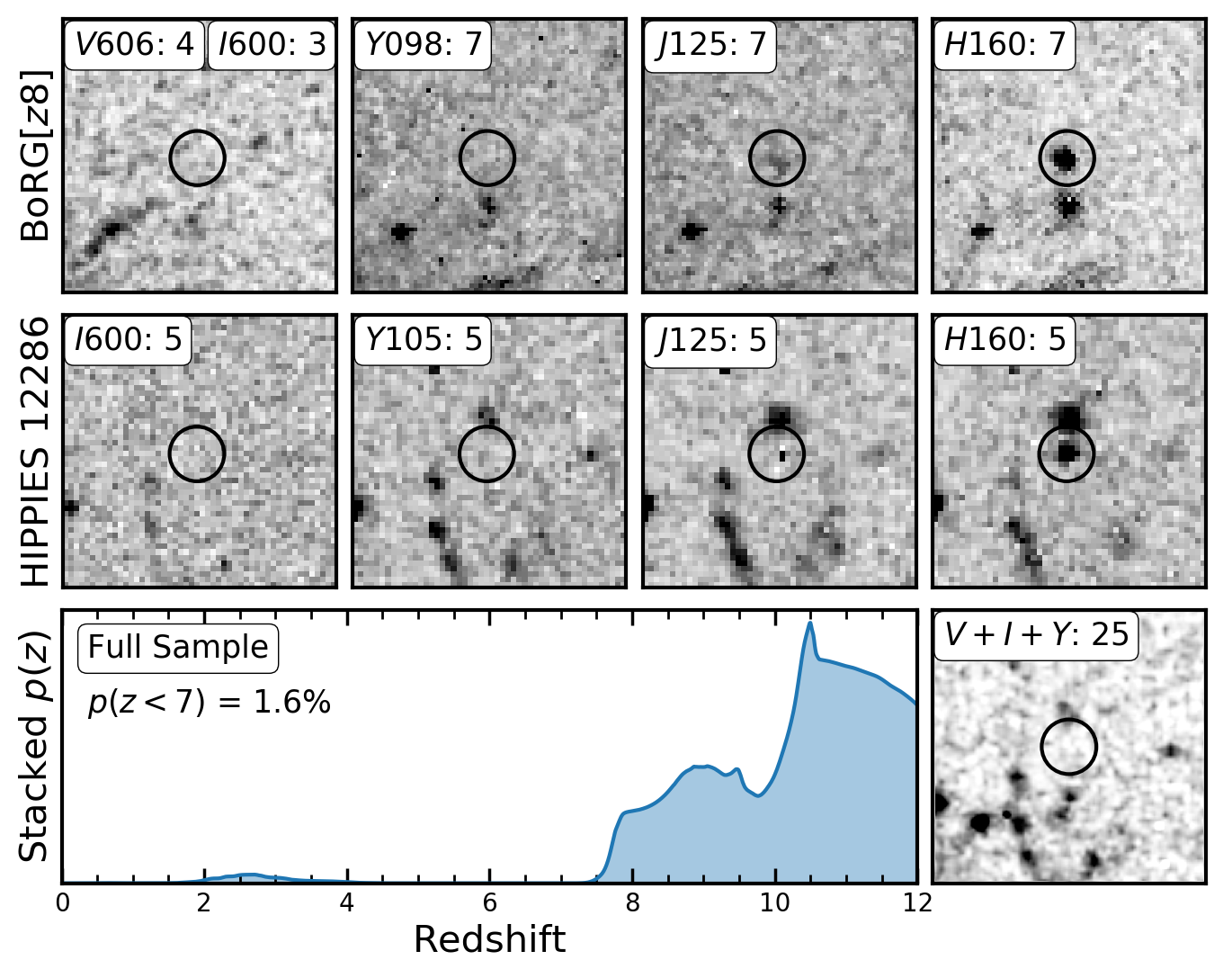}
\caption{Stacks of image postage stamps and the photometric redshift 
probability distributions of all candidates in our sample. 
We separate the stamps by dataset in order to combine images with the 
same pixelscales, showing \borg\ in the top row and \hippies\ in the second 
row. These stamps are 5\arcsec\ on a side, displayed on a zscale interval 
with no image smoothing. The black circle of $r=0\farcs5$ indicates
the expected position of source flux. In the bottom row we show the 
stacked redshift PDF of all sources in our sample (left) and a stack of 
all $V$, $I$, and $Y$ imaging for all candidates (including WISP) 
resampled onto a common 
pixel scale of 0\farcs01/pixel (right). In all cases, including the resampled 
stack, no source flux is visible in the optical and $Y$ filters at the 
expected position, indicating that our sample is not dominated by lower-redshift
contaminants.
\label{fig:contamstacks}}
\end{figure}

The optical imaging in these \hst\ datasets may be too shallow to detect flux 
from individual low-redshift sources, and so we stack postage stamps of 
the candidates in order to check for any optical detections. 
We create several stacks, separating the candidates by dataset to avoid 
introducing any artificial features by combining images on different 
pixel scales. We also separate the imaging by filter, creating one stack for 
each dataset that includes all available images obtained with the WFC3/UVIS
and ACS/WFC cameras 
(\vband\ and \iaband\ for \borg, and \iaband\ for \hippies), a stack 
of all WFC3/IR $Y$ images as an additional dropout band, and stacks of all 
$J$ and $H$ images as the candidate detection bands. 
We do not include the candidate from the WISP survey in these stacks because 
it is the only one on the 0\farcs13 pixel scale. 
In each case, the images
are combined using a weighted mean in order to down-weight any bad pixels.
The image stacks are displayed in the top two rows of 
Figure~\ref{fig:contamstacks}, where we indicate the number of images in each 
filter that contributed to the stack. As can be seen in the left two panels 
of each row, there is no appreciable source flux at the expected position 
in the dropout bands.

We next estimate the detection significance at the center of each stack 
using $r=0\farcs2$ circular apertures.
We place a grid of unique apertures across each stamp, using a segmentation 
map created by combining the individual maps for each input field
to avoid source flux. 
The $1\sigma$ noise in the stamp is then taken as the standard deviation
of a Gaussian distribution fit to the distribution of background aperture 
fluxes. With the aperture fluxs measured at the center of each stamp,
we find a signal to noise of $\mathrm{S/N}<1$ in both sets of $V+I$ and $Y$ 
stamps, confirming that there is no significant optical detection in the
stacked imaging. 
These imaging stacks provide at best an upper limit estimate of the 
low redshift contamination in the sample, suggesting that the majority of 
our sample is not comprised of contaminants.
For reference, the S/N in $H$ ($J$) is 24.1 (3.7) 
and 65.5 (9.3) for the \borg\ and \hippies\ stacks, respectively.

As an extra check, we resample all available $V$, $I$, and $Y$ imaging, 
including the \ibband\ imaging from the WISP candidate, to a much finer 
pixel scale of 0\farcs01/pixel. This resampling then allows us to combine 
all 25 images into a single, summed stack, which is shown in the bottom right 
corner of Figure~\ref{fig:contamstacks}. If a majority of the candidates in 
our sample were in fact lower-redshift interlopers, we would expect to 
detect flux in this combined stack of 25 images. However, there is no 
visible signal in the black circle in the bottom right panel of 
Figure~\ref{fig:contamstacks}, indicating that lower-redshift contamination
does not dominate our sample. 

In the bottom row of Figure~\ref{fig:contamstacks}, we also show the 
stacked redshift PDFs for all 13 sources in our sample. There is a 
non-zero portion of the PDF at $z\sim2-4$, with a peak around $z\sim2.5$.
This low-redshift peak is consistent with the detection of the 
4000\AA\ break in lieu of the \lya\ break between \jband\ and \hband.
Alternatively, if high equivalent width emission line galaxies are 
contributing to that low-redshift peak, it could correspond to strong \oiii\ 
in \hband\ up to $z\sim2.4$, with \oii\ transitioning into the \hband\ filter 
at $z\sim2.75$. 

The portion of the stacked PDF that lies at $z<7$ amounts 
to only 1.6\% of the total integrated PDF. However, as we have not used 
luminosity priors in our \eazy\ runs (see Section~\ref{sec:photz}), the 
redshift PDFs do not take into account the relative abundances of high- and
low-redshift galaxies in the universe. Therefore, we cannot use the 
redshift PDF to estimate the contamination from lower-redshift galaxies in 
our sample. Instead, in the next section we aim to quantify the contamination 
fraction by artifically dimming low-redshift sources and attempting to 
reselect them as high-redshift candidates.

%%%%%%%%%%%%%%%%%%%%%%%%%%%%%%%%%%%%%%%%
\subsection{Catalog-Level Estimate of Faint Lower-redshift Interlopers}\label{sec:faintlowz}
We begin by identifying a high-fidelity sample of galaxies at low redshift. 
From the full \se\ catalog, we select all sources with S/N$_H>5$ and 
$H_{160} < 23.5$ mag. After running \eazy\ on this sample of bright sources, 
we select all those with 70\% of their redshift probability distribution 
contained in the range $1 \leq z \leq 4$. We consider only resolved sources in 
an attempt to avoid including in this analysis any stars that have been 
incorrectly assigned $z>0$ photometric redshifts. 
Using the size measurements presented in Section~\ref{sec:spex}, we select 
sources with a half-light radius larger than the median $r_{1/2}+1\sigma$
measured for stars in the corresponding dataset. There are a total of 
622, 142, and 335 low-redshift sources -- those that are bright, resolved, and 
reliably in the range $1 \leq z \leq 4$ -- in the \borg, \hippies, and 
WISP datasets, respectively. 

For each \hst\ field containing a candidate high-redshift galaxy in our sample, 
we create a catalog of dimmed, low-redshift galaxies by assigning a new \hband\
magnitude to each real source. The dimmed \hband\ magnitudes are pulled 
randomly from a truncated Gaussian distribution with $\mu=26.5$ and a maximum 
at $m=27$, 0.5 magnitudes fainter than the 5$\sigma$ requirement in our 
selection criteria. We dim the fluxes in all filters by the same value, 
$\Delta m =H_{\mathrm{obs}}-H_{\mathrm{dimmed}}$. 
This process maintains the colors of a $z\sim2$ galaxy while moving all 
photometry to fainter magnitudes.
We create multiple realizations of each input low-redshift source, resulting 
in a catalog of 5000 dimmed sources for each field.
We measure the median flux uncertainties in magnitude bins for all sources in 
the \se\ catalog that are detected in the given \hst\ field, and perturb the 
dimmed fluxes by the error for the appropriate magnitude bin.
In this way, we incorporate field-to-field variations in depths while 
accounting for photometric scatter as a function of magnitude.
Finally, we run \eazy\ on the new photometry and reselect sources as those 
with $H_{160} \leq 26.5$ and 70\% of their redshift PDF at $z> 8$.

\begin{figure}
\epsscale{1.1}
\plotone{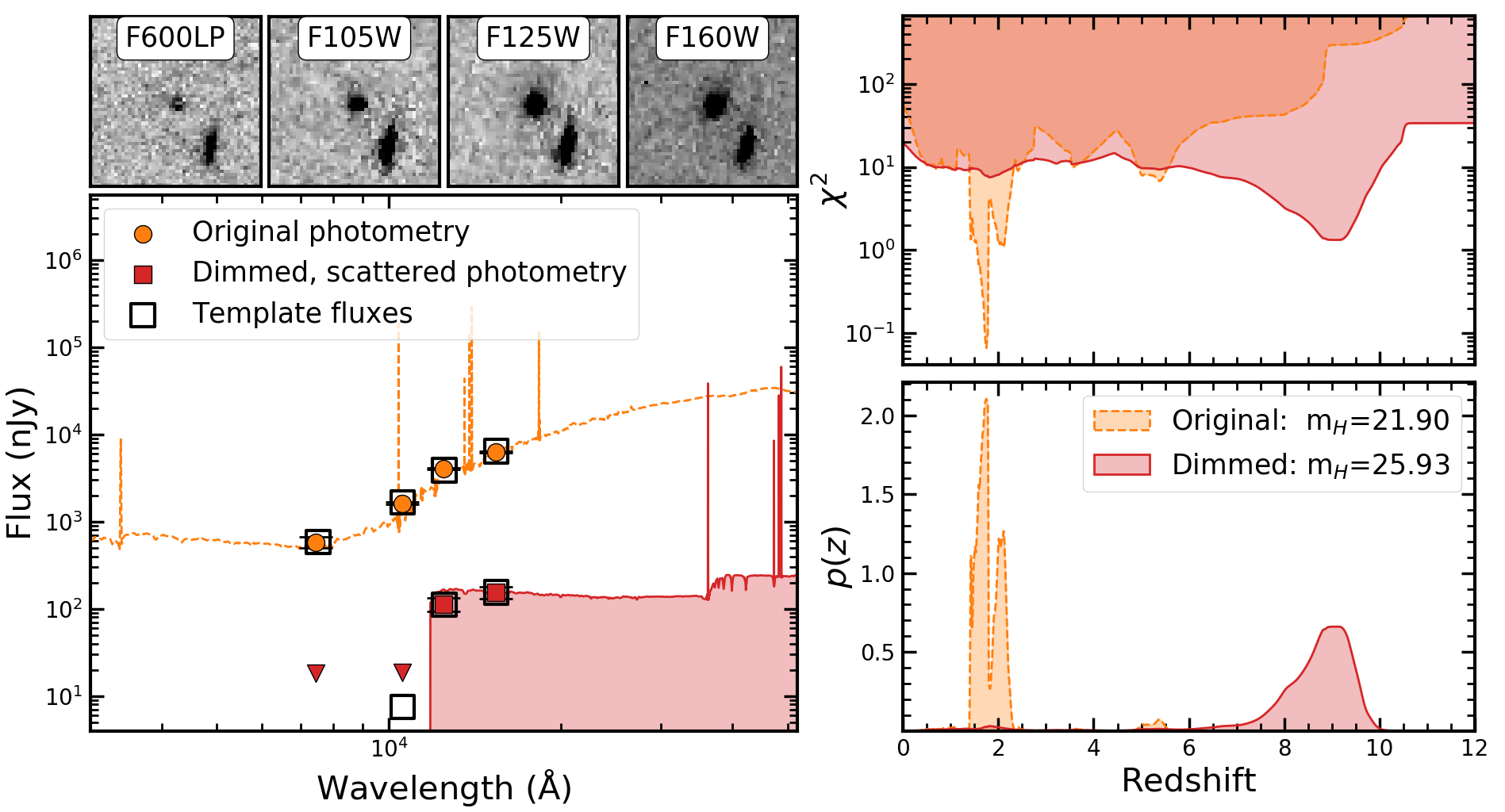}
\caption{An example of a low-redshift source from Par0259+0032 that is
selected as a high-redshift candidate after the fluxes are dimmed as 
described in the text. 
In the large panel on the left, we plot the original and dimmed photometry 
as orange circles and red squares, respectively. The best-fitting \eazy\
templates are also displayed, with the corresponding $\chi^2$ distributions 
in the top right panel and the redshift PDFs in the bottom right panel. 
We show the original (undimmed) \hst\ imaging in 5\arcsec stamps along the 
top left. 
Dimming the fluxes in all bands preserves the source's colors, but results 
in the source dropping out of the bluer bands, and the new photometry is 
better fit at $z\sim9$. 
Through this analysis, we expect low contamination rates ($<$1\%-12\%)
from faint lower-redshift galaxies in our sample.
\label{fig:contamdimmed}}
\end{figure}

For each field, we then look at the total number of the 5000 sources that 
were ``recovered'' as high-redshift candidates. 
There were no dimmed galaxies selected as high-redshift sources in five
of the 12 fields that contain our candidates 
(Par0440-5244, Par0713+7405, Par0756+3043, Par1033+5051 and Par1301+0000),
implying a near-zero contamination rate.
For the remaining fields, we select between $1-4$ dimmed sources. 
Figure~\ref{fig:contamdimmed} shows an example of one such low-redshift source,
with the original (undimmed) \hst\ image stamps of the source as well as the 
original and dimmed photometry plotted as orange squares and red circles, 
respectively. We also show both sets of \eazy\ template fits, redshift PDFs,
and $\chi^2$ distributions. 
As can be seen here, the dimming process maintains the $J-H$ color of the 
source within the photometric scatter, but the bluer fluxes have dropped out 
of the optical bands. 

We use equation 4 from \citet{finkelstein2015a} to estimate the contamination 
fractions in each field that correspond to the number of recovered sources:
\begin{equation}
F = \frac{N_{\mathrm{dim,sel}}}{N_{\mathrm{dim,low}z}} \frac{N_{\mathrm{tot,low}z}}{N_{\mathrm{high}z}}.
\label{eqn:dimmed}
\end{equation}
Here, $N_{\mathrm{dim,sel}}$ is the number of dimmed low-redshift sources 
that were selected as high-redshift galaxies, and $N_{\mathrm{dim,low}z}$ is 
the number of input dimmed sources (5000 in all fields).
The ratio $N_{\mathrm{tot,low}z} / N_{\mathrm{high}z}$ accounts for the 
relative number densities of low- and high-redshift sources in our survey 
volume, where $N_{\mathrm{high}z}$ is the number of high-redshift candidates 
detected in the dataset (i.e., 7 candidates in \borg\ fields, 5 in \hippies, 
and 1 in WISP), and $N_{\mathrm{tot,low}z}$ is the total number of faint, 
low-redshift galaxies identified in the catalog of each dataset. For the 
latter number, we select low-redshift sources with 70\% of their
integrated redshift PDF contained in the range $1 \leq z \leq 4$ and 
magnitudes in the range $24 < m_{H160} \leq 27$. We include sources below 
the threshold of $m_{H160} \geq26.5$ in order to account for the 
prevalence of fainter low-redshift galaxies that may be photometrically 
scattered into our sample magnitude range. In an effort to avoid stars
and spurious detections and to ensure quality photometric redshift fits, we 
also require a half-light radius larger than the median measured for stars in 
the catalog as well as a $\geq$5$\sigma$ detection in \hband. 
We note that these selections, and especially the last two, are likely to 
miss real faint, low-redshift sources in our catalog. However, we find that 
without these requirements, we incorrectly include many stars and 
marginally-detected sources with poorly-constrained redshift PDFs in our 
total for $N_{\mathrm{tot,low}z}$.

As an example, we find that \borg\ field Par0456-2203 has:
$N_{\mathrm{dim,sel}} = 1$ with $N_{\mathrm{tot,low}z} = 3342$ in the \borg\
catalog, and $N_{\mathrm{high}z} = 7$, resulting in a contamination 
fraction of $F=0.095$. 
We find contamination rates of 1.4\% (Par0926+4536), 2.2\% (Par0956-0450 and 
Par0259+0032), 2.9\% (Par0843+4114), 9.5\% (Par0926+4000 and Par0456-2203),
and 12.7\% (Par335). The fields with $N_{\mathrm{dim,sel}} = 0$ have 
contamination rates of $<$1\% (Par0713+7405) and $<$9.5\% (Par1033+5051,
Par0440-5244, Par0756+3043, and Par1301+0000).
We again note that these contamination rates may be underestimated. We have 
limited our analysis to galaxies in the range $1<z<4$ because 
the majority of the low-redshift portions of our candidates' photometric 
redshift PDFs is in this range. Yet we are therefore not 
accounting for low-redshift contamination from sources outside of this range.
Also, the SEDs of sources in the catalog are constrained by only a handful of 
\hst\ photometric measurements, resulting in uncertainties in the 
photometric redshifts measured for the bright, low-redshift sources that 
are chosen to be dimmed, the dimmed sources that 
are then `recovered' as high-redshift galaxies ($N_{\mathrm{dim,sel}}$), 
and the faint, low-redshift sources ($N_{\mathrm{tot,low}z}$).
However, with at most four bands of photometry, this analysis represents 
our best attempt at identifying contamination from faint, low-redshift 
galaxies.

As the estimated contamination rates are small, and as we expect the 
contamination rates measured in the bins of magnitude we use when calculating
the UV luminosity function to be even lower, 
we do not reduce our observed number densities by the fractions reported here.

%%%%%%%%%%%%%%%%%%%%%%%%%%%%%%%%%%%%%%%%%%%%%%%%%%%%%%%%%%%%%%%%%%%%%%%
\section{The Bright End of the Luminosity Function at $z\sim9$} \label{sec:lf}
In this section we calculate the rest-UV luminosity function at 
$8.5 \lesssim z \lesssim 11$. This measurement first requires correcting 
the candidate luminosities for any magnification imparted by nearby 
neighbors (Section~\ref{sec:magnification}), and estimating the completeness 
of our sample as a function of magnitude (Sections~\ref{sec:sims} and 
\ref{sec:veff}).

%%%%%%%%%%%%%%%%%%%%%%%%%%%%%%%%%%%%%%%%
\subsection{Source Magnification} \label{sec:magnification}
Gravitational lensing can have the effect of boosting the brightness of 
sources such that they are detected in a flux-limited sample when 
intrinsically they would be too faint. Lensing can benefit high-redshift 
surveys by bringing the fainter portions of the galaxy population within reach 
of \hst\ surveys \citep[e.g.,][]{ellis2001,coe2013,coe2015,schmidt2014b,atek2015a,bouwens2017b,livermore2017},
providing more constraining power for the faint end slope of the luminosity 
function to higher redshifts than would otherwise be possible. 
However, the magnification of sources can also significantly change the shape 
of the measured luminosity function, especially at the bright end where the 
low number counts are easily boosted by an increase in observed luminosity. 
Many of the candidates in our sample have close neighbors, and five  
in particular have one or more bright neighbors within $\sim$3\arcsec.
We now estimate the magnification that each candidate experiences, following 
the steps laid out by \citet{mason2015a} for strong lensing.

The observed magnitude of a candidate will be increased by a factor 
$-2.5 log_{10}(\mu)$ by a bright, close neighbor. The magnification, $\mu$,
depends on the redshifts of the candidate and the lensing source, the 
velocity dispersion of the lens, and the projected distance between the lens 
and the candidate in the image plane. 
Following \citet{mason2015a}, we assume each of the potential lensing sources 
can be modeled as a singular isothermal sphere, where the Einstein
radius of the lens is proportional to the square of the velocity dispersion 
of the lensing galaxy ($\sigma^2$) \citep[see equation 4 of][]{mason2015a}.

We estimate $\sigma$ using the empirical, redshift-dependent 
Faber-Jackson relationship 
that \citet{mason2015a} dervive using \borg\ fields (see their Table 1). 
For this estimate, we use the \eazy-calculated photometric redshifts for 
the neighboring sources, including the IRAC photometry where available to 
improve on the photometric redshift constraints. 
We obtain IRAC photometry for each 
neighbor as described in Section~\ref{sec:irac}, adopting the \galfit\ model 
magnitudes for detections and using $r=2.5$ pixel aperture fluxes for 
non-detections.
When calculating $\theta_{\mathrm{ER}}$, ideally we would use the apparent 
magnitudes from the same filters as \citet{mason2015a} for each redshift bin.
However, we find that the majority
of the neighboring sources have a significant detection (S/N$\geq$3) only 
in \hband, and so we use \hband\ apparent magnitudes for all neighbors,
regardless of redshift. 
As can be seen in Figures~\ref{fig:sample1} and \ref{fig:sample1b}, some of 
the candidates have multiple bright, close neighbors. We assume that each 
neighboring source is an independent lens, such that the total magnification
$\mu$ for a candidate is the product of the magnification from each 
neighboring source, $\mu_i$. This method assumes that the neighbors
each lens the light from the candidate along the line of sight independently,
an assumption that we discuss in more detail below. 

As in \citet{rojas-ruiz2020}, we use a Monte Carlo approach to calculate $\mu$,
the corrected \hband\ apparent magnitude, and the absolute rest-UV 
magnitude, $M_{\mathrm{UV}}$, of each candidate.
We perform 5000 calculations of $\theta_{\mathrm{ER}}$ and $\mu_i$ for each
neighbor within 5\arcsec\ of a candidate, increasing the number of 
realizations from 1000 in \citet{rojas-ruiz2020} because of the broader 
redshift PDFs in our catalog.
For each realization, we pull the redshift for the candidate and each 
neighbor randomly from their respective redshift PDFs.
We randomly draw the \hband\ fluxes for the candidate and neighbors from 
Gaussian distributions centered on the observed value and with a standard 
deviation equal to the flux uncertainties. In this way, we are including the 
uncertainties in the photometric redshifts and the flux uncertainties
in our calculations of $\mu$ and $M_{\mathrm{UV}}$.
For the rest-UV magnitude, we use the fact that a rest-frame 1500\AA\ 
wavelength falls near the middle of the \hband\ filter in the $z\sim9.5$ 
observed frame. 

\citet{finkelstein2022a} show that including all nearby neighbors in 
these calculation biases the resulting magnification towards higher values of 
$\mu$. Specifically, they show that on average this approach results in 
$\mu\sim1.4$ along randomly-selected lines of sight where the expected 
magnification is $\mu=1$. Following their example, we therefore only 
consider the magnifaction contribution from neighbors with a median 
$\mu_i > 1.4$. The $\mu_i$ of any neighbor with 
$\mu_{i,\mathrm{median}} < 1.4$ is set to unity. 
As a result of our Monte Carlo process, we have a distribution of 
5000 $\mu$'s, corrected \hband\ fluxes, and $M_{\mathrm{UV}}$ values. We take the 
median of each distribution as the parameter estimate, and adopt the 
68\% interval as the uncertainty.

We find that three candidates experience a significant magnification:
Par0713+7405\_95 with $\mu=1.45\pm0.13$, 
Par0259+0032\_194 with $\mu=3.81\pm11.8$, and Par0456-2203\_473
with $\mu=15.68\pm12.50$ (note that this candidate also has the largest 
observed \rhalf). This intermediate-to-strong lensing is 
unsurpising as each of these three candidates has a bright, close neighbor. 
However, we note that the magnifications of the two candidates with the 
strongest lensing are highly uncertain, with $\sigma_{\mu} > \mu$ for 
Par0259+0032\_194. 
We report the median $M_{\mathrm{UV}}$ 
for each candidate in Table~\ref{tab:sample}, adopting the 
magnification-corrected magnitude for the three lensed candidates and 
providing the 68\% intervals 
that account for the uncertainties in photometric redshift, \hband\ fluxes, 
and magnification where applicable.
We use the distribution of $M_{\mathrm{UV}}$ values in calculating 
the luminosity function in Section~\ref{sec:veff}.

%%%%%%%%%%%%%%%%%%%%%%%%%%%%%%%%%%%%%%%%
\subsection{Simulated Sources} \label{sec:sims}
In order to quantify the incompleteness in our sample of high-redshift 
galaxies caused by our selection criteria, we create a catalog of 
simulated sources and add them to the real images from each WFC3 pointing. 
We then perform the same steps of source detection, photometry, and sample 
selection described in Sections~\ref{sec:photometry} and \ref{sec:selection},
treating the images containing simulated sources identically to how we 
treat the original images. 

We create a catalog of $10^5$ sources, with each source assigned a redshift, 
spectral template, observed flux, size, and shape from parameter 
distributions as follows. The source redshifts are pulled randomly from a 
uniform distribution in the range $7 \leq z \leq 12$. Each source is then 
assigned a spectral template from the \cite{bc03} library with either a 
\cite{salpeter1955} or \cite{chabrier2003} initial mass function, and an 
exponentially-declining star formation history with a characteristic 
timescale of either $\tau=0.01$, 0.5, or 5.0 Gyr. These timescales are 
adopted to approximate a single burst of star formation at one extreme 
($\tau=0.01$ Gyr) and a constant star formation history at the other 
($\tau=5$ Gyr).
The template metallicities 
are one of 0.02, 0.2, 0.4, or $1.0Z_{\odot}$, and are pulled randomly from 
a lognormal distribution that peaks at $0.2Z_{\odot}$.
The ages of the stellar populations are similarly taken from a tweaked 
lognormal distribution that peaks at 10 Myr and decreases with an enchanced 
tail towards older ages to ensure that some older ($>400$ Myr) populations 
exist. The ages available for each simulated source are restricted to be 
between 10 Myr and the age of the universe at the given redshift. 
To each spectral template, we then apply dust extinction due to the 
interstellar medium using the \cite{calzetti2000} extinction law and a 
color excess $E(B-V)$ pulled randomly from a truncated Gaussian distribution
with $\mu=0.1$, $\sigma=0.15$, $E(B-V)_{\mathrm{min}}=0$, and 
$E(B-V)_{\mathrm{max}} = 0.5$. We finally apply attenutation through the 
IGM following the prescriptions in \cite{inoue2014}.
The observed \hband\ magnitudes are pulled from a distribution constructed as 
the combination of a uniform distribution in the range $22 \leq H \leq 28$ 
and a truncated Gaussian for fainter magnitudes with $\mu=26.5$, $\sigma=1$, 
$H_{\mathrm{min}} = 22$, and $H_{\mathrm{max}}=28$. We then normalize each 
spectral template to the \hband\ magnitude assigned to the simulated source. 

\begin{figure*}
\plotone{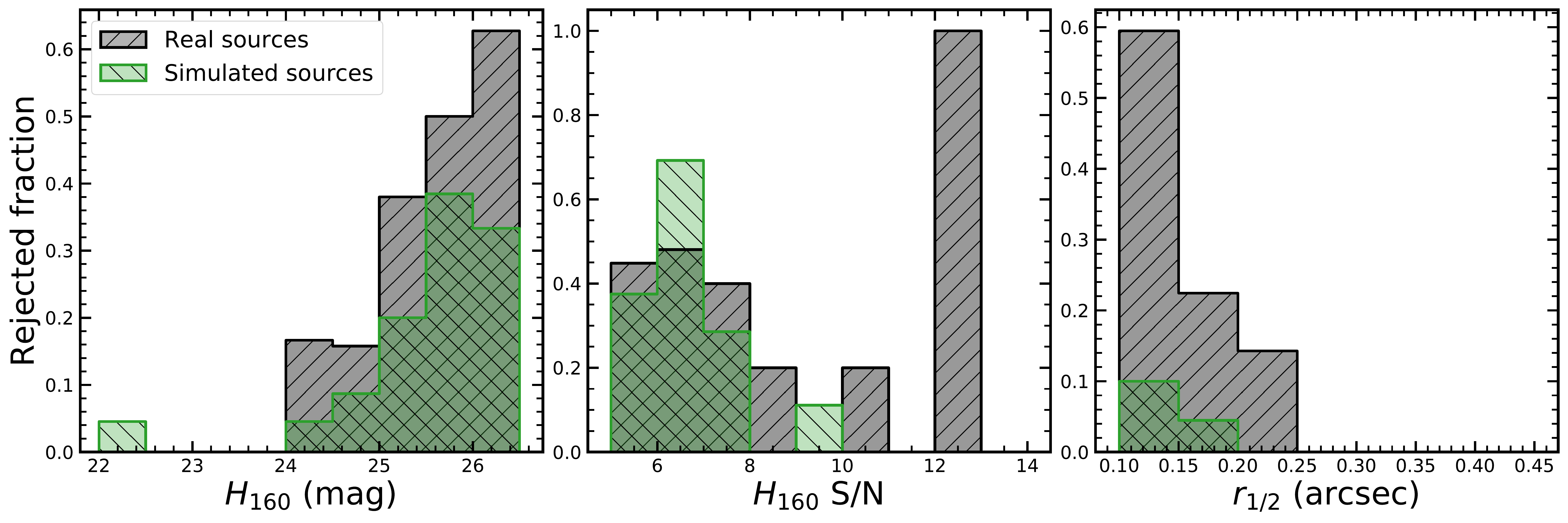}
\caption{
The fraction of real (grey histograms) and simulated (green histograms) 
sources rejected in category 1e during visual inspection 
(Section~\ref{sec:inspection}). This category 
represents the only subjective rejections in our visual classification scheme.
We show the rejection fractions as a function of \hband\ magnitude (left 
panel), \hband\ detection S/N (middle panel), and half-light radius (right 
panel). 
The distributions of rejected real sources approximately follow those for 
the rejected simulated sources, indicating a bias likely introduced by our 
visual inspection that can directly affect our measurement of the luminosity
function. 
We use the left panel of this figure to derive 
a correction factor that accounts for these rejection fractions 
as a function of \hband\ magnitude, and apply it to the calculated 
completeness and effective volumes in Section~\ref{sec:veff}.
\label{fig:rejection}}
\end{figure*}

We generate images of each source using \galfit\ \citep{peng2002} 
as \sersic\ profiles with index $n$ and axis ratio $b/a$ pulled randomly from 
truncated Gaussian distributions with $\mu_n=1$ and $\sigma_n=2.5$
in the range $0.5 \leq n \leq 5$, and $\mu_{b/a}=0.8$ and $\sigma_{b/a}=0.2$, 
$0.1 \leq b/a \leq 1$, respectively. 
The half-light radii are determined by the absolute magnitude of each 
source following the empirical size-luminosity relation presented 
by \cite{kawamata2018}:
\begin{equation}
r_{1/2}(M_{\mathrm{UV}}) = r_0 \times 10^{-0.4 (M_{\mathrm{UV}} - M*)\beta},
\label{eqn:sizes}
\end{equation}
where $r_0$ is the modal radius at $M_{\mathrm{UV}} = -21$,
$M^*$ is the characteristic magnitude of the Schechter function fit to the 
luminosity function, and $\beta$ is the slope of the size-luminosity relation.
We adopt the following parameter values from Table~2 of \citet{kawamata2018}:
\begin{eqnarray}
r_0/\mathrm{kpc} & = & %\left\{ 
  \begin{cases}
    0.94  & ~~~~ \text{for  $z \leq 7.5$}  \\
    0.58  & ~~~~ \text{for  $7.5 <  z \leq 8.5$}  \\
    0.42  & ~~~~ \text{for  $z > 8.5$} 
  \end{cases}
%  \right. 
  \nonumber \\
\beta & = & %\left\{ 
  \begin{cases}
    0.44 & ~~~~ \text{for $z \leq 8.5$} \\
    0.40 & ~~~~ \text{for $z > 8.5$}
  \end{cases} \\
%  \right. \\
M^* & = & %\left\{ 
  \begin{cases}
    -20.56 & \text{for $z \leq 7.5$}  \\
    -19.95 & \text{for $7.5 <  z \leq 8.5$} \\
    -19.80 &  \text{for $z > 8.5$,}
  \end{cases} \nonumber
\end{eqnarray}
and additionally impose $\beta=0.25$ for sources with 
$M_{\mathrm{UV}} \leq -21$, resulting in a size-luminosity relation described 
by a broken power-law such that the slope flattens for brighter galaxies. 
In order to account for intrinsic scatter in the relationship at fixed 
magnitude, we apply a 0.2 dex scatter to the sizes derived from 
equation~\ref{eqn:sizes}.
We note that the \rhalf\ distribution measured in the \se\ catalogs of 
recovered simulated sources traces that of the real sources, 
with the simulated source sizes bracketing those of the real sources,
confirming our choice of parameter values for this relation. 
The \galfit\ model images for 
each simulated source are convolved with the measured PSFs 
(Section~\ref{sec:psfs}) corresponding to each dataset and filter.

Finally, we create 200 realizations of each of the 132 \hst\ pointings 
considered in this paper, each containing $\sim$100 sources randomly placed 
across the image, for a total of $\sim$20000 input sources per 
pointing\footnote{
As can be seen from Table~\ref{tab:obs}, some fields cover significantly 
more or less area than the average $\sim$4.6 arcmin$^2$. In practice, 
we therefore add simulated sources such that the density of sources in each 
pointing is constant at $\sim$22 sources/arcmin$^2$. This 
corresponds to 100 sources for each 4.6 arcmin$^2$ field. However, we add 
fewer sources per realization and produce more realizations for fields 
that have been partially masked as described in Section~\ref{sec:borg}.
The larger fields are the result of combining two partially-overlapping 
\hst\ pointings into a single mosaic, and to these we add more sources 
per realization. We perform at least 150 realizations per field, and so the 
total number of simulated sources added to these larger fields exceeds 
20000.}.
We then run our full sample selection process on these fields, including source 
detection and photometry (Section~\ref{sec:photometry}), 
photometric redshift fitting (Section~\ref{sec:photz}), and the selection 
citeria used to select real high-redshift candidates 
(Section~\ref{sec:selection}).
The one exception is that we do not add Galactic 
extinction to the fluxes of the simulated sources, and so we skip the 
Galactic extinction correction step described in Section~\ref{sec:photometry}.
Finally, we match the catalogs of recovered sources to the simulated input 
source positions using a matching radius of 0\farcs5.

%%%%%%%%%%%%%%%%%%%%%%%%%%%%%%%%%%%%%%%%
\subsubsection{Quantifying Effects of Visual Inspection} \label{sec:sims_inspection}

As described in Section~\ref{sec:inspection}, the visual inspection 
process can affect sample compeleteness by removing sources in a subjective 
way that is not replicable with simulated sources. 
In an attempt to quantify this completeness effect, we included the recovered
simulated sources in our visual inspections, inspecting an equal number of 
real and simulated sources.  
For each field, we randomly selected the same number of simulated sources 
as there were real initial candidates. In this way, the field-to-field filter 
coverages and noise properties of the simulated sources 
matched those of the real candidates.
Importantly, this inspection was ``anonymized'', or performed without 
knowing which sources were simulated.
This is the first time such a method -- crucial to correctly calculating
sample completeness -- has been used in high-redshift galaxy
selection.

Of the 166 real candidates that were rejected through visual inspection, 
we focus here on the 81 rejections that were based on a subjective opinion 
(category~1e, ``a hot pixel, cosmic ray, or source with a strange 
morphology'', Section~\ref{sec:inspection}). 
In Figure~\ref{fig:rejection}, we show the rejected fraction of the real and 
simulated sources that were rejected in category~1e. The fraction of rejected 
simulated sources increases towards fainter \hband\ magnitude (left panel),
lower detection S/N in \hband\ (middle), and smaller half-light radius
(right). Additionally, the fraction of rejected real sources largely follows 
the same distribution as that for simulated sources, indicating that this trend
represents a bias in our visual inspection. However, we do not visually 
inspect the many thousands of simulated sources that meet our selection 
criteria.
As we use the recovery of simulated sources to derive the effective volume 
probed in each field, this bias imparted through our visual inspection can 
therefore directly affect our measurement of the luminosity function. 
In order to account for the rejected fraction of simulated sources, we 
use the left panel of Figure~\ref{fig:rejection} to apply a correction factor 
to the measured completeness in bins of \hband\ magnitude. For example, we 
decrease the number of recovered sources in each field with 
$25.5 < m_{H160} < 26.0$ by 38\%, assuming that this percentage of sources 
would have been rejected during a complete visual inspection. 
We discuss this correction further in Section~\ref{sec:veff}.

\begin{figure*}
\epsscale{1.1}
\plotone{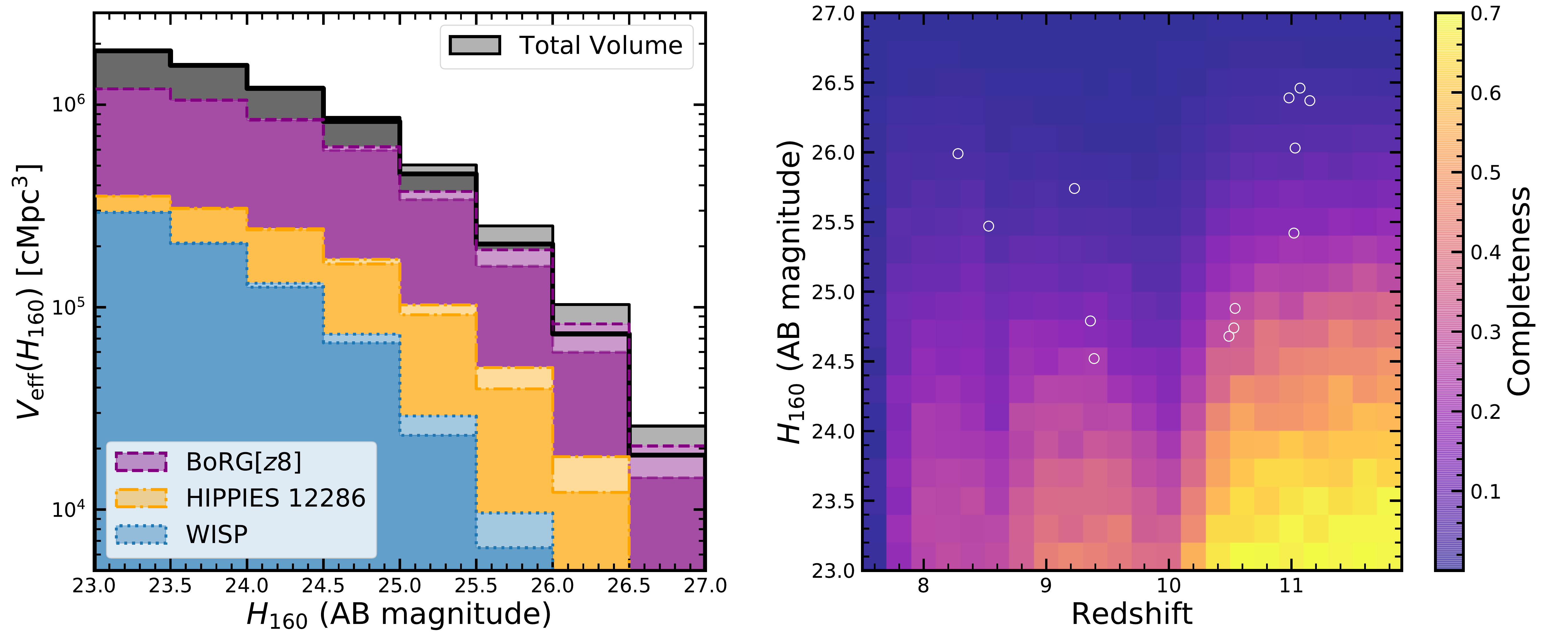}
\caption{The completeness and effective volume calculated for our selection 
criteria. 
We plot the effective volume as a function of \hband\ 
magnitude in the left panel, showing each dataset separately with the total 
volume in black. For all datasets, the lighter shaded region shows the 
effective volume calculated directly from the completeness simulations, while
the darker region includes the correction for reduced completeness 
due to sources removed through visual inspection. As expected given the 
distribution of rejected fractions in the left panel of 
Figure~\ref{fig:rejection}, the effect of these corrections increases towards 
fainter magnitudes. The thick black line indicates the total, corrected 
volume for all three combined datasets.
We show the results of our completeness simulations as a function 
of redshift and \hband\ magnitude in the right panel. The white circles 
indicate the redshifts and magnitudes of our 13 candidates. The completeness is 
highest for bright sources and falls off towards fainter magnitude. It also
peaks at $z>10$, when the \lya\ break redshifts out of the $J$ band 
and sources are \hband-only single detections. At these redshifts, 
non-detections in all three blue filters result in larger portions of the 
redshift PDF at $z>8$, and the recovery
fraction in the WISP fields increases significantly and boosts 
the overall completeness. 
\label{fig:veffs}}
\end{figure*}

%%%%%%%%%%%%%%%%%%%%%%%%%%%%%%%%%%%%%%%%
\subsection{Effective Volume}\label{sec:veff}
We calculate the number density of UV-bright galaxies at $z\sim9$ using 
the effective volume method \citep{felten1977}. This method
involves determining the maximum volume within which a galaxy of a 
given luminosity would be detected by these \hst\ observations and selected
with our specific set of criteria.
The effective volume, $V_{\mathrm{eff}}$, is given by
\begin{equation}
V_{\mathrm{eff}} (M_{\mathrm{UV}}) = \int \frac{\mathrm{d}V}{\mathrm{d}z} P(M_{\mathrm{UV}},z) \: \mathrm{d}z,
\label{eqn:veff}
\end{equation}
where $\mathrm{d}V/\mathrm{d}z$ is the differential comoving volume element,
and $P(M_{\mathrm{UV}},z)$ is the probability that a source of absolute
magnitude $M_{\mathrm{UV}}$ and redshift $z$ will make it into our sample 
given our selection function.
We estimate $P(M,z)$ using the set of field-specific simulations described 
in Section~\ref{sec:sims}. 

We calculate the completeness as the ratio of the number of recovered
sources to input sources in bins of input magnitude and redshift.
In order to account for the simulated sources that we rejected during our 
visual inspection, we decrease the number of recovered sources in each 
magnitude bin by the fraction of rejected simulated sources shown in 
Figure~\ref{fig:rejection}. For this correction, we scale the magnitude 
bins to the 5$\sigma$ \hband\ magnitude limit of each field. In this way 
we approximate rejecting an increasing number of simulated sources as 
we approach the field depth. We do not include the non-zero fraction of 
bright ($m_{H160} \sim 22.3$) rejected simulated sources, as there are no 
rejected real sources in that magnitude bin. 

We show the resulting completeness as a function of redshift and \hband\ 
magnitude in the right panel of Figure~\ref{fig:veffs}. There are almost no 
recovered sources at $z<7.5$, where the \lya\ break falls close to the 
center of the \yaband\ and \ybband. In these cases, the sources are detected 
in these filters with S/N$>$2 and therefore do not satisfy our selection 
criteria. As the \yaband\ is narrower than the \ybband, simulated sources 
at $z\sim7.5$ are recovered in \borg\ fields, while the recovery fraction in 
\hippies\ fields starts to increase for $z>8$. 
The completeness dips slightly at $z\sim8.6$, the redshift
at which the \lya\ break transitions almost entirely out of the 
\ybband\ and sources in \borg\ and HIPPIES fields become two-band detections. 
There is another dip at $z\sim10$, as the \lya\ break shifts out of the 
\jband\ and \jbband\ filters and sources are detected in \hband\ only. 
After each initial dip, the completeness increases with redshift as more 
filters lie blueward of the break, ruling out lower-redshift solutions and 
increasing the portion of the redshift PDF that lies at $z>8$. 
The recovery in WISP fields is very low at $z\lesssim10$ because the broad 
\jbband\ filter results in poorly constrained photometric redshift PDFs for
sources from $z\sim7$ to $z\sim10$. The completeness is therefore 
highest at $z>10.5$, when the WISP recovery fractions increase again for 
\hband-only detections.

We calculate the effective volume probed in each field using 
equation~\ref{eqn:veff}. In the left panel of Figure~\ref{fig:veffs},
we show the effective volume calculated as a function of \hband\ apparent
magnitude.
For each dataset, the darker shaded region shows the $V_{\mathrm{eff}}$ that 
has been corrected for the bias introduced by our visual inspections, and 
the lighter shaded region shows the original, uncorrected $V_{\mathrm{eff}}$.
It can be seen here how this correction increasingly reduces the effective 
volume for fainter \hband\ magnitudes. The thick solid black line indicates 
the total, corrected volume of all three combined datasets.
We use the effective volume calculated as a function of absolute UV magnitude 
for our calculation of number densities in Section~\ref{sec:densities}, 
adopting a rest-wavelength of 1500\AA\ that corresponds to approximately 
the center of the \hband\ filter at $z\sim9.5$. 

%%%%%%%%%%%%%%%%%%%%%%%%%%%%%%%%%%%%%%%%
\subsection{Volume Number Densities}\label{sec:densities}
We calculate the observed number density of galaxies in the redshift 
range $8.5\lesssim z \lesssim 11$ following the methods described in 
\citet{finkelstein2022a}. We describe our steps here but refer the reader
to \citet{finkelstein2022a} for more details. 
We use a Markov Chain Monte Carlo (MCMC) analysis with the Cash 
statistic \citep{cash1979,ryan2011} as the goodness-of-fit measure. 
For a given assumed number density, the Cash statistic 
describes the likelihood that the observed number of galaxies matches the
expected number, but models the probability distribution as a Poisson 
rather than a Gaussian distribution. This choice of statistic is therefore 
appropriate for our small sample size. 

Additionally, we use the ``pseudo-binning'' technique developed by 
\citet{finkelstein2022a}, which reduces the biases imparted by choice of 
absolute magnitude bin center and width on the recovered shape of the 
luminosity function. 
At each step of the MCMC chain, the number density is calculated for a 
given $M_{\mathrm{UV}}$ in a bin with a width randomly pulled from the 
range $0.2-1.5$ mag.
The volume associated with this bin is determined using the completeness 
simulations described in Section~\ref{sec:veff} and equation~\ref{eqn:veff}. 
We pull an absolute magnitude for each source randomly from the distributions
calculated in Section~\ref{sec:magnification}, using the magnification-corrected
distributions for the three strongly-lensed sources. We then determine the 
observed number of sources in the bin given these randomly-drawn quantities. 
We perform this calculation in magnitude 
steps of $\Delta M_{\mathrm{UV}} = 0.1$ in the range 
$-24 < M_{\mathrm{UV}} < -20$. In this way, the number density posterior 
distribution incorporates the uncertainties on the absolute magnitude of 
each candidate -- which in turn fold in the uncertainties in redshift, 
\hband\ observed magnitude, and magnification correction -- as well as 
reducing the dependence on choice of magnitude bins.
We use 20 walkers with a burn in of $10^5$ steps, after which we discard the 
chains and use an additional $10^4$ steps to calculate the posterior 
distribution.

We present the calculated number densities in Table~\ref{tab:lf} in the 
Appendix and display our results in Figure~\ref{fig:lfs}, plotting the 
68\% and 95\% ranges of the density posterior as blue shaded regions. The 
shading extends from $M_{\mathrm{UV}}=-23.05$, the brightest bin that contains 
a candidate, to $M_{\mathrm{UV}}=-20.5$, the absolute magnitude at which our 
average completeness drops below 20\%. 
We plot vertical blue error bars throughout the shaded region to indicate 
the uncertainty in our measurements due to cosmic variance. We estimate this 
uncertainty using the calculator from \citet{bhowmick2020}. At all magnitudes,
the uncertainty on the number density due to cosmic variance is very low.
The overall uncertainties are dominated by Poisson noise.
For comparison, we also show a number of results from the literature, 
including the pseudo-binning results from 
\citet{finkelstein2022a} for their 11 candidates in the same redshift range
(purple shaded region). We discuss the implications of our results in 
the context of other similar high-redshift searches in 
Section~\ref{sec:discussion}.

\begin{figure*}
\plotone{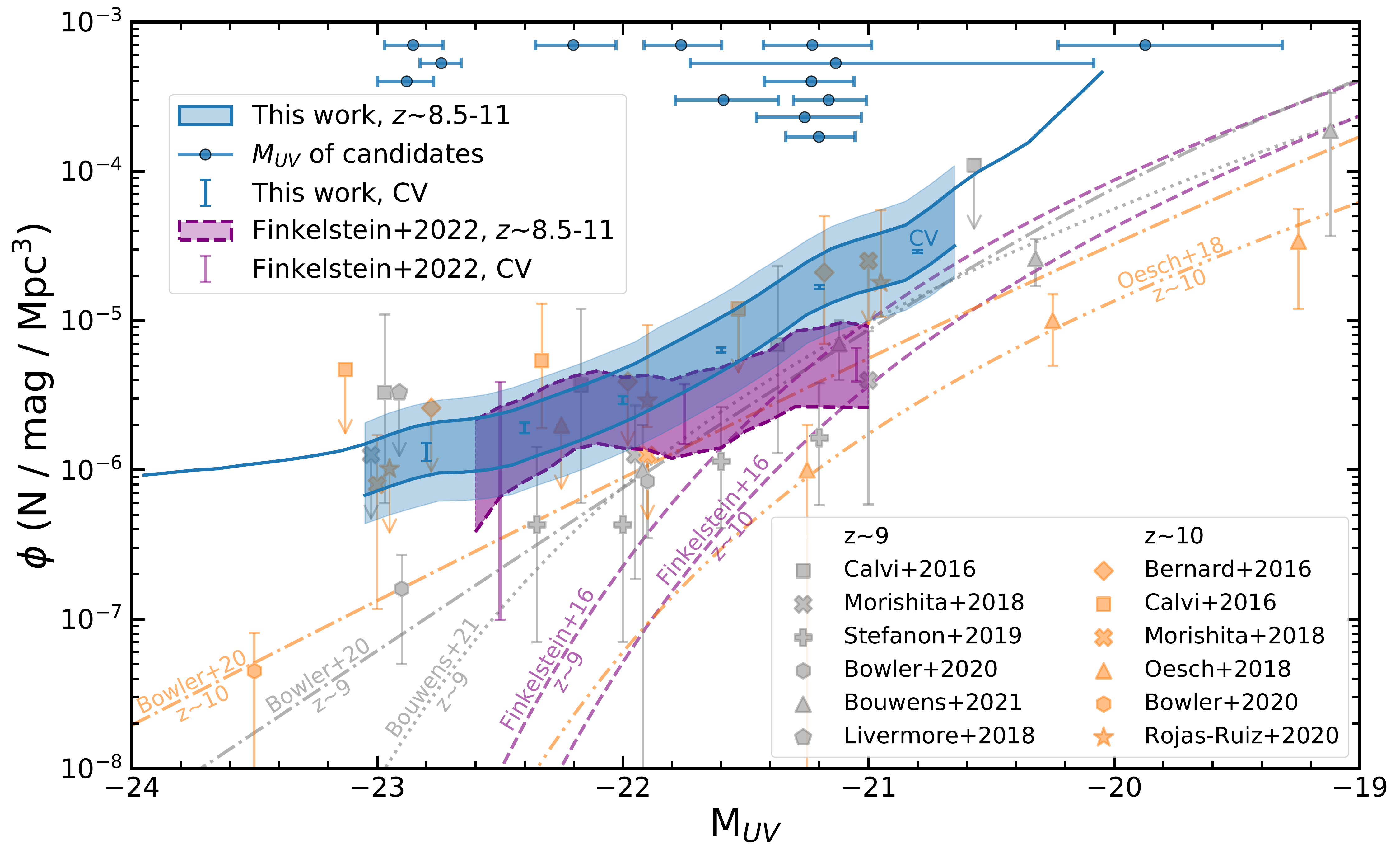}
\caption{The UV luminosity function (blue shaded region) in the 
range $8.5 \lesssim z \lesssim 11$ as measured in the \borg, \hippies\ and 
WISP datasets using the pseudo-binning
technique developed by \citet{finkelstein2022a}. 
The inner (outer) blue shaded region is the Poisson uncertainty determined 
by the 68\% (95\%) range of the number density posterior, extending from 
the brightest bin that 
contains a candidates to the magnitude at which our average completeness drops 
below 20\%. The solid
blue line shows the upper 68\% range at brighter and fainter magnitudes,
and the blue vertical errorbars indicate the uncertainty due to cosmic variance
based on the calculator from \citet{bhowmick2020}. 
We show the absolute magnitudes of each of our 13 candidates at the top as 
blue circles, with errorbars representing their
$M_{\mathrm{UV}}$ posterior distribution 68\% ranges.
We compare our results with those from \citet{finkelstein2022a} (purple
shaded region), as well as other studies using \hst\ parallel fields
\citep{bernard2016,calvi2016,livermore2018,morishita2018,rojas-ruiz2020},
ground-based imaging in the COSMOS/UltraVISTA field
\citep{stefanon2019,bowler2020}, and \hst\ imaging in the extragalactic
legacy fields \citep{oesch2018,bouwens2021a}.
For reference, we also show the Schechter function fits from 
\citet{oesch2018} and \citet{bouwens2021a}, 
the Schechter function predictions from \citet{finkelstein2016},
and the double power law 
luminosity functions from \citet{bowler2020}.
We find a high density of bright galaxies, in good agreement with other 
\hst\ parallel studies. Our luminosity function is consistent for 
$M_{\mathrm{UV}} \lesssim -21.5$ with that 
of \citet{finkelstein2022a}, which was calculated using the same method 
but in a handful of fields.
The uncorrelated nature of the pure-parallel observations we include in our 
analysis significantly reduces the effects of cosmic variance, allowing 
for a more complete sampling of the population of bright galaxies at \zsamp. 
\label{fig:lfs}}
\end{figure*}

%%%%%%%%%%%%%%%%%%%%%%%%%%%%%%%%%%%%%%%%%%%%%%%%%%%%%%%%%%%%%%%%%%%%%%%
\section{Discussion} \label{sec:discussion}

%%%%%%%%%%%%%%%%%%%%%%%%%%%%%%%%%%%%%%%%
\subsection{Comparison with Previous Luminosity Function Calculations}
We begin by comparing our results to other studies in similar redshift 
ranges. In Figure~\ref{fig:lfs}, we include number density measurements from 
high-redshift galaxy searches performed in \hst\ parallel fields
\citep{bernard2016,calvi2016,livermore2018,morishita2018,rojas-ruiz2020}, 
ground-based imaging in the COSMOS/UltraVISTA field 
\citep{stefanon2019,bowler2020}, the CANDELS fields
\citep{finkelstein2022a} as well as those that included the 
Hubble Frontier Fields \citep{oesch2018} and the Hubble Ultra Deep 
Field \citep{bouwens2021a}.
Collectively, these surveys cover a large dynamic range in both magnitude and 
volume, yet individually they have complementary strengths. The 
Hubble extragalactic legacy fields have the depth needed to probe to 
fainter magnitudes, yet include only a handful of lines of sight and are 
therefore more susceptible to cosmic variance. On the other hand, the 
\hst\ pure parallel programs include hundreds of independent pointings, 
and the ground-based COSMOS/UltraVISTA field covers a very large area at 
a shallow depth.
These surveys are more likely to detect the rare, brightest galaxies, but 
at the expense of lower sample completeness for 
$M_{\mathrm{UV}} \gtrsim -21.5$.

Our results are consistent with other searches performed in 
\hst\ parallel surveys. 
\citet{calvi2016}, \citet{morishita2018} and \citet{rojas-ruiz2020}
measured the rest-UV luminosity function in the 
BoRG[$z910$] survey, which includes imaging in \jhband. The second detection
filter helps protect against the selection of spurious sources and improves 
the photometric redshift by more fully constraining the location of the 
\lya\ break. The 132 \hst\ fields we include in our results are less
optimized for the selection of \zsamp\ galaxies. 
Nevertheless, the number densities from these works at
$z\sim9$ (grey symbols), $z\sim10$ (orange symbols), and across the full 
redshift range $8.4 < z < 10.6$ as in \citet[][orange stars]{rojas-ruiz2020}
lie approximately at the median density we calculate across the full 
absolute magnitude range that we consider. \citet{bernard2016} explore 
some of the same fields that we do, yet our two samples of high-redshift 
galaxy candidates do not overlap. However, the number densities reported 
by \citet{bernard2016} at $z\sim10$ are in good agreement with our 
measurements. We note that while the UV luminosity function from the 
SuperBoRG fields has not been published at the time of this writing, 
\citet{roberts-borsani2021b} report the identification of 49 candidates in 
the range $z\sim8-12$, likely resulting in a high number density in agreement 
with these other searches in pure-parallel data.

Overall, the results using the \hst\ pure parallel surveys find number 
densities above those measured in the \hst\ legacy or ground-based imaging. 
The two brightest magnitude bins at $z\sim9$ from \citet{bouwens2021a} 
are consistent with our measurement within the uncertainties, yet the 
\citet{oesch2018} measurements at $z\sim10$ are much lower. 
Both studies include fields deep enough to probe the $z\sim9-10$ galaxy 
population down to $M_{\mathrm{UV}} > -18$, yet do not find galaxies at 
$M_{\mathrm{UV}}<-22$.
The brightest $z\sim10$ candidate contributing to the luminosity 
function of \citet{oesch2018} is GN-z11 \citep{oesch2016} with a 
magnitude of $m_{\mathrm{UV}} = -21.6$.
The bright end of the luminosity function measured by \citet{stefanon2019}
is also lower but consistent with our measurements within the uncertainties.

We next focus on the comparison between our results and those of 
\citet{finkelstein2022a}, as we have performed very similar sample selections, 
completeness analyses, and have used the same method to measure the UV 
luminosity function. 
Our luminosity functions are fully consistent down to 
$M_{\mathrm{UV}} \sim -21.8$, below which the purple region decreases 
and is more consistent with the Schechter functions of 
\citet[][predicted for a smooth decline in the luminosity function at 
$z>8$]{finkelstein2016}, \citet[][showing a slower evolution in the bright end of the luminosity function at $z>8$]{bowler2020} and 
\citet[][proposing an accelerated decline in the luminosity function]{bouwens2021a}.
The high-redshift sample presented by \citet{finkelstein2022a} include an 
overdensity in the EGS field, which may serve to increase their number 
densities over similar studies in the \hst\ legacy fields.
At the bright end, however, the uncertainties of \citet{finkelstein2022a} are
dominated by cosmic variance, finding a fractional uncertainty of 0.95 at
$M_{\mathrm{UV}}=-22.5$. With over 130 independent pointings included in 
our analysis, we find a fractional uncertainty of 0.14 at 
$M_{\mathrm{UV}}=-22.8$ (blue vertical errorbars). This indicates that the 
high density of bright galaxies as reported here and in 
\citet{finkelstein2022a} is robust against cosmic variance, though we note
that the Poisson 
uncertainties in our measurements are large.
As described in Section~\ref{sec:faintlowz}, we expect our 
contamination from lower-redshift interlopers to increase towards fainter 
magnitudes. This increasing contamination could explain why we calculate 
a higher luminosity function than that presented by \citet{finkelstein2022a}
for $M_{\mathrm{UV}} \gtrsim -21.8$.

These results indicate that wide-area or pure parallel programs
provide valuable insight about the bright end of the rest-UV luminosity 
function. Their large areas or 
many uncorrelated pointings significantly reduce the effects of cosmic 
variance and allow for a more complete sampling of the population of 
bright, \zsamp\ galaxies.
While there is significant scatter and Poisson uncertainty for 
$M_{\mathrm{UV}} \lesssim -22$, there is a broad consensus among these programs
that there is an excess of bright galaxies at \zsamp\ compared to a Schechter
function parameterization of the luminosity function. 
By $M_{\mathrm{UV}} \sim -21.5$, results 
from these shallow observations may be more prone to contamination, and we
should turn to deeper fields for measurements of the luminosity function. 

While many factors affect the evolving abundance of galaxies with redshift, 
the two dominant processes influencing the bright end of the observed 
UV luminosity function at $z \sim 9$ are likely star formation efficiencies 
and dust attenuation. \citet{somerville2015b} and \citet{yung2019a}
showed that the predicted number density of
high redshift galaxies ($z \gtrsim 4$) is very sensitive to the
efficiency with which molecular hydrogen is converted into stars. This
is because at these redshifts, the molecular gas consumption time (a few 
Gyr in nearby galaxies) becomes comparable to the age of the
Universe. Specifically, in their fiducial model, \citet{yung2019a}
adopt a double power law for the relationship between molecular gas
surface density and star formation rate surface density, where the slope 
of the power law becomes steeper above a critical molecular gas surface 
density. Galaxies at early times are more compact than those at lower 
redshifts, and so most star formation is likely occurring in gas at densities 
above this critical density. The net effect is that gas depletion times are 
shorter, and star formation efficiencies higher, at high redshift.
The authors find that with a slope of $\sim 2$ at high gas densities 
\citep[a bit steeper than the slope of unity measured for nearby spiral galaxies;][]{bigiel2008},
their semi-analytic model can reproduce the observed number density of 
bright galaxies out to $z\sim 8$. A model with a single power law star 
formation relation with a slope of either unity or 1.5 
\citep[as in the canonical Kennicutt-Schmidt relation][]{schmidt1959,schmidt1963,kennicutt1989,kennicutt1998} 
significantly underproduces bright galaxies even by $z\sim 6$. 

Dust attenuation can also play a major role in determining the shape of the 
bright end of the UV luminosity function \citep[e.g.,][]{vogelsberger2020a}.
The high density of UV-bright galaxies in Figure~\ref{fig:lfs} may indicate 
low or negligible dust content in massive galaxies at these redshifts.
The shallow evolution in the bright end at high redshift may be cause by a
redshift and/or mass dependence in dust production and dust-to-metal ratios.
\citet{finkelstein2022a} shows that the semi-analytic model from
\citet{yung2019a} can best reproduce their observed density of bright
galaxies at $z\sim9$ when there is no dust attenuation included
\citep[yet SED fitting for the galaxies in their sample indicate
non-negligible dust contents][]{tacchella2022a}. It is also important to note 
that the effects on the bright end of lower dust attenuation
are degenerate with those of higher star formation efficiencies, and
detailed observations with \jwst\ NIRCam and MIRI will be required to
separate these effects in bright, high-redshift galaxies.

This observed excess could also be due to a high fraction of
faint active galactic nuclei (AGNs) in our sample. While there have been
only a handful of AGN detected at $z>7$, massive
accreting black holes must exist at earlier times to explain the
masses of the quasars observed at $z\sim6-7$ \citep{banados2018}.
Additionally, there is a growing body of evidence that some of the brightest
known $z > 7$ star-forming galaxies may in fact host detectable AGN activity,
typically detected via significant \ion{N}{5} emission
\citep[e.g.,][]{tilvi2016,laporte2017b,hu2017} including one of the
highest-redshift spectroscopically-confirmed galaxies at $z=8.68$
\citep{zitrin2015,mainali2018}.
We cannot rule out the possibility that the luminosity of these
bright galaxies is partially due to an AGN contribution. Spectroscopic
followup with NIRSpec's medium and high resolution gratings could help shed 
light on this question.

%%%%%%%%%%%%%%%%%%%%%%%%%%%%%%%%%%%%%%%%
\subsection{Bright Galaxies during the Epoch of Reionization with \jwst}\label{sec:brightgals}
This sample of 13 galaxy candidates at $8.5\lesssim z \lesssim 11$ represent 
the very bright end of the rest-UV luminosity function, with 
$M_{\mathrm{UV}} \lesssim -21.2$. 
Such massive galaxies likely formed in regions of relative overdensity, 
where the build-up of mass began at earlier times. These regions are of 
particular interest in understanding the early stages of the reionization 
of the IGM. 
Whether the progression of the epoch of reionization at early times 
was dominated by ionizing emission from massive galaxies with high 
escape fractions \citep[e.g.,][]{naidu2020,naidu2021,matthee2021}, or 
from the more numerous lower-mass population \citep[e.g.,][]{finkelstein2019a},
these overdense regions would have been among the first to ionize. 
Bright galaxies may therefore reside in 
ionized bubbles even as early as $z\gtrsim8$. Indeed, \lya\ has been detected
in the spectra of two galaxies at $z\sim8.7$ in the EGS field
\citep{zitrin2015,larson2022}, potentially 
indicating ionized regions large enough for the \lya\ photons to 
redshift out of resonance before encountering the neutral IGM.

The \zsamp\ candidates we present in this paper may trace overdense regions 
at early times, making them excellent targets for follow-up with \jwst.
Deep NIRCam imaging can be used to detect the population of fainter 
galaxies as demonstrated by \citet[][at $z\sim7$]{castellano2016,castellano2018}, and 
deep NIRSpec spectroscopy can be used to probe reionization in these 
regions via \lya\ and \ion{Mg}{2} \citep[e.g.,][]{henry2018}.
The NIRSpec prism also provides the opportunity to characterize 
the ionizing efficiencies of bright galaxies. 
The building narrative surrounding reionization hints at a chaotic
early universe, with galaxies undergoing intense bursts of star-formation, 
producing massive, bright stars that power strong nebular emission lines 
\citep[e.g.,][]{stark2016,finkelstein2016,endsley2020b}.
The relative strength and ratios of these lines, such
as that of \oiii$\lambda\lambda4959,5007$/\oii$\lambda\lambda3727,3729$
\citep[=O32; e.g.,][]{steidel2016}, can therefore reveal significant detail 
about the strength of the ionizing continuum in galaxies at $z>9$
\citep[e.g.,][]{shapley2003,erb2010,steidel2016,ravindranath2020}. 
Observational evidence shows that 
high O32 correlates with an increasing ionizing efficiency 
\citep[e.g.,][]{tang2018}. For galaxies during the epoch of reionization, 
large measured ionizing efficiencies may indicate sources that are capable
of ionizing a big bubble in the IGM. 
The combination of NIRSpec spectroscopy and deep NIRCam imaging will 
test whether reionization is driven by the most massive galaxies, or by 
the smaller, fainter sources.

With \jwst\ GO-2426 (Co-PIs: Bagley \& Rojas-Ruiz), we will follow-up five 
sources from this paper (Par0456-2203\_473, Par0756+3043\_92, 
Par0956-0450\_684, Par335\_251, and Par2346-0021\_164\footnote{While
we removed Par2346-0021\_164 from consideration when calculating the 
luminosity function at $z\sim8.5-10$, $>$60\% of the integrated 
redshift PDF is at $z>8$. This candidate may therefore still be a 
high-redshift galaxy. Alternatively, spectroscopic characterization of 
a low-redshift interloper is also valuable in evaluating high-redshift 
galaxy selection techniques.}) as well as six additional candidates 
presented in \citet{rojas-ruiz2020}. With approximately one hour of 
exposure time per target with the NIRSpec prism in fixed slit mode,
the observations will be sensitive enough to detect the \lya\ break 
for redshift confirmation as well as nebular emission lines such as 
\oii, \hb, and \oiii. There are a number of other \jwst\ programs that 
will be exploring bright galaxies during the epoch of reionization,
including GO 1740 (PI: Harikane), GO 1747 (PI: Roberts-Borsani),
GO 1933 (PI: Matthee), GO 2279 (PI: Naidu), and GO 2659 (PI: Weaver).
Together, these Cycle 1 \jwst\ observations will significantly expand our 
understanding of massive galaxies at early times and their contribution 
to the reionization of the IGM.

%%%%%%%%%%%%%%%%%%%%%%%%%%%%%%%%%%%%%%%%%%%%%%%%%%%%%%%%%%%%%%%%%%%%%%%
\section{Summary} \label{sec:summary}
We have presented a search for bright \zsamp\ candidate galaxies in 132 \hst\ 
fields covering 620 arcmin$^2$. The fields were observed in parallel as
part of the \borg, \hippies, and WISP surveys. The independent and 
uncorrelated nature of these pointings significantly reduces the effect of 
cosmic variance compared to observations covering larger contiguous areas.
While the three datasets we include in our analysis use slightly different 
filter sets, all provide imaging in \hband, a $J$-band filter 
(either \jband\ or \jbband), and a subset of \vband, \iaband, \ibband, 
\yaband\ and \ybband. This filter suite allows for two-band detections of 
galaxies at \zsamp\ and \hband-only detections at $z\gtrsim10$. 
The majority of the \hst\ pointings were also observed by \spitzer/IRAC at 
3.6\micron\ and (for a smaller subset) 4.5\micron. The \hst\ imaging
in two bluer filters is needed to detect the redshifted \lya\ break, and 
the IRAC imaging at longer wavelengths is crucial for ruling out 
contamination by passive or dusty star-forming galaxies at $z\sim2-3$ and 
low mass stars or brown dwarfs.

We performed a careful $z\gtrsim8.5$ sample selection using 
criteria involving detection significance (S/N$_{160}>5$) 
and photometric redshift fitting. 
We used \eazy\ to measure photometric 
redshifts and selected sources with at least 70\% of their redshift 
probability distribution function at $z>8$. We also enforced a lower limit 
for source half-light radius, using a random forest algorithm to set the 
radius thresholds for each dataset.
We conducted a thorough check for image persistence,
exploring all observations obtained in the 24 hours before each image of 
a candidate field. We performed an additional 
screening for self-persistence in WISP images that is caused by zeroth orders 
in the grism exposures. 

Even the most conservative selection criteria cannot remove all spurious 
detections. 
We therefore visually inspected all candidates, rejecting diffraction spikes, 
sources along image edges, bad pixels, and sources with morphologies that 
appeared similar to hot pixels, cosmic rays, or other artifacts. Importantly,
we performed this visual inspection on both real and simulated candidates, 
randomizing their order and anonymizing their nature. This technique 
allowed us to include the effects of these subjective rejections in 
our completeness corrections. 

We next measured the photometry of each candidate in all available 
\spitzer/IRAC imaging, using \galfit\ to model the light from the 
candidates and any neighboring sources in the IRAC images. We refit the 
\hst+\spitzer\ photometry with \eazy\ and rejected one source with a redshift 
probability distribution that no longer satisfied our selection criteria. 
Additionally, we rejected one candidate 
with a low $\Delta \chi^2$ 
that indicated the high-redshift solution was not preferred at high 
significance. Finally, we explored whether any of the candidates were 
likely to be stars by fitting the \hst+\spitzer\ photometry with 
spectral models of M, L, and T dwarf stars. All candidates were better 
fit by a high-redshift galaxy template. Our final sample was
thirteen candidates from $8.3 \lesssim z \lesssim 11$, including the first 
$z>8$ candidate identified from the WISP survey.

We presented our sample in Section~\ref{sec:sample}, including a comparison 
to previous studies in the same fields. While our selection criteria are not
sensitive to the $z\sim7-8$ samples previously identified in the \borg\ fields, 
we did recover two candidates presented by \citet{bradley2012} that likely 
lie at the high end of their redshift distribution. We did not recover any of 
the three \zsamp\ candidates presented by \citet{bernard2016}, highlighting
the dependence of high-redshift candidate samples on the methods used to 
select them.
We then explored the possibility of low-redshift contamination in our 
high redshift sample in two ways. First, we stacked the imaging in all filters 
and did not detect significant flux in any of the dropout bands. Next, we 
estimated the contamination from faint low-redshift galaxies by dimming the 
fluxes of bright $z=1-4$ galaxies and reselecting them with our selection 
criteria. We found low contamination rates of $\sim$1-3\%
in almost all fields (with the highest rate of 12.9\% in the WISP field). 

After applying a correction for magnification along the line of sight to
the three sources with bright, close neighbors, we calculated the UV 
absolute magnitudes of each candidate. Our magnitude uncertainties included 
the uncertainties in the photometric redshifts, \hband\ fluxes, and 
magnification corrections. We used simulated sources to quantify our 
sample selection incompleteness and determine the effective volume of the 
combined 132 pointings. We then calculated the rest-UV luminosity function 
using an MCMC analysis that incorporates our magnitude uncertainties 
and the pseudo-binning technique presented by 
\citet{finkelstein2022a}. We found number densities that are higher than those 
presented in most searches in \hst\ legacy fields, but are 
in good agreement with other 
searches performed in \hst\ parallel fields. Our luminosity function is also 
consistent with that of \citet{finkelstein2022a} down to 
$M_{\mathrm{UV}} \sim -21.8$. Taken together, these results demonstrate the 
power of pure parallel programs in measuring the bright end of the 
luminosity function with a much smaller uncertainty due to cosmic variance. 
However, for fainter magnitudes, the shallow parallel observations may be 
more prone to contamination. 

These results also indicate evidence for an excess of bright 
($M_{\mathrm{UV}} \sim -22$) \zsamp\ galaxies compared to expectations for a 
smoothly evolving Schechter function from lower redshifts. We concluded by 
discussing possible implications, 
including the potential for increased star formation 
efficiencies and decreased dust attenutation to explain the observations. We 
also discussed what can be learned by studying bright galaxies during the 
epoch of reionization.
These galaxies likely trace overdense regions that may have been 
among the first to ionize. They are also excellent targets for efficient
followup with \jwst, bright enough for detailed spectroscopic analysis of 
their ionizing power and contribution to reionization. Five of the sources 
presented in this paper will be observed with the NIRSpec prism as part of 
\jwst\ GO-2426, and many other bright high-redshift galaxies will be observed 
during Cycle 1. Our understanding of the early universe stands to be 
transformed in the coming year of observations.

%%%%%%%%%%%%%%%%%%%%%%%%%%%%%%%%%%%%%%%%%%%%%%%%%%%%%%%%%%%%%%%%%%%%%%%
%\begin{acknowledgements}
\vspace{5mm}

MBB, SLF, and KDF acknowledge that they work at an institution, the 
University of Texas at Austin, that sits on indigenous land. 
The Tonkawa lived in central Texas and the
Comanche and Apache moved through this area. We pay our
respects to all the American Indian and Indigenous Peoples
and communities who have been or have become a part of
these lands and territories in Texas. We are grateful to be
able to live, work, collaborate, and learn on this piece of
Turtle Island.

In this paper we refer to the \textit{James Webb Space Telescope}
using only the acronym \jwst,
reflecting our choice to celebrate the promise of this telescope without 
acknowledging the public official for whom it is named.
This individual has been implicated in anti-LGBTQI+ attitudes that do
not reflect the authors' values related to inclusion in science.

The authors are grateful to Mira Mechtley for developing the comprehensive 
pipeline use to reduce the \hippies\ dataset presented in this paper.

This research was partially supported by NASA through ADAP awards NNX16AN47G 
(linked to Spitzer GO program 11121) and 80NSSC18K0954.
MBB acknowledges support from a NASA Keck PI Data Award, PID 83/2019B\_N127, 
administered by the NASA Exoplanet Science Institute. 
SRR acknowledges financial support from the International Max Planck 
Research School for Astronomy and Cosmic Physics at the University of 
Heidelberg (IMPRS--HD).
RSS is supported by the Simons Foundation.
%.
YSD acknowledges the support from the NSFC grants 11933003, and the China 
Manned Space Project with No. CMS-CSST-2021-A05. 

This research is based on observations made with the NASA/ESA 
\textit{Hubble Space Telescope} obtained from the Space Telescope Science 
Institute, which is operated by the Association of Universities for Research 
in Astronomy, Inc., under NASA contract NAS 5–26555. These observations are 
associated with programs 11519, 11520, 11524, 11528, 11530, 11533, 11534,
11541, 11700, 11702, 12024, 12025, 12283, 12286, 12572, 12902, 12905, 
13352, 13517 and 14178. \\

This research has made use of the NASA/IPAC Infrared Science Archive, which 
is funded by the National Aeronautics and Space Administration and operated by 
the California Institute of Technology.

%%%%%%%%%%%%%%%%%%%%%%%%%%%%%%%%%%%%%%%%%%%%%%%%%%%%%%%%%%%%%%%%%%%%%%%
\facilities{\textit{HST} (WFC3),
            \textit{Spitzer} (IRAC), 
            MAST, IRSA}

\software{Astropy \citep{astropy},
          Photutils \citep[v0.6;][]{bradley2019},
          SciPy \citep{scipy},
          Scikit-learn \citep{scikit-learn},
          Source Extractor \citep{bertin1996},
          \galfit\ \citep{peng2010},
          \eazy\ \citep{brammer2008}
          }

%\end{acknowledgements}
%%%%%%%%%%%%%%%%%%%%%%%%%%%%%%%%%%%%%%%%%%%%%%%%%%%%%%%%%%%%%%%%%%%%%%%
\bibliography{z9}

%%%%%%%%%%%%%%%%%%%%%%%%%%%%%%%%%%%%%%%%%%%%%%%%%%%%%%%%%%%%%%%%%%%%%%%
\appendix

%%%%%%%%%%%%%%%%%%%%%%%%%%%%%%%%%%%%%%%%%%
\section{Candidate Photometry and Luminosity Function Measurements} \label{app:phot}
In Table~\ref{tab:samplephot}, we provide the photometry of each candidate 
in all available filters in units of nJy. In Table~\ref{tab:lf}, we report 
the median and 68\% range of the number density posterior distribution 
plotted in Figure~\ref{fig:lfs}.

\begin{deluxetable}{cccccccccccc}
\tablecaption{Sample Photometry \label{tab:samplephot}}
\movetabledown=5mm
\tablehead{
\colhead{Par} & \colhead{ID} & \colhead{\vband} & \colhead{\yaband} & \colhead{\jband} & \colhead{\hband} & \colhead{3.6\micron} & \colhead{4.5\micron} \\
\colhead{} & \colhead{} & \colhead{(nJy)} & \colhead{(nJy)} & \colhead{(nJy)} & \colhead{(nJy)} & \colhead{(nJy)} & \colhead{(nJy)}
}
\startdata
Par1033+5051 & 116 & $-0.35\pm4.71$ & $0.11\pm9.47$ & $153.80\pm11.21$ & $145.76\pm15.83$ & $420.31\pm188.84$ & $1386.37\pm744.31$ \\
Par0456-2203 & 473 & $-26.35\pm21.49$ & $-10.55\pm16.58$ & $176.79\pm31.70$ & $563.72\pm50.93$ & $2964.83\pm668.70$ & $2987\pm1079.72$ \\
Par0926+4000 & 369 & $23.34\pm27.31$ & $-6.60\pm11.81$ & $51.28\pm36.70$ & $487.13\pm35.67$ & $162.72\pm356.68$ & \nodata \\
Par0756+3043 & 92 & $-1.17\pm12.04$ & $1.22\pm15.25$ & $-9.08\pm11.69$ & $140.53\pm25.49$ & $276.56\pm455.82$ & $389.83\pm1186.16$ \\[2mm]
\hline
Par & ID & \iaband\ & \yaband\ & \jband\ & \hband\ & 3.6\micron\ & 4.5\micron\ \\
\hline
Par0440-5244 & 497 & $-9.97\pm18.86$ & $-6.55\pm9.49$ & $193.34\pm18.95$ & $235.66\pm29.11$ & $333.83\pm1767.52$ & $738.38\pm623.41$ \\
Par1301+0000 & 37 & $-24.32\pm43.33$  & $-3.41\pm16.46$ & $-5.51\pm24.00$ & $246.95\pm33.66$ & $-340.46\pm437.01$ & $1542.47\pm1349.34$ \\
Par0440-5244 & 593 & $-2.27\pm7.16$ & $7.23\pm5.21$  & $-0.97\pm4.72$ & $94.47\pm10.81$ & $153.45\pm290.15$ & $-318.25\pm1150.19$ \\[2mm]
\hline
Par & ID & \iaband\ & \ybband\ & \jband\ & \hband\ & 3.6\micron\ & 4.5\micron\ \\
\hline
Par0713+7405 & 95 & $38.80\pm28.05$   & $-3.33\pm14.72$ & $107.11\pm21.64$ & $183.20\pm23.86$ & $251.30\pm217.45$ & $277.66\pm661.93$ \\
Par0259+0032 & 194 & $-3.20\pm20.21$ & $-9.25\pm12.88$ & $148.17\pm23.77$\tablenotemark{a} & $440.55\pm40.74$ & \nodata & \nodata \\
Par0843+4114 & 120 & $-20.46\pm13.94$ & $21.41\pm16.82$  & $-1.83\pm9.23$ & $101.05\pm19.05$ & $-190.14\pm521.57$ & $-138.04\pm1263.79$ \\
Par0956-0450 & 684 & $-6.77\pm17.48$ & $-8.50\pm11.00$ & $10.04\pm12.52$ & $404.85\pm18.50$ & $16.36\pm481.12$ & $335.27\pm707.27$ \\
Par0926+4536 & 155 & $-4.68\pm12.92$ & $15.20\pm15.14$  & $-16.90\pm8.77$ & $102.65\pm18.64$ & $144.67\pm159.01$ & $240.19\pm746.29$ \\[2mm]
\hline
Par & ID & \vband\ & \ibband\ & \jbband\ & \hband\ & 3.6\micron\ & 4.5\micron\ \\
\hline
Par335 & 251 & \nodata & $-1.64\pm14.63$ & $14.54\pm17.31$ & $460.15\pm34.90$ & $77.47\pm719.45$ & \nodata \\
\enddata
\tablenotetext{a}{The \jband\ photometry quoted here for 
Par0259+0032\_194 is that measured after masking out the bad pixels as 
described in Section~\ref{sec:cand0259_194}.}
\tablecomments{The candidates are ordered by filter set, eather than in
redshift-order as in Table~\ref{tab:sample}.
}
\end{deluxetable}

\begin{deluxetable}{ccc}
\tablecaption{Measured Number Densities for $z\sim8.5-11$ \label{tab:lf}}
\movetabledown=5mm
\tablehead{
\colhead{UV Magnitude} & \colhead{Number Density} & \colhead{68\% Range} \\
\colhead{} & \colhead{($10^{-6}$ Mpc$^{-3}$)} & 
\colhead{($10^{-6}$ Mpc$^{-3}$)}}
\startdata
-23.15 & $<0.91$ &  \nodata \\
-23.05 & 1.02 & $0.68-1.49$ \\
-22.95 & 1.18 & $0.77-1.72$ \\
-22.85 & 1.33 & $0.88-1.95$ \\
-22.75 & 1.44 & $0.96-2.10$ \\
-22.65 & 1.48 & $0.97-2.16$ \\
-22.55 & 1.54 & $1.00-2.27$ \\
-22.45 & 1.66 & $1.08-2.49$ \\
-22.35 & 1.92 & $1.25-2.85$ \\
-22.25 & 2.21 & $1.41-3.27$ \\
-22.15 & 2.53 & $1.63-3.76$ \\
-22.05 & 2.93 & $1.91-4.37$ \\
-21.95 & 3.48 & $2.26-5.16$ \\
-21.85 & 4.22 & $2.74-6.32$ \\
-21.75 & 5.17 & $3.33-7.71$ \\
-21.65 & 6.35 & $4.09-9.46$ \\
-21.55 & 7.86 & $5.07-11.70$ \\
-21.45 & 9.98 & $6.48-14.83$ \\
-21.35 & 12.98 & $8.37-19.12$ \\
-21.25 & 16.87 & $11.00-24.73$ \\
-21.15 & 20.40 & $13.16-30.36$ \\
-21.05 & 23.44 & $15.17-34.63$ \\
-20.95 & 25.95 & $16.75-38.65$ \\
-20.85 & 28.98 & $18.64-43.52$ \\
-20.75 & 37.59 & $23.70-56.79$ \\
-20.65 & 50.39 & $31.65-76.55$ 
\enddata
\tablecomments{The number densities are taken as the median of the MCMC 
posterior at each magnitude step using the ``pseudo-binning'' technique 
of \citet{finkelstein2022a}. The 68\% range of the posterior distribution
listed in the third column is plotted as the dark blue shaded region in 
Figure~\ref{fig:lfs}. 
}
\end{deluxetable}

%%%%%%%%%%%%%%%%%%%%%%%%%%%%%%%%%%%%%%%%%%
\section{Rejected candidates} \label{app:rejections}

\begin{figure*}
\gridline{\fig{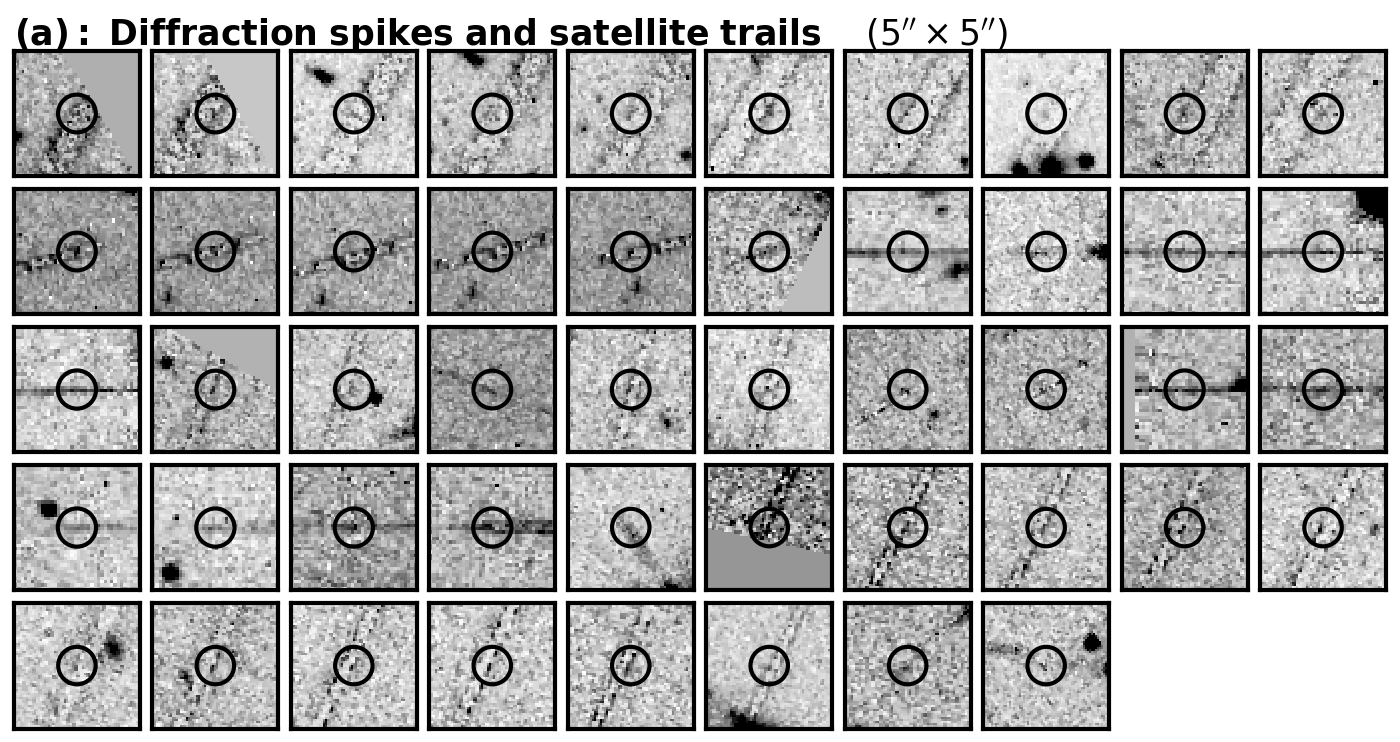}{0.75\textwidth}{}}
\vspace{-9mm}
\gridline{\fig{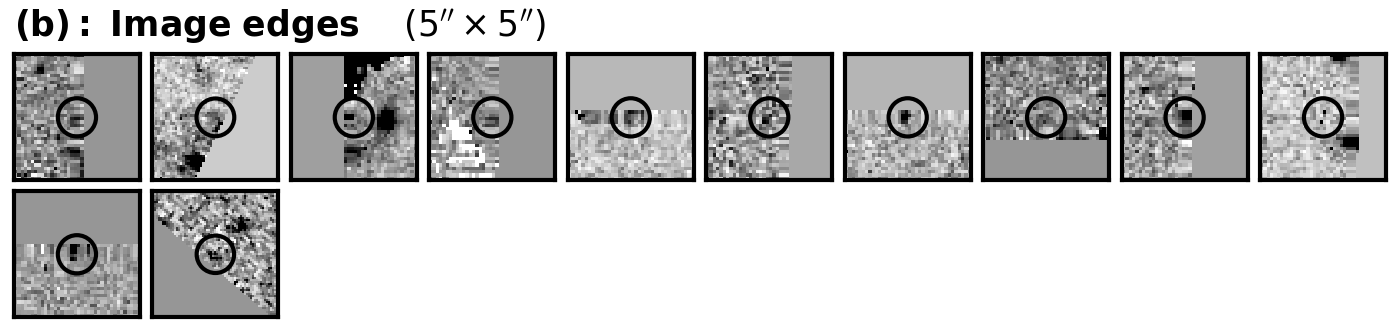}{0.75\textwidth}{}}
\vspace{-9mm}
\gridline{\fig{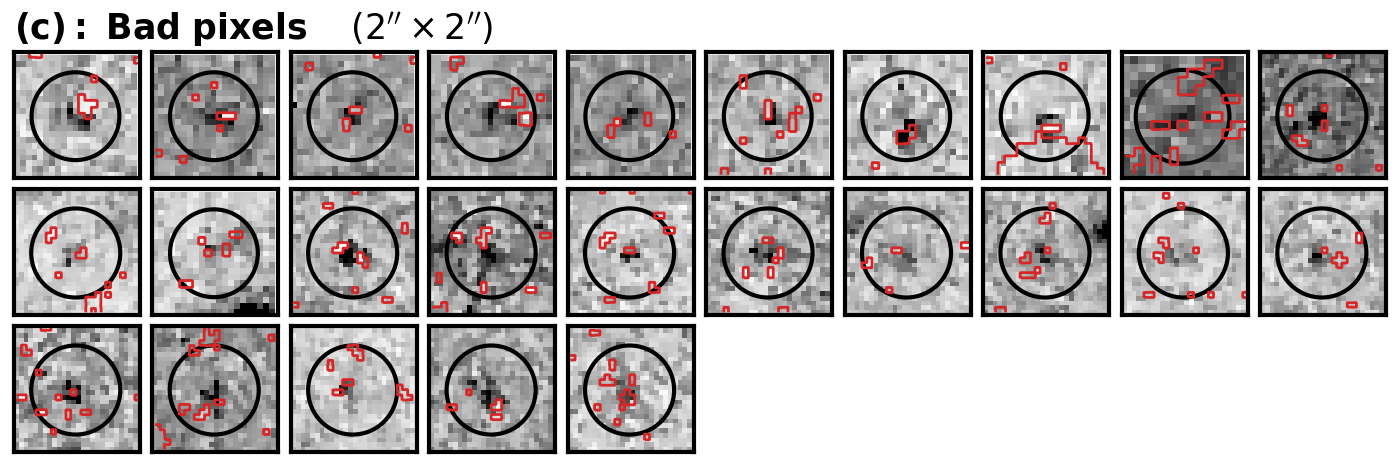}{0.75\textwidth}{}}
\vspace{-9mm}
\gridline{\fig{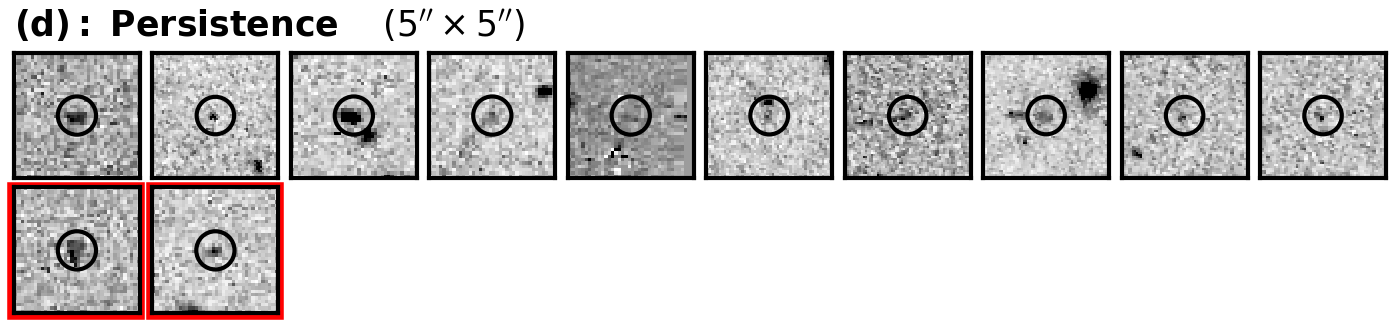}{0.75\textwidth}{}}
\caption{The \hband\ postage stamps of the 97 spurious sources rejected 
in the non-subjective categories during our source vetting 
(Sections~\ref{sec:inspection} and \ref{sec:persistence}). 
Group (a) shows $5\arcsec \times 5\arcsec$ stamps of the 48 sources that 
were rejected as satellite trail remnants or diffraction spikes through 
visual inspection.
Group (b) shows $5\arcsec \times 5\arcsec$ stamps of the 12 sources that
 were rejected along image edges or near 
image edges in regions of increased noise.  
Group (c) shows stamps of the 25 sources rejected 
as either bad pixels or sources with 
photometry contaminated by bad pixels through visual inspection of the RMS 
maps at the source positions.
We adpot a threshold of $RMS>0.1$, equivalent to weights of $<100$.
The bad pixels are identified by red contours. Each stamp is 2\arcsec\ on 
a side, zooming in on the source position to
highlight the source shapes.
Group (d) shows $5\arcsec \times 5\arcsec$ stamps of the 12 sources that 
were rejected as image persistence. 
Ten were identified based on analysis of counts in FLTs observed in
the 24 hours preceding the observation of the FLTs that contributed to the
mosaic. The remaining two (surrounded by red boxes) were identified in the
WISP survey based on the source position relative to a bright source.
Each stamp is displayed on a zscale interval (calculated using the pixels in 
the 5\arcsec\ stamp) with no image smoothing or
pixel interpolation. The $r=0\farcs75$ circles show the position of the
rejected candidate as a reference.
\label{fig:rejects}}
\end{figure*}

In this section, we show \hband\ postage stamps for all 166
rejected sources. These candidates were selected via the criteria discussed 
in Section~\ref{sec:selection} and then later rejected through 
persistence checks (Section~\ref{sec:persistence}), or visual 
inspection (Section~\ref{sec:inspection}).
All stamps are displayed on a linear stretch with a zscale interval that 
highlights pixel values near the image median. These stamp normalizations 
are calculated using the pixels in a $5\arcsec \times 5\arcsec$ box centered
on the source position. We note that on the reverse colormap adopted here, 
very negative values are displayed as white pixels. We do not apply any 
pixel interpolation or smoothing to these stamps, and use $r=0\farcs75$ 
circles to indicate the source positions.

\begin{figure}
\plotone{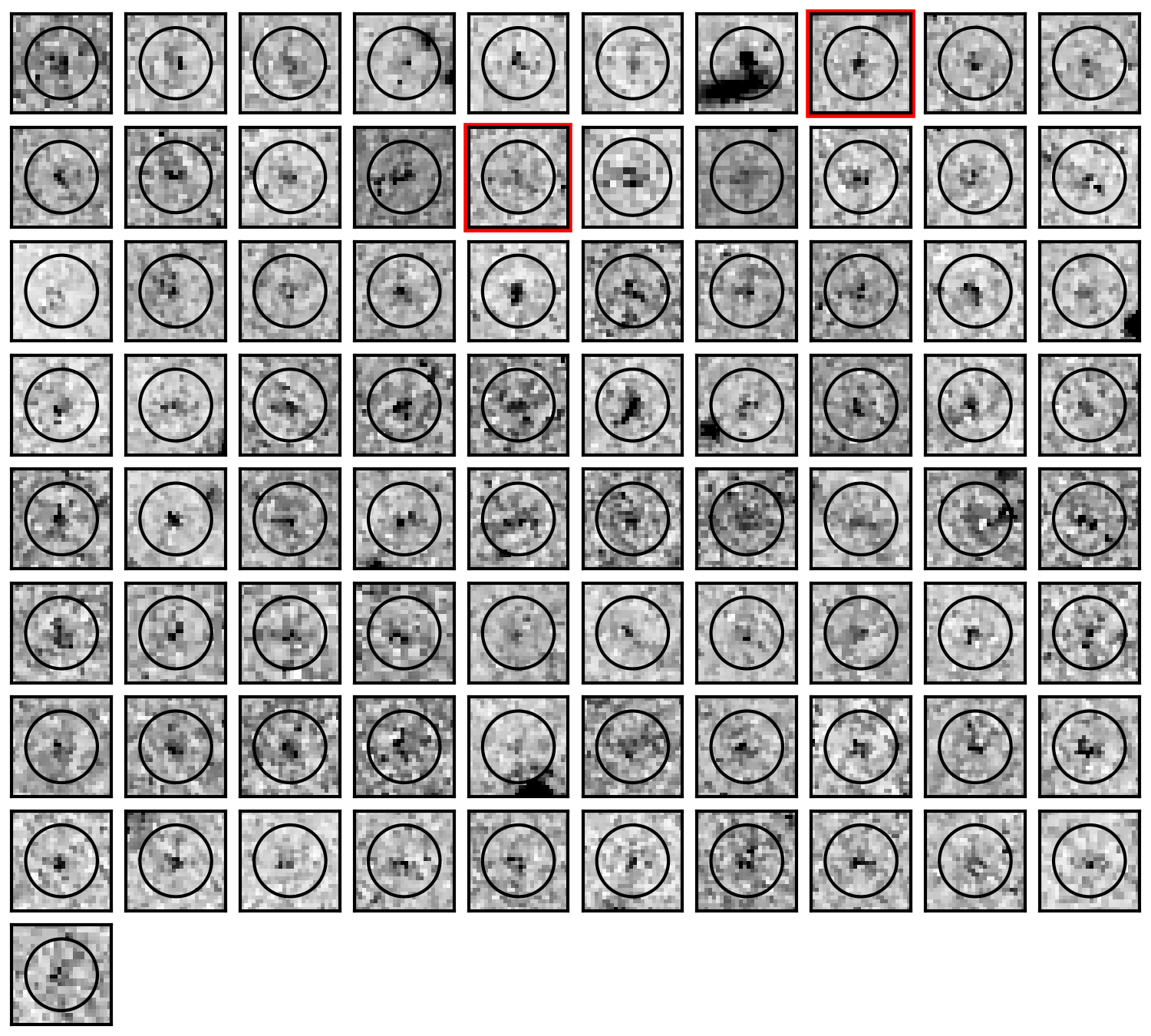}
\caption{The \hband\ postage stamps ($2\arcsec \times 2\arcsec$)
of the 81 sources that were rejected 
as having unreliable source shapes or morphologies. 
We note that this rejection is the only subjective part of the visual 
inspection and rejection process, and we quantify the biases introduced 
through these rejections using Figure~\ref{fig:rejection}.
The stamps are displayed in the same manner as those in 
Figure~\ref{fig:rejects}. The two candidates from \citetalias{bernard2016} 
that we rejected during visual inspection are identified by red boxes:
borg\_0240-1857\_25 in the top row (third from the right) and 
borg\_0456-2203\_1091 in the second row (fifth from the left).
\label{fig:reject_1d1e}}
\end{figure}

Of the 193 candidates that pass our selection criteria, the majority of 
them are spurious detections that we classify visually in four categories.
First, we reject 48 sources as satellite remnants or diffraction spikes 
(Figure~\ref{fig:rejects}a). As can be seen in 
Figure~\ref{fig:rejects}a, many of the sources in this category 
are detected as part of the same set of satellite trails.
Next, we reject 12 sources that are detected on or very close to an image edge
(Figure~\ref{fig:rejects}b). The sources in this category suffer from 
incomplete source extraction and unreliable photometry. 
Through inspection of the RMS maps at the position of each candidate, we 
identify 25 sources that are bad pixels in the RMS map (RMS$>$0.1, equivalent 
to a weight less than 100) or are contaminated by bad pixels (i.e., 
bad pixels are included in the isophotal area of the source, therefore 
impacting the photometry).
We show the central $2\arcsec \times 2\arcsec$ around each of 
these candidates in Figure~\ref{fig:rejects}c to zoom in on the 
morphologies of these spurious or contaminated sources, and indicate the 
locations of the bad pixels with red contours.
The final category of visually rejected sources are those that have been 
identified as hot pixels, detector artifacts, or sources with strange 
morphologies. This category is the only subjective part of our 
visual inspection, and we attempt to account for these rejections in our 
calculate of the effective volume (see Section~\ref{sec:veff}). We show the 
81 sources rejected in this category in Figure~\ref{fig:reject_1d1e}, 
zooming in on the celtral $2\arcsec \times 2\arcsec$ to highlight the 
morphologies.

Finally, we reject 12 candidates as contamination by image persistence. 
In these cases, a previous observation of a bright source left a residual 
charge on the detector that fades with time. If this leftover flux is 
detected in the \hband\ and is absent from the other filters, the persistence
will be selected as a high-redshift source. We have identified 10 
cases of persistence based on the counts registered on the detector at the 
position of the candidates prior to the candidate's observation. An additional 
2 WISP sources are identified as persistence due to their location 
relative to a bright source and the survey strategy of the WISP observations
(see Section~\ref{sec:persistence}).
We show all 12 cases of persistence in Figure~\ref{fig:rejects}d.

\end{document}